\documentclass[11pt,a4paper]{article}
\usepackage[margin=25mm]{geometry}
\usepackage{amsmath,amssymb}
\usepackage{graphicx}
\usepackage{booktabs,longtable,array,calc}
\usepackage[protrusion=false,expansion=false]{microtype}
\usepackage[T1]{fontenc}\usepackage{mathptmx}\usepackage{courier}
\usepackage[utf8]{inputenc}
\usepackage{textcomp}
\usepackage{newunicodechar}
\newunicodechar{−}{--}\newunicodechar{≥}{\ensuremath{\geq}}\newunicodechar{≤}{\ensuremath{\leq}}\newunicodechar{≈}{\ensuremath{\approx}}\newunicodechar{×}{\ensuremath{\times}}\newunicodechar{→}{\ensuremath{\rightarrow}}\newunicodechar{±}{\ensuremath{\pm}}\newunicodechar{Σ}{\ensuremath{\Sigma}}\newunicodechar{λ}{\ensuremath{\lambda}}\newunicodechar{μ}{\ensuremath{\mu}}\newunicodechar{σ}{\ensuremath{\sigma}}\newunicodechar{δ}{\ensuremath{\delta}}\newunicodechar{γ}{\ensuremath{\gamma}}\newunicodechar{β}{\ensuremath{\beta}}\newunicodechar{ε}{\ensuremath{\varepsilon}}\newunicodechar{τ}{\ensuremath{\tau}}\newunicodechar{Δ}{\ensuremath{\Delta}}\newunicodechar{ŷ}{\ensuremath{\hat{y}}}\newunicodechar{∈}{\ensuremath{\in}}\newunicodechar{⇔}{\ensuremath{\Leftrightarrow}}\newunicodechar{←}{\ensuremath{\leftarrow}}\newunicodechar{…}{\ldots}\newunicodechar{’}{'}\newunicodechar{–}{--}\newunicodechar{—}{---}\newunicodechar{“}{``}\newunicodechar{”}{''}
\usepackage[hidelinks,breaklinks]{hyperref}

\title{Input-Layer Starvation: Why Per-Layer Pruning Breaks IoT Intrusion Detectors}
\author{Md Anas Biswas\\ \small School of Computing, University of Portsmouth, Portsmouth, United Kingdom\\ \small ORCID 0009-0009-0113-5816 \quad up2082724@myport.ac.uk\\ \small\itshape Preprint, September 2026.\\ \small Code and results: \url{https://github.com/anasbiswas1/iot-trust-compression}}
\date{}
\begin{document}
\maketitle
\begin{abstract}

Intrusion detectors for small Internet-of-Things (IoT) devices are usually compressed by pruning and judged by overall accuracy. We show that this hides a severe class-level failure, find its cause, and give low-overhead prevention and repair. On CICIoT2023, a two-layer convolutional detector pruned with uniform layer-wise magnitude pruning at 80\% sparsity loses 16 points of accuracy but half of its macro-F1, the mean per-class F1 (0.542 to 0.271 over five independently trained models); 17 of 34 classes are materially damaged. Remaining weight count does not explain it: a perceptron and a transformer pruned to the same or fewer weights lose at most 0.096. The first layer does. It has 192 weights; uniform pruning leaves 38, 46\% of its 64 filters lose every input weight, and fine-tuning under that starvation leaves the running means of the first normalisation layer displaced by up to 0.8 standard deviations in a few surviving channels, on which the deployed model collapses. Protecting those 192 weights, or pruning globally at the same sparsity, prevents the collapse (loss 0.013); recomputing the normalisation statistics on unlabelled training data, with no weight changed, repairs it (loss 0.039) and returns the false-alert rate to 33\% (dense 29\%). Damage shows a strong increasing dose-response in first-layer sparsity, starving a perceptron\textquotesingle s input layer reproduces the collapse, and the pattern holds on TON\_IoT. The failure is misattribution and false alerts, not silent evasion: on validation-selected blind spots, uniformly pruned detectors misattribute 72\% of the traffic, against 50\% with the first layer protected and 47\% for the dense model.

\end{abstract}


\section*{1. Introduction}

Intrusion detection systems (IDS) watch network traffic and raise an alert when they see an attack. On small Internet-of-Things (IoT) devices the detector must be small too, so a trained network is usually pruned before deployment: most of its weights are set to zero and the rest are fine-tuned. The usual safety check is overall accuracy before and after. This paper shows that the check misses a severe failure in the detector most often used on IoT devices, finds the cause, and gives low-overhead prevention and repair.

Our detector is a one-dimensional convolutional network (a CNN that slides small filters along the list of traffic features) with two convolutional layers, trained on CICIoT2023 to tell 34 traffic types apart. Pruned with uniform layer-wise magnitude pruning, in which every layer loses the same fraction of its weights, its accuracy drops moderately while its macro-F1 halves and half of the classes are materially damaged. Attacks are still flagged, but as the wrong attack, and benign traffic is flooded with false alerts. Accuracy shows a degraded model, not which classes were lost or where the errors went.

Earlier work found that pruning damages rare classes and treated this as the price of compression. We asked whether it is the compression or the recipe. A multilayer perceptron and a small transformer pruned to the same or fewer remaining weights do not collapse; the CNN does. The difference is its first layer: a single-input-channel convolution with 192 weights. Uniform pruning at 80\% leaves it 38, disconnects almost half of its filters from the input, and starves everything downstream. We call this input-layer starvation. It is a partial form of the layer collapse described by Tanaka et al. \cite{r1}, occurring at an ordinary sparsity because the layer is tiny in absolute terms.

The cause was established by intervention. Protecting the 192 weights, pruning globally, sweeping first-layer sparsity, and starving a second architecture\textquotesingle s input layer all point the same way, and recomputing the normalisation statistics of a collapsed model, with no weight changed, repairs it. The contributions are:

\begin{enumerate}
\def\labelenumi{\arabic{enumi}.}
\item
  A causal account of per-class collapse in pruned intrusion detectors: remaining weight count does not explain it, and input-layer starvation under uniform allocation is established by protection, global pruning, a monotone dose-response and a positive control that induces the collapse in a resistant architecture, for one-shot and gradual pruning alike.
\item
  What the onset depends on: across first layers of 32, 64 and 128 filters, equal surviving-weight counts give very different damage, and the fraction of connected filters describes it better; uniform 80\% pruning leaves about half connected in every width. Live filters mediate the effect in the CNN; in the MLP neither live units nor covered features alone do.
\item
  The proximate failure: after fine-tuning under starvation, a few first-layer normalisation channels carry wrong running means, and the model degrades in proportion to how wrong they are. Re-estimating them from unlabelled training data recovers most of the CNN loss, less in the perceptron and on TON\_IoT, and changes protected or globally pruned models by nothing.
\item
  The security consequence, measured out of sample: on validation-selected blind spots the uniformly pruned detector\textquotesingle s misattribution exceeds the dense model\textquotesingle s by 25 points where the fixed recipes exceed it by 2 to 4, and its false-alert rate is double the dense rate, while its attack-to-benign rate does not rise.
\item
  Low-overhead prevention and repair: prevention by allocation (protect the first layer, or prune globally; size and latency unchanged, macro-F1 recovered 0.26, run-to-run variance from 0.164 to 0.012) and repair by recalibration (one pass of unlabelled training data through a deployed model).
\end{enumerate}

Every number in this paper is produced by a committed notebook and stored in a public repository, with pass criteria written before each experiment ran. Section 2 positions the work; Section 3 gives the data, models and protocol; Sections 4 and 5 report the collapse and its mechanism; Section 6 tests a second dataset; Section 7 measures the security consequence; Section 8 reports the fixes and their cost; Sections 9 and 10 discuss limitations and conclude.

\section*{2. Related work}

\subsection*{2.1 Pruning and the allocation of sparsity across layers}

Pruning removes weights from a trained network and then fine-tunes what remains. The simplest and most widely used rule keeps the weights with the largest absolute values and removes the rest \cite{r2,r3}. Two allocation choices exist. Per-layer (uniform) pruning removes the same fraction from every layer; global pruning applies one magnitude threshold across the whole network, so layers are pruned at different rates. Systematic comparisons find that global allocation usually gives better accuracy at the same total sparsity \cite{r4,r5}. It is also common practice to exempt or lightly prune the first and last layers, and sparse-training methods encode the same intuition by giving small layers a lower target sparsity \cite{r6}. We do not claim these heuristics as new. What we add is a causal account of what goes wrong when they are not followed on a detector whose first layer is tiny, and a measurement of the security consequence.

\subsection*{2.2 Pruning damages some classes more than others}

Hooker et al. \cite{r7,r8} showed that pruned image classifiers keep their overall accuracy while a small set of classes and atypical examples bear most of the loss, and that these losses concentrate on rare and hard cases. Tran et al. \cite{r9} gave a theoretical account in terms of gradient norms and distance to the decision boundary. Paganini \cite{r10} reported the same imbalance across many models and classes, and Good et al. \cite{r11} proved that recall distortion is inherent to pruning even on balanced data. This literature treats disparate damage as an intrinsic price of removing capacity. Its experiments differ from ours in ways that matter for interpretation: the models are large image networks whose first convolution has thousands of weights, the pruning is gradual and applied during training with pruned weights allowed to recover, and the standard implementation leaves the first layer dense. Our finding does not overturn those results. It shows that in small tabular detectors under a per-layer recipe, one-shot or gradual, most of the damage has a different and removable cause, and it separates that recipe-induced component from the smaller intrinsic one.

\subsection*{2.3 Layer collapse and conservation}

Tanaka et al. \cite{r1} named layer collapse, the failure in which pruning removes an entire layer and leaves a network that cannot be trained, and showed that it follows from a conservation law for synaptic saliency, a gradient-weighted importance score: the smallest layer has the largest score per weight. The conserved quantity is not raw magnitude and the argument changes under normalisation layers, so the theorem does not by itself say how a magnitude threshold treats a small layer. Two facts are nevertheless established: global magnitude pruning at very high sparsity can itself remove whole layers, which Gupta et al. \cite{r12} fix with a per-layer minimum retained count; and a rule that removes the same fraction from every layer pushes the smallest layer towards collapse first, because a fixed fraction of a tiny layer is very few weights. Later sparse-training work describes the same danger at the neuron level, a unit that keeps its outgoing weights but loses every incoming one, and repairs it by re-adding random connections \cite{r13}. Two normalisation results are close to ours. Saikumar and Varghese \cite{r14} identify signal collapse, a loss of activation variance across layers after one-shot pruning, as the reason pruned networks fail, and repair it with REFLOW without updating any trainable weight; Wu and Johnson \cite{r15} analyse how batch normalisation\textquotesingle s running estimates diverge from batch statistics and when the estimates mislead at inference. Our recalibration result belongs to that family and does not claim to invent it. The contribution here is upstream and downstream of it: which allocation rule starves which layer, and what the resulting failure costs in intrusion-detection terms.

\subsection*{2.4 Compression in intrusion detection}

Lightweight intrusion detection for IoT and edge devices is an active area, and one-dimensional convolutional networks over flow features are among the most common designs \cite{r16,r17,r18}. These studies report overall accuracy, weighted F1, model size or latency after compression, and rarely report per-class results.

Of the recent IoT intrusion-detection papers that prune a neural detector and state how, Broggi et al. \cite{r19} set each hidden layer\textquotesingle s pruning as a percentage of that layer, mention no exemption and report only a macro-averaged F1; Dehrouyeh et al. \cite{r20} apply the gradual schedule of Zhu and Gupta \cite{r3} to a 65\% target, which the standard toolkit applies layer by layer, do not state how the input layer is treated and report frequency-weighted metrics; Park et al. \cite{r21} apply structured pruning without stating the allocation and report binary attack recall. The toolkit itself demonstrates per-module pruning first and global pruning separately \cite{r22}, and its own global example leaves the first convolution almost untouched under a 20\% budget. In this small sample no paper reports exempting the input layer and none reports per-class results; not reporting is not the same as not doing (Section 9.4). Our work connects two observations already in the literature, that aggregate accuracy hides minority-attack failures and that per-layer allocation is what the toolkit teaches first.

\subsection*{2.5 Security of compressed models}

Work on the security of compression has studied backdoors that activate after quantisation or pruning \cite{r23} and adversarial inputs that exploit the gap between a full model and its compressed copy \cite{r24}; both need an attacker who modifies the model or perturbs the input. The failure we study needs neither: ordinary traffic of a type the uniformly pruned detector cannot attribute is misfiled. It is a weaker attacker with a weaker goal, attribution failure rather than evasion, and we report it as such.

\subsection*{2.6 Position of this paper}

The novelty is not that first layers deserve protection, which practitioners already know, nor that pruning hurts rare classes, which is established. Gupta et al. \cite{r12} already show that global magnitude pruning can collapse a layer and that a per-layer minimum retained count prevents it. Repairing a pruned network through its normalisation state without touching weights is also known \cite{r14}. What is new here is the causal chain that joins these facts to a security consequence in intrusion detection, listed in the contributions of Section 1: elimination, intervention, dose-response, positive control, replication, decomposition, identification of the displaced statistics that carry the deployed failure, and a measured consequence with low-overhead prevention and repair. A companion study by the author (SABER, in preparation) reports a related normalisation-state failure at deployment in a distillation pipeline. The two studies share infrastructure: SABER reuses this study\textquotesingle s frozen CICIoT2023 split and its shallow CNN baselines as teachers. They do not share pruned models, interventions or result tables; every normalisation result reported here comes from notebooks 46 to 48 of this paper\textquotesingle s repository, and SABER\textquotesingle s findings are not used as evidence for any claim in this paper.

\section*{3. Materials and method}

\subsection*{3.1 Datasets}

CICIoT2023 \cite{r25} is the primary dataset. After cleaning and the repository\textquotesingle s stratified subsampling, we use 3,661,696 flow records with 39 numeric features and 34 classes: benign traffic and 33 attack types spanning DDoS, DoS, reconnaissance, spoofing, credential, web and Mirai-botnet attacks. TON\_IoT \cite{r26,r27,r28} is the second dataset, used to test whether the mechanism generalises: 93,644 records with 30 features after integer encoding of nominal fields, and 10 classes (normal traffic and nine attack types). Labels are used exactly as published.

\subsection*{3.2 Splits and leakage discipline}

Random splits of flow data leak information, because neighbouring flows from the same capture are near-duplicates. We therefore split CICIoT2023 by provenance order within each capture: earlier records train, later records validate and test, in the proportions 70/15/15 (2,563,172 / 549,253 / 549,271 records). Feature standardisation is fitted on the training portion only. TON\_IoT ships as a single curated file whose original capture order is not guaranteed, so its split follows the file\textquotesingle s row order within each class; we state this because it is weaker than a true temporal split. The same split is frozen for every model and every seed.

\subsection*{3.3 Architectures}

Three architectures are used (Table 1). The primary detector is the CNN1D common in lightweight IoT intrusion detection: the 39 features are treated as a sequence of length 39 with one input channel, and two convolutional layers with kernel size 3 are followed by global pooling and a linear classification head. Its first layer, conv.0, maps one input channel to 64 filters with kernel 3 and therefore has exactly 192 weights; this number is central to the paper. The multilayer perceptron (MLP) is a fully connected network with hidden widths 256 and 128; its input layer has $39 \times 256$ = 9,984 weights. The FT-Transformer embeds each feature as a 96-dimensional token and applies three attention layers; its architecture settings were recovered from the trained checkpoint and confirmed by reproducing its validation macro-F1 (0.524 to 0.532 across seeds, Section 4.2). All models use batch normalisation or layer normalisation as appropriate and are trained with square-root inverse-frequency class weights, which give rare classes a mild boost without hiding their fragility.

\begingroup\scriptsize\setlength{\tabcolsep}{3pt}\renewcommand{\arraystretch}{1.05}\begin{longtable}[]{@{}
  >{\raggedright\arraybackslash}p{(\columnwidth - 6\tabcolsep) * \real{0.1816}}
  >{\raggedright\arraybackslash}p{(\columnwidth - 6\tabcolsep) * \real{0.4487}}
  >{\raggedright\arraybackslash}p{(\columnwidth - 6\tabcolsep) * \real{0.1389}}
  >{\raggedright\arraybackslash}p{(\columnwidth - 6\tabcolsep) * \real{0.2308}}@{}}
\toprule\noalign{}
\begin{minipage}[b]{\linewidth}\raggedright
\textbf{Architecture}
\end{minipage} & \begin{minipage}[b]{\linewidth}\raggedright
\textbf{Layers (weights per layer)}
\end{minipage} & \begin{minipage}[b]{\linewidth}\raggedright
\textbf{Total parameters}
\end{minipage} & \begin{minipage}[b]{\linewidth}\raggedright
\textbf{Role in the paper}
\end{minipage} \\
\begin{minipage}[b]{\linewidth}\raggedright
CNN1D (primary)
\end{minipage} & \begin{minipage}[b]{\linewidth}\raggedright
conv.0: Conv1d 1 to 64, kernel 3 (192); conv.3: Conv1d 64 to 128, kernel 3 (24,576); head: Linear 128 to 34 (4,352)
\end{minipage} & \begin{minipage}[b]{\linewidth}\raggedright
29,730
\end{minipage} & \begin{minipage}[b]{\linewidth}\raggedright
The detector that collapses under the uniform layer-wise recipe
\end{minipage} \\
\begin{minipage}[b]{\linewidth}\raggedright
MLP
\end{minipage} & \begin{minipage}[b]{\linewidth}\raggedright
body.0: Linear 39 to 256 (9,984); body.3: Linear 256 to 128 (32,768); head: Linear 128 to 34 (4,352)
\end{minipage} & \begin{minipage}[b]{\linewidth}\raggedright
about 50,000
\end{minipage} & \begin{minipage}[b]{\linewidth}\raggedright
Co-equal arm (validation gate passed); capacity control; positive control by input starvation
\end{minipage} \\
\begin{minipage}[b]{\linewidth}\raggedright
FT-Transformer (small)
\end{minipage} & \begin{minipage}[b]{\linewidth}\raggedright
token embedding (96 per feature), 3 attention layers with 8 heads, feed-forward width 192, classification head
\end{minipage} & \begin{minipage}[b]{\linewidth}\raggedright
227,938
\end{minipage} & \begin{minipage}[b]{\linewidth}\raggedright
Capacity control without a locality prior; weaker baseline, reported with that caveat
\end{minipage} \\
\midrule\noalign{}
\endhead
\bottomrule\noalign{}
\endlastfoot
\end{longtable}\endgroup

\emph{Table 1. Architectures and layer sizes. The CNN\textquotesingle s first layer has 192 weights, 0.7\% of its prunable weights.}

\subsection*{3.4 Training, pruning and fine-tuning}

Baselines are trained with Adam (learning rate 0.001, batch size 4,096) for up to 40 epochs with early stopping on validation macro-F1 (patience 6). Pruning follows the standard toolkit call, \path{torch.nn.utils.prune.l1_unstructured}, applied to the weight matrix of every convolutional and linear layer; biases and normalisation parameters are never pruned. After pruning, the network is fine-tuned for 8 epochs (learning rate 0.0005, batch size 4,096) with gradient masks that keep pruned weights at zero, and a final hard mask is applied so that saved models contain exact zeros. Table 2 lists the recipes. The uniform layer-wise recipe is the per-layer rule; the two fixed recipes either exempt conv.0 or replace the per-layer rule with a single global threshold at the same total sparsity. For the capacity sweeps of the MLP and the transformer, the classification head (and the transformer\textquotesingle s token embedding) was held at 80\% while the body was pruned at the stated sparsity, so that extreme sparsity was never applied to tiny layers by accident; the transformer was pruned with the same code extended to its attention projection matrices. Achieved per-layer sparsity was checked after every run and is reported with the results.

In symbols, with $W_l$ the weight tensor of layer l and $Q_s(\cdot)$ the s-quantile of the absolute values, the mask, the achieved sparsity and the gradual schedule are

\begin{equation*}m_{l,i}\  = \ 1\left\lbrack |w_{l,i}|\  > \ \tau_{l} \right\rbrack,\ \ \ \tau_{l}\  = \ Q_{s}\left( |W_{l}| \right)\ \ \ (per - \text{layer}),\ \ \ \tau_{l}\  = \ Q_{s}\left( |W| \right)\ \ \ (\text{global})\tag{1}\end{equation*}
\begin{equation*}s_{l}\  = \ 1\  - \ \frac{\Sigma_{i}m_{l,i}}{n_{l}},\ \ \ n_{l}\  = \ \text{number}\ of\ \text{weights}\ in\ \text{layer}\ l\tag{2}\end{equation*}
\begin{equation*}s_{t}\  = \ s_{f}\left( 1\  - \ {(1\  - \ t/T)}^{3} \right),\ \ \ t\  = \ 0,\ \ldots,\ T\tag{3}\end{equation*}
so that per-layer pruning uses one threshold per layer and global pruning one threshold for the whole network, and the gradual schedule reaches the final sparsity $s_f$ at step T along a cubic curve.

\begingroup\scriptsize\setlength{\tabcolsep}{3pt}\renewcommand{\arraystretch}{1.05}\begin{longtable}[]{@{}
  >{\raggedright\arraybackslash}p{(\columnwidth - 4\tabcolsep) * \real{0.2457}}
  >{\raggedright\arraybackslash}p{(\columnwidth - 4\tabcolsep) * \real{0.5235}}
  >{\raggedright\arraybackslash}p{(\columnwidth - 4\tabcolsep) * \real{0.2308}}@{}}
\toprule\noalign{}
\begin{minipage}[b]{\linewidth}\raggedright
\textbf{Recipe}
\end{minipage} & \begin{minipage}[b]{\linewidth}\raggedright
\textbf{What is pruned and how}
\end{minipage} & \begin{minipage}[b]{\linewidth}\raggedright
\textbf{Total sparsity reached}
\end{minipage} \\
\begin{minipage}[b]{\linewidth}\raggedright
Default, per-layer uniform 80\%
\end{minipage} & \begin{minipage}[b]{\linewidth}\raggedright
Each weight matrix (every Linear and Conv1d) loses its 80\% smallest-magnitude weights independently; biases and normalisation layers untouched
\end{minipage} & \begin{minipage}[b]{\linewidth}\raggedright
78.4\% of all parameters
\end{minipage} \\
\begin{minipage}[b]{\linewidth}\raggedright
First layer protected
\end{minipage} & \begin{minipage}[b]{\linewidth}\raggedright
Same as default, but conv.0 is left dense (192 weights)
\end{minipage} & \begin{minipage}[b]{\linewidth}\raggedright
77.8\%
\end{minipage} \\
\begin{minipage}[b]{\linewidth}\raggedright
Global magnitude 80\%
\end{minipage} & \begin{minipage}[b]{\linewidth}\raggedright
One magnitude threshold across all weight matrices so that 80\% of prunable weights are removed in total; small layers with larger weights keep more
\end{minipage} & \begin{minipage}[b]{\linewidth}\raggedright
78.4\% (identical to default)
\end{minipage} \\
\begin{minipage}[b]{\linewidth}\raggedright
Dose sweep (Section 5)
\end{minipage} & \begin{minipage}[b]{\linewidth}\raggedright
conv.0 pruned at 0/20/50/80/90/95\% while conv.3 and head stay at 80\%
\end{minipage} & \begin{minipage}[b]{\linewidth}\raggedright
varies by design
\end{minipage} \\
\begin{minipage}[b]{\linewidth}\raggedright
Hand-built masks (Section 5)
\end{minipage} & \begin{minipage}[b]{\linewidth}\raggedright
conv.0 mask constructed at a fixed number of surviving weights with the survivors spread across many filters or concentrated in few
\end{minipage} & \begin{minipage}[b]{\linewidth}\raggedright
varies by design
\end{minipage} \\
\begin{minipage}[b]{\linewidth}\raggedright
Gradual with recovery (Section 5.2)
\end{minipage} & \begin{minipage}[b]{\linewidth}\raggedright
Sparsity follows the cubic schedule of Zhu and Gupta from 0 to 80\% over eight fine-tuning epochs, the mask is recomputed from current weights every 100 steps and training is dense between updates so pruned weights can regrow (a regrowth-enabled variant; the original method masks gradients); two fixed-mask epochs follow. Per-layer and global variants.
\end{minipage} & \begin{minipage}[b]{\linewidth}\raggedright
80\%
\end{minipage} \\
\midrule\noalign{}
\endhead
\bottomrule\noalign{}
\endlastfoot
\end{longtable}\endgroup

\emph{Table 2. Pruning recipes. All one-shot recipes share the same masked fine-tuning; the gradual recipe trains as described in its row.}

\subsection*{3.5 Paired independent seeds}

Every compression result is reported over five independently trained baseline models (seeds 0 to 4), each pruned from its own weights, so that each pair of baseline and pruned model is independent of the others. The split is frozen; only initialisation and data order vary. Masks from different seeds share only about 12\% of their surviving weights, so agreement across seeds is agreement across genuinely different networks. Means and standard deviations (sd) across the five seeds are reported throughout, and seeds are the unit of replication.

\subsection*{3.6 What counts as a damaged class}

Per-class recall (the fraction of a class\textquotesingle s test records the model labels correctly) is compared between each baseline and its pruned version. A class is materially damaged in one seed if its recall falls by at least 0.10 and by more than twice the seed-to-seed standard deviation of its baseline validation recall, so that near-ceiling classes with tiny natural variation are not counted on a trivial drop. A class is counted as damaged for a recipe if this happens in at least three of the five seeds. With TP, FP and FN the per-class true positives, false positives and false negatives on the test partition, C the number of classes and σ\_c\^{}val the seed-to-seed standard deviation of class c\textquotesingle s baseline validation recall:

\begin{equation*}r_{c}\  = \ \frac{{TP}_{c}}{{TP}_{c}\  + \ {FN}_{c}},\ \ \ P_{c}\  = \ \frac{{TP}_{c}}{{TP}_{c}\  + \ {FP}_{c}},\ \ \ {F1}_{c}\  = \ \frac{2\ P_{c}r_{c}}{P_{c}\  + \ r_{c}}\tag{4}\end{equation*}
\begin{equation*}\text{macro} - F1\  = \ \frac{1}{C}\Sigma_{c}{F1}_{c}\tag{5}\end{equation*}
\begin{equation*}\Delta_{c}\  = \ r_{c}^{dense}\  - \ r_{c}^{pruned};\ \ \ \text{damaged}\ in\ a\ \text{seed}\ \  \Leftrightarrow \ \ \Delta_{c}\  \geq \ 0.10\ \ and\ \ \Delta_{c}\  > \ 2\ \sigma_{c}^{val}\tag{6}\end{equation*}
\subsection*{3.7 Capacity axis and mechanism tests}

To compare architectures at equal remaining capacity, every pruned model records its number of remaining non-zero prunable weights, and all models are placed on that axis. Four manipulations of the input layer follow. The dose-response sweeps conv.0 sparsity with the rest of the network at 80\%. The positive control prunes the MLP\textquotesingle s input layer to the CNN\textquotesingle s surviving counts (1,997, 998, 192, 96 and 38) with the rest at 80\%. Hand-built masks fix the surviving conv.0 weights at 38 or 96 and either spread them over as many filters as possible (each filter\textquotesingle s largest weight first, then the largest remaining weights) or concentrate them in as few as possible (whole filters in order of total magnitude, with any remainder in one partial filter); a second family keeps a K-unit by F-feature block of the MLP\textquotesingle s input layer at a fixed count T = $K \times F$ (T = 96 and 192). The width test repeats the conv.0 sweep on 32- and 128-filter versions of the CNN with overlapping surviving counts.

Connected channels are the filters or units with at least one surviving input weight. For a layer of C channels with k weights each (N = Ck) of which T survive, the expected number of connected channels under uniformly random survival, the probability that one filter stays connected under uniform pruning at sparsity s, and two measures of effective dimensionality (how many independent directions a representation uses, computed from the eigenvalues $\lambda_i$ of its covariance on 20,000 validation records for 160 runs) are

\begin{equation*}E\lbrack L\rbrack\  = \ C\ \left( 1\  - \ \frac{C(N\  - \ k,\ T)}{C(N,\ T)} \right),\ \ \ C(n,\ t)\ the\ \text{binomial coefficient}\tag{7}\end{equation*}
\begin{equation*}p_{connected}(s)\  \approx \ 1\  - \ s^{k},\ \ \ k\  = \ \text{kernel width}\ (3\ \text{here})\tag{8}\end{equation*}
\begin{equation*}PR\  = \ \frac{\left( \Sigma_{i}\lambda_{i} \right)\ ^{2}}{\Sigma_{i}\lambda_{i}^{2}},\ \ \ {rank}_{99}\  = \ min\{\ k\ :\ \Sigma_{i \leq k}\lambda_{i}\  \geq \ 0.99\ \Sigma_{i}\lambda_{i}\ \}\tag{9}\end{equation*}
Representation probes test whether class information survives inside the pruned network even when its output collapses: a logistic-regression classifier for each class is fitted on the penultimate-layer features of training and validation records (capped at 8,000 per class) and evaluated on the untouched test partition; the score is the area under the ROC curve (AUC). Head refitting trains a fresh dense linear classifier on the frozen pruned body using validation data. Normalisation recalibration is defined in Section 3.11.

\subsection*{3.8 Security semantics and threat model}

Each of the 34 classes is assigned to one of eight alert families (benign, DoS, DDoS, Mirai and other malware, reconnaissance, spoofing/MITM, credential, web application; full mapping in Appendix S3), a transparent proxy for how an analyst would triage an alert. DoS and DDoS are kept separate because they call for different responses. For attack records we report the misattribution rate (predicted class is not the true class), the attack-to-benign rate (an attack labelled benign, which is silent evasion), and the cross-family rate (an attack labelled as an attack of a different family, which misdirects the response). For benign records we report the benign-to-attack rate, the false-alert load. With A the attack records, B the benign records, b the benign class, f(.) the family of a class and ŷ the prediction:

\begin{equation*}\text{misattribution}\  = \ \frac{1}{|A|}\Sigma_{i \in A}\ 1\left\lbrack ŷ_{i}\  \neq \ y_{i} \right\rbrack,\ \ \ \text{attack} - to - \text{benign}\  = \ \frac{1}{|A|}\Sigma_{i \in A}\ 1\left\lbrack ŷ_{i}\  = \ b \right\rbrack\tag{10}\end{equation*}
\begin{equation*}\text{cross} - \text{family}\  = \ \frac{1}{|A|}\Sigma_{i \in A}\ 1\left\lbrack f(ŷ_{i})\  \neq \ f(y_{i}),\ \ ŷ_{i}\  \neq \ b \right\rbrack,\ \ \ \text{benign} - to - \text{attack}\  = \ \frac{1}{|B|}\Sigma_{i \in B}\ 1\left\lbrack ŷ_{i}\  \neq \ b \right\rbrack\tag{11}\end{equation*}
\begin{equation*}E_{recipe}\  = \ {misattribution}_{recipe}\  - \ {misattribution}_{dense}\ \ \ (\text{identical blind} - \text{spot traffic})\tag{12}\end{equation*}
The excess in Eq. (12) is the quantity reported in Section 7 (Section 7.3 explains why). The attacker in Section 7 knows only that the deployed detector is a CNN pruned with the uniform layer-wise recipe; their blind-spot set is estimated on attacker-side models (classes that dense detectors catch on validation records, mean recall at least 0.5, but uniformly pruned detectors miss there, mean recall below 0.2) and evaluated on untouched test records of held-out models over all ten leave-two-seeds-out folds. The same traffic is evaluated on held-out models of every recipe, so the attacker\textquotesingle s advantage is measured conditionally on the recipe; an earlier version that selected on test records is retained in the supplement as the leaked variant. Because the cross-family rates depend on the mapping, the evaluation is repeated under three alternative mappings.

\subsection*{3.9 Deployment cost}

Model size is reported as the serialized state-dictionary size in bytes and as an index-value sparse payload estimate: a tensor that contains any zero is charged 8 bytes per non-zero weight (4-byte value plus 4-byte index) and a fully dense tensor 4 bytes per weight; biases and normalisation parameters are dense. Because the convention matters when a layer is left fully dense, the uniform 8-byte figure is also reported where it differs. Latency and throughput are measured on a single CPU thread with 25 warm-up runs and 100 timed repeats at batch sizes 1, 32, 256 and 1,024, on the dense PyTorch backend actually used, without assuming that unstructured zeros give any speed-up.

\subsection*{3.10 Statistics and pre-stated gates}

Inference rests on effect sizes across seeds rather than significance tests. Key contrasts are reported with 95\% intervals from a seed-level bootstrap (20,000 resamples of the five paired seeds) together with the per-seed minimum and maximum. Dose-response monotonicity is summarised by the Spearman rank correlation between sparsity and loss, computed over dose means and over all individual runs; the interval for the latter resamples whole seed trajectories, since runs from one seed are not independent. The threat-model folds reuse the same five models, so their spread describes variation across overlapping folds, not independent deployments. Every experiment in Sections 4 to 8 was run by a notebook whose pass criteria were written into the notebook before it ran; each notebook writes a verdict file that records the outcome whichever way it fell, and the criteria that were not met are reported in the text. All notebooks, result tables and verdict files are in the public repository.

\subsection*{3.11 Normalisation statistics}

Each batch-normalisation layer standardises every channel with a running mean $\mu$ and variance $\sigma^2$ and re-scales with learned γ and β (Eq. 13). During training the running statistics are exponential moving averages of the batch statistics with momentum m (Eq. 14, default m = 0.1). Recalibration replaces them by a cumulative average over T = 200 batches of 4,096 training records passed through the fine-tuned model in training mode, with no weight updated and no labels used. The sensitivity analysis of Section 5.9 moves the statistics linearly from their saved to their recalibrated values (Eq. 15) and reports, per channel, the saved mean\textquotesingle s displacement in units of the recalibrated standard deviation (Eq. 16).

\begin{equation*}y\  = \ \gamma\ \frac{x\  - \ \mu}{\sqrt{\sigma^{2}\  + \ \varepsilon}}\  + \ \beta\tag{13}\end{equation*}
\begin{equation*}EMA:\ \ \mu\  \leftarrow \ (1\  - \ m)\ \mu\  + \ m\ \mu_{B};\ \ \ \ \text{cumulative}:\ \ \mu\  = \ \frac{1}{T}\Sigma_{t}\mu_{B,t}\ \ \ (and\ \text{likewise}\ for\ \sigma^{2})\tag{14}\end{equation*}
\begin{equation*}\mu(t)\  = \ (1\  - \ t)\ \mu_{saved}\  + \ t\ \mu_{recal},\ \ \ \sigma^{2}(t)\  = \ (1\  - \ t)\ \sigma_{saved}^{2}\  + \ t\ \sigma_{recal}^{2},\ \ \ t\  \in \ \lbrack 0,\ 1\rbrack\tag{15}\end{equation*}
\begin{equation*}\delta_{c}\  = \ \frac{\mu_{saved,c}\  - \ \mu_{recal,c}}{\sqrt{\sigma_{recal,c}^{2}\  + \ \varepsilon}}\tag{16}\end{equation*}
\section*{4. The collapse, and what it is not}

4.1 The uniform layer-wise recipe destroys half of the detector\textquotesingle s macro-F1 while accuracy falls by 16 points

Across five independently trained CNN baselines, macro-F1 (Eq. 5) is 0.542 +/- 0.006. After pruning with the uniform layer-wise recipe at 80\% sparsity and fine-tuning, it is 0.271 +/- 0.164, and 17 of the 34 classes are materially damaged in at least three of five seeds. Overall accuracy falls from 0.708 to 0.552; what the aggregate does not show is that the damage is concentrated: some classes lose almost all of their recall (for example DoS-UDP\_Flood from 0.68 to 0.00 and DoS-HTTP\_Flood from 0.87 to 0.21 in the anchor model) while large classes are barely affected. The seed-to-seed standard deviation of 0.164 is itself a finding: with the uniform layer-wise recipe, the outcome of pruning the same architecture on the same data depends on the training seed (best seed 0.405, worst seed 0.009).

The damage is not silent evasion. The attack-to-benign rate, the fraction of attack records the detector labels as benign, falls from 4.35\% in the dense model to 0.88\% after pruning. What rises is misattribution: the fraction of attack records given the exact right label falls from 70.9\% to 57.4\%, substitutions across attack families rise from 15.7\% to 26.2\%, and the fraction of benign records flagged as attacks rises from 32.1\% to 82.1\%. A binary malicious-versus-benign evaluation inverts the verdict entirely: the pruned model\textquotesingle s binary attack F1 (0.972) is slightly higher than the dense model\textquotesingle s (0.969), even as balanced accuracy falls from 0.818 to 0.585. An evaluator who checks only whether attacks are detected would rate the collapsed model an improvement.

\subsection*{4.2 It is not lost capacity}

The natural explanation is that 80\% of the weights was too many to remove. Table 3 shows that remaining weight count alone cannot be the explanation. The MLP, a co-equal arm whose validation band (0.567 to 0.593) overlaps the CNN\textquotesingle s (0.561 to 0.575), pruned to 5,145 remaining weights (body at 90\%, head at 80\%), slightly fewer than the CNN\textquotesingle s 5,823 under uniform 80\% pruning, loses 0.013 macro-F1 with no class damaged; at 3,007 weights (body at 95\%), about half the CNN\textquotesingle s remainder, it loses 0.022 and still no class is damaged. The transformer, at 7,305 and 3,987 weights, loses 0.070 and 0.096 with a gentle slope. The CNN loses 0.271 at 5,823. The transformer is the weaker classifier of the three (validation 0.524 to 0.532, below the CNN\textquotesingle s band), which argues against the capacity explanation rather than for it: a weaker model with fewer remaining weights should be the more fragile one. At its matched-capacity cell its probes stay at 0.93 to 0.99 and a head refit recovers only 43\% of its small loss, so its degradation is not concentrated in the final layer as the CNN\textquotesingle s is.

\begingroup\scriptsize\setlength{\tabcolsep}{3pt}\renewcommand{\arraystretch}{1.05}\begin{longtable}[]{@{}
  >{\raggedright\arraybackslash}p{(\columnwidth - 10\tabcolsep) * \real{0.1603}}
  >{\raggedright\arraybackslash}p{(\columnwidth - 10\tabcolsep) * \real{0.1603}}
  >{\raggedright\arraybackslash}p{(\columnwidth - 10\tabcolsep) * \real{0.1816}}
  >{\raggedright\arraybackslash}p{(\columnwidth - 10\tabcolsep) * \real{0.1496}}
  >{\raggedright\arraybackslash}p{(\columnwidth - 10\tabcolsep) * \real{0.1603}}
  >{\raggedright\arraybackslash}p{(\columnwidth - 10\tabcolsep) * \real{0.1880}}@{}}
\toprule\noalign{}
\begin{minipage}[b]{\linewidth}\raggedright
\textbf{Architecture}
\end{minipage} & \begin{minipage}[b]{\linewidth}\raggedright
\textbf{Cell}
\end{minipage} & \begin{minipage}[b]{\linewidth}\raggedright
\textbf{Remaining non-zero weights}
\end{minipage} & \begin{minipage}[b]{\linewidth}\raggedright
\textbf{Macro-F1 (mean)}
\end{minipage} & \begin{minipage}[b]{\linewidth}\raggedright
\textbf{Loss vs own baseline}
\end{minipage} & \begin{minipage}[b]{\linewidth}\raggedright
\textbf{Classes damaged (\textgreater=3/5 seeds)}
\end{minipage} \\
\begin{minipage}[b]{\linewidth}\raggedright
CNN1D
\end{minipage} & \begin{minipage}[b]{\linewidth}\raggedright
uniform 50\%
\end{minipage} & \begin{minipage}[b]{\linewidth}\raggedright
14,560
\end{minipage} & \begin{minipage}[b]{\linewidth}\raggedright
0.499
\end{minipage} & \begin{minipage}[b]{\linewidth}\raggedright
0.042
\end{minipage} & \begin{minipage}[b]{\linewidth}\raggedright
4
\end{minipage} \\
\begin{minipage}[b]{\linewidth}\raggedright
CNN1D
\end{minipage} & \begin{minipage}[b]{\linewidth}\raggedright
uniform 80\%
\end{minipage} & \begin{minipage}[b]{\linewidth}\raggedright
5,823
\end{minipage} & \begin{minipage}[b]{\linewidth}\raggedright
0.271
\end{minipage} & \begin{minipage}[b]{\linewidth}\raggedright
0.271
\end{minipage} & \begin{minipage}[b]{\linewidth}\raggedright
17
\end{minipage} \\
\begin{minipage}[b]{\linewidth}\raggedright
MLP
\end{minipage} & \begin{minipage}[b]{\linewidth}\raggedright
body 50\%
\end{minipage} & \begin{minipage}[b]{\linewidth}\raggedright
22,246
\end{minipage} & \begin{minipage}[b]{\linewidth}\raggedright
0.551
\end{minipage} & \begin{minipage}[b]{\linewidth}\raggedright
0.002
\end{minipage} & \begin{minipage}[b]{\linewidth}\raggedright
0
\end{minipage} \\
\begin{minipage}[b]{\linewidth}\raggedright
MLP
\end{minipage} & \begin{minipage}[b]{\linewidth}\raggedright
body 80\%
\end{minipage} & \begin{minipage}[b]{\linewidth}\raggedright
9,421
\end{minipage} & \begin{minipage}[b]{\linewidth}\raggedright
0.546
\end{minipage} & \begin{minipage}[b]{\linewidth}\raggedright
0.008
\end{minipage} & \begin{minipage}[b]{\linewidth}\raggedright
0
\end{minipage} \\
\begin{minipage}[b]{\linewidth}\raggedright
MLP
\end{minipage} & \begin{minipage}[b]{\linewidth}\raggedright
body 90\%
\end{minipage} & \begin{minipage}[b]{\linewidth}\raggedright
5,145
\end{minipage} & \begin{minipage}[b]{\linewidth}\raggedright
0.540
\end{minipage} & \begin{minipage}[b]{\linewidth}\raggedright
0.013
\end{minipage} & \begin{minipage}[b]{\linewidth}\raggedright
0
\end{minipage} \\
\begin{minipage}[b]{\linewidth}\raggedright
MLP
\end{minipage} & \begin{minipage}[b]{\linewidth}\raggedright
body 95\%
\end{minipage} & \begin{minipage}[b]{\linewidth}\raggedright
3,007
\end{minipage} & \begin{minipage}[b]{\linewidth}\raggedright
0.531
\end{minipage} & \begin{minipage}[b]{\linewidth}\raggedright
0.022
\end{minipage} & \begin{minipage}[b]{\linewidth}\raggedright
0
\end{minipage} \\
\begin{minipage}[b]{\linewidth}\raggedright
FT-Transformer
\end{minipage} & \begin{minipage}[b]{\linewidth}\raggedright
body 50\%
\end{minipage} & \begin{minipage}[b]{\linewidth}\raggedright
112,272
\end{minipage} & \begin{minipage}[b]{\linewidth}\raggedright
0.497
\end{minipage} & \begin{minipage}[b]{\linewidth}\raggedright
-0.004
\end{minipage} & \begin{minipage}[b]{\linewidth}\raggedright
0
\end{minipage} \\
\begin{minipage}[b]{\linewidth}\raggedright
FT-Transformer
\end{minipage} & \begin{minipage}[b]{\linewidth}\raggedright
body 80\%
\end{minipage} & \begin{minipage}[b]{\linewidth}\raggedright
44,907
\end{minipage} & \begin{minipage}[b]{\linewidth}\raggedright
0.474
\end{minipage} & \begin{minipage}[b]{\linewidth}\raggedright
0.019
\end{minipage} & \begin{minipage}[b]{\linewidth}\raggedright
1
\end{minipage} \\
\begin{minipage}[b]{\linewidth}\raggedright
FT-Transformer
\end{minipage} & \begin{minipage}[b]{\linewidth}\raggedright
body 90\%
\end{minipage} & \begin{minipage}[b]{\linewidth}\raggedright
22,791
\end{minipage} & \begin{minipage}[b]{\linewidth}\raggedright
0.458
\end{minipage} & \begin{minipage}[b]{\linewidth}\raggedright
0.034
\end{minipage} & \begin{minipage}[b]{\linewidth}\raggedright
3
\end{minipage} \\
\begin{minipage}[b]{\linewidth}\raggedright
FT-Transformer
\end{minipage} & \begin{minipage}[b]{\linewidth}\raggedright
body 95\%
\end{minipage} & \begin{minipage}[b]{\linewidth}\raggedright
11,733
\end{minipage} & \begin{minipage}[b]{\linewidth}\raggedright
0.437
\end{minipage} & \begin{minipage}[b]{\linewidth}\raggedright
0.056
\end{minipage} & \begin{minipage}[b]{\linewidth}\raggedright
4
\end{minipage} \\
\begin{minipage}[b]{\linewidth}\raggedright
FT-Transformer
\end{minipage} & \begin{minipage}[b]{\linewidth}\raggedright
body 97\%
\end{minipage} & \begin{minipage}[b]{\linewidth}\raggedright
7,305
\end{minipage} & \begin{minipage}[b]{\linewidth}\raggedright
0.423
\end{minipage} & \begin{minipage}[b]{\linewidth}\raggedright
0.070
\end{minipage} & \begin{minipage}[b]{\linewidth}\raggedright
7
\end{minipage} \\
\begin{minipage}[b]{\linewidth}\raggedright
FT-Transformer
\end{minipage} & \begin{minipage}[b]{\linewidth}\raggedright
body 98.5\%
\end{minipage} & \begin{minipage}[b]{\linewidth}\raggedright
3,987
\end{minipage} & \begin{minipage}[b]{\linewidth}\raggedright
0.397
\end{minipage} & \begin{minipage}[b]{\linewidth}\raggedright
0.096
\end{minipage} & \begin{minipage}[b]{\linewidth}\raggedright
8
\end{minipage} \\
\midrule\noalign{}
\endhead
\bottomrule\noalign{}
\endlastfoot
\end{longtable}\endgroup

\emph{Table 3. The capacity axis. Remaining non-zero counts are read from the saved masks. The MLP and transformer cells sweep the body with the head (and the transformer\textquotesingle s token embedding) held at 80\%. Macro-F1 loss is measured against each architecture\textquotesingle s own five-seed dense baseline.}

\includegraphics[width=0.85\linewidth]{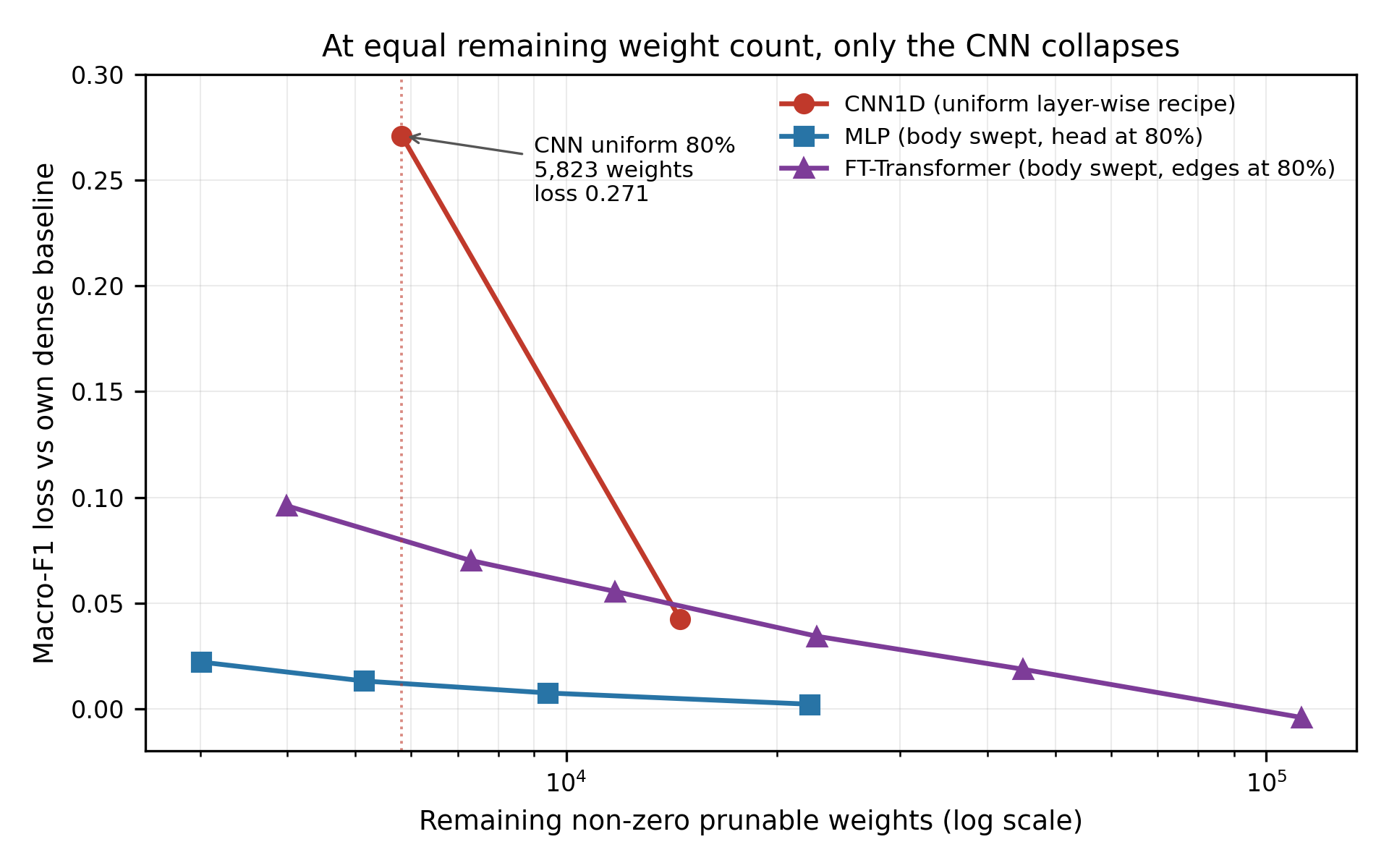}

\emph{Figure 1. The capacity axis. Macro-F1 loss against remaining non-zero prunable weights for three architectures; the dotted line marks the remainder of the uniformly pruned CNN (5,823 weights).}

\subsection*{4.3 It is not lost information}

The pruned CNN still contains the information it fails to use. For each class that collapses, a linear classifier fitted on the pruned network\textquotesingle s penultimate-layer features (training and validation records only) and evaluated on the untouched test partition reaches an AUC above 0.908 in every case, with a mean drop of 0.007 across the 13 measurable classes relative to the same probe on the dense network. Refitting only the final linear layer on the frozen pruned body restores macro-F1 to 0.526 and improves calibration beyond the dense model. Recomputing the normalisation statistics recovers nothing when done on the pruned model before fine-tuning (0.008 to 0.006), but, as Section 5.9 shows, recovers most of the loss when done on the fine-tuned model. The output collapses; the weights that produce it are largely sound. Two other compressions of the same model, int8 quantisation of the linear layer and float16 storage, leave every metric at the dense value (macro-F1 0.546 and 0.545 against 0.545): the failure is specific to what pruning does to conv.0, which the rest of the paper establishes.

\section*{5. Mechanism: input-layer starvation}

\subsection*{5.1 A 192-weight layer}

The CNN\textquotesingle s first layer, conv.0, has $64 \times 1 \times 3$ = 192 weights (Table 1), 0.7\% of the network\textquotesingle s prunable weights, while conv.3 has 24,576 and the head 4,352. The uniform layer-wise recipe removes 80\% of every layer independently, so conv.0 keeps 38 weights. The 38 survivors are spread over the 64 filters much as random survival would spread them, so roughly 35 filters keep at least one weight and about 29 are disconnected from the input. Every downstream computation sees the input only through those 35 filters, most of them reduced to a single weight. Sections 5.2 to 5.7 test whether this is the cause.

\subsection*{5.2 Protecting the layer, or pruning globally, prevents the collapse at the same total sparsity}

Table 4 compares the recipes on the same five baselines. Leaving conv.0 untouched while pruning conv.3 and the head at 80\% reduces the loss from 0.271 (95\% interval 0.16 to 0.41; per-seed range 0.14 to 0.53) to 0.013 (0.00 to 0.03) and the damaged classes from 17 to 2, at a total sparsity of 77.8\% against 78.4\%; the recovered macro-F1 is 0.258 (95\% interval 0.14 to 0.41), and the damaged-class count is not sensitive to the practical-loss threshold (22, 17 and 14 classes at 0.05, 0.10 and 0.15 for uniform pruning; 2, 2 and 1 protected). Global magnitude pruning at exactly 80\% of all prunable weights, with no layer exempted, gives 0.014 and no damaged class: the single threshold keeps 71\% of conv.0 on its own, removes 89\% of conv.3 and 29\% of the head. Both fixes collapse the seed-to-seed spread from 0.164 to 0.012 and 0.007; under both, probes on the collapsed classes stay above 0.91 and head refitting reaches 0.548 and 0.541. Why global pruning protected the small layer here is an empirical observation, not a theorem; Gupta et al. \cite{r12} show that at much higher sparsity a global threshold can still remove a small layer entirely. First-layer protection is the cleaner intervention, at the cost of keeping 154 more weights than uniform pruning (5,977 against 5,823 non-zero prunable weights, 2.6\% more).

The same holds when pruning is gradual and weights are allowed to recover. The gradual recipe of Table 2 (Eq. 3, regrowth-enabled) was run with per-layer and global allocation. Under the per-layer schedule conv.0 still ends at 80\% sparsity with 34 connected filters and the detector still collapses (loss 0.165, 95\% interval 0.12 to 0.23; 8 classes damaged); the global schedule keeps 69\% of conv.0 and loses 0.029 with no class damaged (last two rows of Table 4). Regrowth occurred in conv.3 (8 to 19 weights per run) but no pruned conv.0 weight re-entered the mask in any run; a per-layer rule does not forbid re-entry, so this is an empirical finding. The gradual schedule reduces the damage by about 40\% relative to one-shot pruning: schedule affects severity, it does not remove the allocation effect.

\begingroup\scriptsize\setlength{\tabcolsep}{3pt}\renewcommand{\arraystretch}{1.05}\begin{longtable}[]{@{}
  >{\raggedright\arraybackslash}p{(\columnwidth - 8\tabcolsep) * \real{0.2778}}
  >{\raggedright\arraybackslash}p{(\columnwidth - 8\tabcolsep) * \real{0.2564}}
  >{\raggedright\arraybackslash}p{(\columnwidth - 8\tabcolsep) * \real{0.1816}}
  >{\raggedright\arraybackslash}p{(\columnwidth - 8\tabcolsep) * \real{0.0962}}
  >{\raggedright\arraybackslash}p{(\columnwidth - 8\tabcolsep) * \real{0.1880}}@{}}
\toprule\noalign{}
\begin{minipage}[b]{\linewidth}\raggedright
\textbf{Recipe}
\end{minipage} & \begin{minipage}[b]{\linewidth}\raggedright
\textbf{Non-zero prunable weights (conv.0 / conv.3 / head)}
\end{minipage} & \begin{minipage}[b]{\linewidth}\raggedright
\textbf{Macro-F1 (mean +/- sd)}
\end{minipage} & \begin{minipage}[b]{\linewidth}\raggedright
\textbf{Loss}
\end{minipage} & \begin{minipage}[b]{\linewidth}\raggedright
\textbf{Classes damaged}
\end{minipage} \\
\begin{minipage}[b]{\linewidth}\raggedright
Uniform layer-wise 80\%
\end{minipage} & \begin{minipage}[b]{\linewidth}\raggedright
5,823 (38 / 4,915 / 870)
\end{minipage} & \begin{minipage}[b]{\linewidth}\raggedright
0.271 +/- 0.164
\end{minipage} & \begin{minipage}[b]{\linewidth}\raggedright
0.271
\end{minipage} & \begin{minipage}[b]{\linewidth}\raggedright
17
\end{minipage} \\
\begin{minipage}[b]{\linewidth}\raggedright
First layer protected, rest 80\%
\end{minipage} & \begin{minipage}[b]{\linewidth}\raggedright
5,977 (192 / 4,915 / 870)
\end{minipage} & \begin{minipage}[b]{\linewidth}\raggedright
0.529 +/- 0.012
\end{minipage} & \begin{minipage}[b]{\linewidth}\raggedright
0.013
\end{minipage} & \begin{minipage}[b]{\linewidth}\raggedright
2
\end{minipage} \\
\begin{minipage}[b]{\linewidth}\raggedright
Global magnitude 80\%
\end{minipage} & \begin{minipage}[b]{\linewidth}\raggedright
5,824 (136 / 2,598 / 3,090)
\end{minipage} & \begin{minipage}[b]{\linewidth}\raggedright
0.527 +/- 0.007
\end{minipage} & \begin{minipage}[b]{\linewidth}\raggedright
0.014
\end{minipage} & \begin{minipage}[b]{\linewidth}\raggedright
0
\end{minipage} \\
\begin{minipage}[b]{\linewidth}\raggedright
Gradual per-layer schedule, regrowth-enabled
\end{minipage} & \begin{minipage}[b]{\linewidth}\raggedright
5,823 (38 / 4,915 / 870)
\end{minipage} & \begin{minipage}[b]{\linewidth}\raggedright
0.377 +/- 0.070
\end{minipage} & \begin{minipage}[b]{\linewidth}\raggedright
0.165
\end{minipage} & \begin{minipage}[b]{\linewidth}\raggedright
8
\end{minipage} \\
\begin{minipage}[b]{\linewidth}\raggedright
Gradual global schedule, regrowth-enabled
\end{minipage} & \begin{minipage}[b]{\linewidth}\raggedright
5,824 (132 / 2,732 / 2,960)
\end{minipage} & \begin{minipage}[b]{\linewidth}\raggedright
0.512 +/- 0.051
\end{minipage} & \begin{minipage}[b]{\linewidth}\raggedright
0.029
\end{minipage} & \begin{minipage}[b]{\linewidth}\raggedright
0
\end{minipage} \\
\midrule\noalign{}
\endhead
\bottomrule\noalign{}
\endlastfoot
\end{longtable}\endgroup

\emph{Table 4. Five recipes on the same five CNN baselines. Counts are five-seed means read from the saved masks; global counts are rounded to the nearest weight. Whatever the schedule, uniform allocation starves conv.0 and the detector collapses; protecting the layer, or using a global threshold, avoids it at the same total sparsity within 0.6\%.}

\subsection*{5.3 Dose-response}

If starvation of conv.0 is the cause, damage should rise smoothly with conv.0 sparsity while the rest of the network is held fixed. Table 5 sweeps conv.0 sparsity from 0\% to 95\% (192 down to 10 surviving weights) with conv.3 and the head at 80\% throughout. Damage is flat to 154 weights, begins at 96 (loss 0.055, four classes), and then climbs steeply: 38 weights 0.271, 19 weights 0.468, 10 weights 0.526, at which point the detector is essentially dead (macro-F1 0.016). These weight counts belong to the 64-filter geometry; Section 5.5 shows what carries over to other widths. The Spearman rank correlation between conv.0 sparsity and loss is 1.0 over the six dose means; over the 30 individual runs it is 0.90 (seed-level bootstrap 95\% interval 0.86 to 0.95).

\subsection*{5.4 Positive control: starving the MLP reproduces the collapse}

The MLP resisted 95\% pruning of its body with the head at 80\% (Table 3). If input-layer starvation is the cause rather than something peculiar to convolution, then starving the MLP\textquotesingle s input layer to the same absolute number of surviving weights should make it collapse too. It does (Table 5, lower rows). With 1,997 or 998 surviving input weights the MLP is intact (loss below 0.01). At 192 weights the loss is 0.058 with four classes damaged, at 96 weights 0.178 with 13, and at 38 weights 0.323 with 18 of 34 classes damaged. The onset region, roughly 100 to 200 surviving input weights, coincides with the 64-filter CNN\textquotesingle s; over the 25 individual MLP runs the Spearman correlation between input-layer sparsity and loss is 0.93 (seed-level bootstrap 95\% interval 0.90 to 0.96), and over the five dose means 0.90. The starvation is harsher for the MLP, whose input weights each connect one feature to one unit, whereas a convolutional weight is shared across all 39 positions; even so the curve has the same shape and the same catastrophe.

\begingroup\scriptsize\setlength{\tabcolsep}{3pt}\renewcommand{\arraystretch}{1.05}\begin{longtable}[]{@{}
  >{\raggedright\arraybackslash}p{(\columnwidth - 10\tabcolsep) * \real{0.1389}}
  >{\raggedright\arraybackslash}p{(\columnwidth - 10\tabcolsep) * \real{0.1603}}
  >{\raggedright\arraybackslash}p{(\columnwidth - 10\tabcolsep) * \real{0.1709}}
  >{\raggedright\arraybackslash}p{(\columnwidth - 10\tabcolsep) * \real{0.1816}}
  >{\raggedright\arraybackslash}p{(\columnwidth - 10\tabcolsep) * \real{0.0962}}
  >{\raggedright\arraybackslash}p{(\columnwidth - 10\tabcolsep) * \real{0.2521}}@{}}
\toprule\noalign{}
\begin{minipage}[b]{\linewidth}\raggedright
\textbf{Architecture}
\end{minipage} & \begin{minipage}[b]{\linewidth}\raggedright
\textbf{Input-layer sparsity}
\end{minipage} & \begin{minipage}[b]{\linewidth}\raggedright
\textbf{Surviving input weights}
\end{minipage} & \begin{minipage}[b]{\linewidth}\raggedright
\textbf{Macro-F1 (mean +/- sd)}
\end{minipage} & \begin{minipage}[b]{\linewidth}\raggedright
\textbf{Loss}
\end{minipage} & \begin{minipage}[b]{\linewidth}\raggedright
\textbf{Classes damaged}
\end{minipage} \\
\begin{minipage}[b]{\linewidth}\raggedright
CNN1D
\end{minipage} & \begin{minipage}[b]{\linewidth}\raggedright
0\%
\end{minipage} & \begin{minipage}[b]{\linewidth}\raggedright
192
\end{minipage} & \begin{minipage}[b]{\linewidth}\raggedright
0.529 +/- 0.012
\end{minipage} & \begin{minipage}[b]{\linewidth}\raggedright
0.013
\end{minipage} & \begin{minipage}[b]{\linewidth}\raggedright
2
\end{minipage} \\
\begin{minipage}[b]{\linewidth}\raggedright
CNN1D
\end{minipage} & \begin{minipage}[b]{\linewidth}\raggedright
20\%
\end{minipage} & \begin{minipage}[b]{\linewidth}\raggedright
154
\end{minipage} & \begin{minipage}[b]{\linewidth}\raggedright
0.527 +/- 0.013
\end{minipage} & \begin{minipage}[b]{\linewidth}\raggedright
0.014
\end{minipage} & \begin{minipage}[b]{\linewidth}\raggedright
1
\end{minipage} \\
\begin{minipage}[b]{\linewidth}\raggedright
CNN1D
\end{minipage} & \begin{minipage}[b]{\linewidth}\raggedright
50\%
\end{minipage} & \begin{minipage}[b]{\linewidth}\raggedright
96
\end{minipage} & \begin{minipage}[b]{\linewidth}\raggedright
0.487 +/- 0.063
\end{minipage} & \begin{minipage}[b]{\linewidth}\raggedright
0.055
\end{minipage} & \begin{minipage}[b]{\linewidth}\raggedright
4
\end{minipage} \\
\begin{minipage}[b]{\linewidth}\raggedright
CNN1D
\end{minipage} & \begin{minipage}[b]{\linewidth}\raggedright
80\% (default)
\end{minipage} & \begin{minipage}[b]{\linewidth}\raggedright
38
\end{minipage} & \begin{minipage}[b]{\linewidth}\raggedright
0.271 +/- 0.164
\end{minipage} & \begin{minipage}[b]{\linewidth}\raggedright
0.271
\end{minipage} & \begin{minipage}[b]{\linewidth}\raggedright
17
\end{minipage} \\
\begin{minipage}[b]{\linewidth}\raggedright
CNN1D
\end{minipage} & \begin{minipage}[b]{\linewidth}\raggedright
90\%
\end{minipage} & \begin{minipage}[b]{\linewidth}\raggedright
19
\end{minipage} & \begin{minipage}[b]{\linewidth}\raggedright
0.074 +/- 0.096
\end{minipage} & \begin{minipage}[b]{\linewidth}\raggedright
0.468
\end{minipage} & \begin{minipage}[b]{\linewidth}\raggedright
25
\end{minipage} \\
\begin{minipage}[b]{\linewidth}\raggedright
CNN1D
\end{minipage} & \begin{minipage}[b]{\linewidth}\raggedright
95\%
\end{minipage} & \begin{minipage}[b]{\linewidth}\raggedright
10
\end{minipage} & \begin{minipage}[b]{\linewidth}\raggedright
0.016 +/- 0.014
\end{minipage} & \begin{minipage}[b]{\linewidth}\raggedright
0.526
\end{minipage} & \begin{minipage}[b]{\linewidth}\raggedright
25
\end{minipage} \\
\begin{minipage}[b]{\linewidth}\raggedright
MLP
\end{minipage} & \begin{minipage}[b]{\linewidth}\raggedright
80\%
\end{minipage} & \begin{minipage}[b]{\linewidth}\raggedright
1,997
\end{minipage} & \begin{minipage}[b]{\linewidth}\raggedright
0.546 +/- 0.006
\end{minipage} & \begin{minipage}[b]{\linewidth}\raggedright
0.008
\end{minipage} & \begin{minipage}[b]{\linewidth}\raggedright
0
\end{minipage} \\
\begin{minipage}[b]{\linewidth}\raggedright
MLP
\end{minipage} & \begin{minipage}[b]{\linewidth}\raggedright
90\%
\end{minipage} & \begin{minipage}[b]{\linewidth}\raggedright
998
\end{minipage} & \begin{minipage}[b]{\linewidth}\raggedright
0.547 +/- 0.008
\end{minipage} & \begin{minipage}[b]{\linewidth}\raggedright
0.006
\end{minipage} & \begin{minipage}[b]{\linewidth}\raggedright
0
\end{minipage} \\
\begin{minipage}[b]{\linewidth}\raggedright
MLP
\end{minipage} & \begin{minipage}[b]{\linewidth}\raggedright
98.1\%
\end{minipage} & \begin{minipage}[b]{\linewidth}\raggedright
192
\end{minipage} & \begin{minipage}[b]{\linewidth}\raggedright
0.495 +/- 0.018
\end{minipage} & \begin{minipage}[b]{\linewidth}\raggedright
0.058
\end{minipage} & \begin{minipage}[b]{\linewidth}\raggedright
4
\end{minipage} \\
\begin{minipage}[b]{\linewidth}\raggedright
MLP
\end{minipage} & \begin{minipage}[b]{\linewidth}\raggedright
99.0\%
\end{minipage} & \begin{minipage}[b]{\linewidth}\raggedright
96
\end{minipage} & \begin{minipage}[b]{\linewidth}\raggedright
0.375 +/- 0.065
\end{minipage} & \begin{minipage}[b]{\linewidth}\raggedright
0.178
\end{minipage} & \begin{minipage}[b]{\linewidth}\raggedright
13
\end{minipage} \\
\begin{minipage}[b]{\linewidth}\raggedright
MLP
\end{minipage} & \begin{minipage}[b]{\linewidth}\raggedright
99.6\%
\end{minipage} & \begin{minipage}[b]{\linewidth}\raggedright
38
\end{minipage} & \begin{minipage}[b]{\linewidth}\raggedright
0.230 +/- 0.072
\end{minipage} & \begin{minipage}[b]{\linewidth}\raggedright
0.323
\end{minipage} & \begin{minipage}[b]{\linewidth}\raggedright
18
\end{minipage} \\
\midrule\noalign{}
\endhead
\bottomrule\noalign{}
\endlastfoot
\end{longtable}\endgroup

\emph{Table 5. Dose-response in the input layer. Upper rows: CNN, conv.0 swept with conv.3 and head at 80\%. Lower rows: MLP, input layer starved to the same absolute counts with the rest at 80\%. Spearman correlations between input-layer sparsity and loss: over dose means, 1.00 (CNN) and 0.90 (MLP, one inversion between the two flat doses); over individual runs, 0.90 (CNN, seed-level bootstrap 95\% interval 0.86 to 0.95) and 0.93 (MLP, 0.90 to 0.96).}

\includegraphics[width=0.85\linewidth]{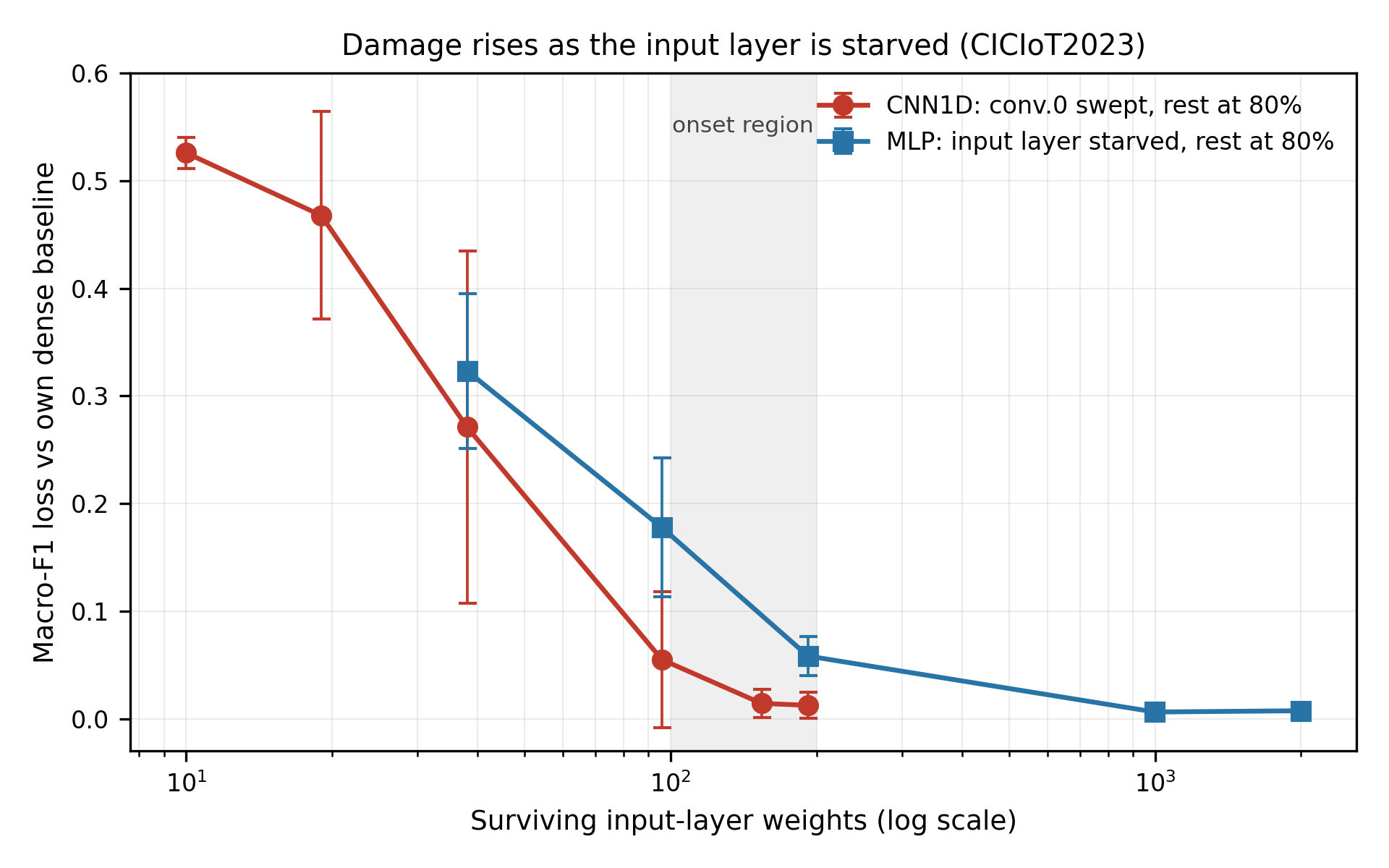}

\emph{Figure 2. Dose-response in the input layer on CICIoT2023. CNN: conv.0 swept with the rest at 80\%. MLP: input layer starved to matched absolute counts with the rest at 80\%. Bars are seed standard deviations; the shaded band marks 100 to 200 surviving weights.}

\subsection*{5.5 The onset is not an absolute weight count}

Is the onset a property of the 192-weight layer or of first layers in general? The same CNN was trained with 32 filters (96 first-layer weights) and 128 filters (384 weights), five seeds each, both validation-gated against the 64-filter band, and conv.0 was swept so that surviving-weight counts overlap across widths (Table 6). At 38 surviving weights the widths differ: the 32-filter network loses 0.047 with 28 of 32 filters still connected, the 64-filter network 0.271 with 35 of 64, the 128-filter network 0.318 with 38 of 128. The pre-stated absolute-count reading fails and is reported as failed. Across all sixteen cells the pooled Spearman correlations with loss are -0.94 for surviving weights, -0.91 for the fraction of connected filters and -0.80 for their number; a pooled correlation can be strong while the predictor fails at a matched operating point, which is what the 38-weight comparison shows. Within the conditions tested, the connected fraction separates a low-loss region (above about 85\% connected, losses of 0.005 to 0.055) from a high-loss region (below about 65\%, losses of 0.11 to 0.53); no condition falls between 65\% and 85\%, so no sharp threshold is located.

The practical consequence is that widening the first layer does not protect against uniform pruning. Under random survival a three-tap filter stays connected with probability 1 - s\^{}3 at sparsity s (Eq. 8), so uniform 80\% pruning leaves about half the filters connected whatever the width. Measured losses under uniform 80\% pruning are 0.436 for 32 filters, 0.271 for 64 and 0.191 for 128 (77 surviving weights, 8 classes damaged; the cell lies between the 75\% and 90\% cells, 0.106 and 0.318). The narrowest layer is hit hardest and the widest is still damaged.

\begingroup\scriptsize\setlength{\tabcolsep}{3pt}\renewcommand{\arraystretch}{1.05}\begin{longtable}[]{@{}
  >{\raggedright\arraybackslash}p{(\columnwidth - 10\tabcolsep) * \real{0.1282}}
  >{\raggedright\arraybackslash}p{(\columnwidth - 10\tabcolsep) * \real{0.1603}}
  >{\raggedright\arraybackslash}p{(\columnwidth - 10\tabcolsep) * \real{0.1603}}
  >{\raggedright\arraybackslash}p{(\columnwidth - 10\tabcolsep) * \real{0.1389}}
  >{\raggedright\arraybackslash}p{(\columnwidth - 10\tabcolsep) * \real{0.1709}}
  >{\raggedright\arraybackslash}p{(\columnwidth - 10\tabcolsep) * \real{0.1068}}@{}}
\toprule\noalign{}
\begin{minipage}[b]{\linewidth}\raggedright
\textbf{Filters}
\end{minipage} & \begin{minipage}[b]{\linewidth}\raggedright
\textbf{Surviving weights}
\end{minipage} & \begin{minipage}[b]{\linewidth}\raggedright
\textbf{Connected filters (fraction)}
\end{minipage} & \begin{minipage}[b]{\linewidth}\raggedright
\textbf{Macro-F1 (mean +/- sd)}
\end{minipage} & \begin{minipage}[b]{\linewidth}\raggedright
\textbf{Loss}
\end{minipage} & \begin{minipage}[b]{\linewidth}\raggedright
\textbf{Classes damaged}
\end{minipage} \\
\begin{minipage}[b]{\linewidth}\raggedright
32
\end{minipage} & \begin{minipage}[b]{\linewidth}\raggedright
96
\end{minipage} & \begin{minipage}[b]{\linewidth}\raggedright
32.0 (1.00)
\end{minipage} & \begin{minipage}[b]{\linewidth}\raggedright
0.516 +/- 0.020
\end{minipage} & \begin{minipage}[b]{\linewidth}\raggedright
0.021
\end{minipage} & \begin{minipage}[b]{\linewidth}\raggedright
0
\end{minipage} \\
\begin{minipage}[b]{\linewidth}\raggedright
32
\end{minipage} & \begin{minipage}[b]{\linewidth}\raggedright
48
\end{minipage} & \begin{minipage}[b]{\linewidth}\raggedright
29.6 (0.93)
\end{minipage} & \begin{minipage}[b]{\linewidth}\raggedright
0.482 +/- 0.044
\end{minipage} & \begin{minipage}[b]{\linewidth}\raggedright
0.055
\end{minipage} & \begin{minipage}[b]{\linewidth}\raggedright
3
\end{minipage} \\
\begin{minipage}[b]{\linewidth}\raggedright
32
\end{minipage} & \begin{minipage}[b]{\linewidth}\raggedright
38
\end{minipage} & \begin{minipage}[b]{\linewidth}\raggedright
27.8 (0.87)
\end{minipage} & \begin{minipage}[b]{\linewidth}\raggedright
0.490 +/- 0.026
\end{minipage} & \begin{minipage}[b]{\linewidth}\raggedright
0.047
\end{minipage} & \begin{minipage}[b]{\linewidth}\raggedright
2
\end{minipage} \\
\begin{minipage}[b]{\linewidth}\raggedright
32
\end{minipage} & \begin{minipage}[b]{\linewidth}\raggedright
19
\end{minipage} & \begin{minipage}[b]{\linewidth}\raggedright
18.0 (0.56)
\end{minipage} & \begin{minipage}[b]{\linewidth}\raggedright
0.101 +/- 0.080
\end{minipage} & \begin{minipage}[b]{\linewidth}\raggedright
0.436
\end{minipage} & \begin{minipage}[b]{\linewidth}\raggedright
21
\end{minipage} \\
\begin{minipage}[b]{\linewidth}\raggedright
32
\end{minipage} & \begin{minipage}[b]{\linewidth}\raggedright
10
\end{minipage} & \begin{minipage}[b]{\linewidth}\raggedright
9.8 (0.31)
\end{minipage} & \begin{minipage}[b]{\linewidth}\raggedright
0.065 +/- 0.052
\end{minipage} & \begin{minipage}[b]{\linewidth}\raggedright
0.472
\end{minipage} & \begin{minipage}[b]{\linewidth}\raggedright
20
\end{minipage} \\
\begin{minipage}[b]{\linewidth}\raggedright
64
\end{minipage} & \begin{minipage}[b]{\linewidth}\raggedright
192
\end{minipage} & \begin{minipage}[b]{\linewidth}\raggedright
64.0 (1.00)
\end{minipage} & \begin{minipage}[b]{\linewidth}\raggedright
0.529 +/- 0.012
\end{minipage} & \begin{minipage}[b]{\linewidth}\raggedright
0.013
\end{minipage} & \begin{minipage}[b]{\linewidth}\raggedright
2
\end{minipage} \\
\begin{minipage}[b]{\linewidth}\raggedright
64
\end{minipage} & \begin{minipage}[b]{\linewidth}\raggedright
154
\end{minipage} & \begin{minipage}[b]{\linewidth}\raggedright
63.6 (0.99)
\end{minipage} & \begin{minipage}[b]{\linewidth}\raggedright
0.527 +/- 0.013
\end{minipage} & \begin{minipage}[b]{\linewidth}\raggedright
0.014
\end{minipage} & \begin{minipage}[b]{\linewidth}\raggedright
1
\end{minipage} \\
\begin{minipage}[b]{\linewidth}\raggedright
64
\end{minipage} & \begin{minipage}[b]{\linewidth}\raggedright
96
\end{minipage} & \begin{minipage}[b]{\linewidth}\raggedright
58.4 (0.91)
\end{minipage} & \begin{minipage}[b]{\linewidth}\raggedright
0.487 +/- 0.063
\end{minipage} & \begin{minipage}[b]{\linewidth}\raggedright
0.055
\end{minipage} & \begin{minipage}[b]{\linewidth}\raggedright
4
\end{minipage} \\
\begin{minipage}[b]{\linewidth}\raggedright
64
\end{minipage} & \begin{minipage}[b]{\linewidth}\raggedright
38
\end{minipage} & \begin{minipage}[b]{\linewidth}\raggedright
34.6 (0.54)
\end{minipage} & \begin{minipage}[b]{\linewidth}\raggedright
0.271 +/- 0.164
\end{minipage} & \begin{minipage}[b]{\linewidth}\raggedright
0.271
\end{minipage} & \begin{minipage}[b]{\linewidth}\raggedright
17
\end{minipage} \\
\begin{minipage}[b]{\linewidth}\raggedright
64
\end{minipage} & \begin{minipage}[b]{\linewidth}\raggedright
19
\end{minipage} & \begin{minipage}[b]{\linewidth}\raggedright
18.6 (0.29)
\end{minipage} & \begin{minipage}[b]{\linewidth}\raggedright
0.074 +/- 0.096
\end{minipage} & \begin{minipage}[b]{\linewidth}\raggedright
0.468
\end{minipage} & \begin{minipage}[b]{\linewidth}\raggedright
25
\end{minipage} \\
\begin{minipage}[b]{\linewidth}\raggedright
64
\end{minipage} & \begin{minipage}[b]{\linewidth}\raggedright
10
\end{minipage} & \begin{minipage}[b]{\linewidth}\raggedright
10.0 (0.16)
\end{minipage} & \begin{minipage}[b]{\linewidth}\raggedright
0.016 +/- 0.014
\end{minipage} & \begin{minipage}[b]{\linewidth}\raggedright
0.526
\end{minipage} & \begin{minipage}[b]{\linewidth}\raggedright
25
\end{minipage} \\
\begin{minipage}[b]{\linewidth}\raggedright
128
\end{minipage} & \begin{minipage}[b]{\linewidth}\raggedright
384
\end{minipage} & \begin{minipage}[b]{\linewidth}\raggedright
128.0 (1.00)
\end{minipage} & \begin{minipage}[b]{\linewidth}\raggedright
0.533 +/- 0.015
\end{minipage} & \begin{minipage}[b]{\linewidth}\raggedright
0.005
\end{minipage} & \begin{minipage}[b]{\linewidth}\raggedright
0
\end{minipage} \\
\begin{minipage}[b]{\linewidth}\raggedright
128
\end{minipage} & \begin{minipage}[b]{\linewidth}\raggedright
192
\end{minipage} & \begin{minipage}[b]{\linewidth}\raggedright
115.4 (0.90)
\end{minipage} & \begin{minipage}[b]{\linewidth}\raggedright
0.507 +/- 0.015
\end{minipage} & \begin{minipage}[b]{\linewidth}\raggedright
0.031
\end{minipage} & \begin{minipage}[b]{\linewidth}\raggedright
3
\end{minipage} \\
\begin{minipage}[b]{\linewidth}\raggedright
128
\end{minipage} & \begin{minipage}[b]{\linewidth}\raggedright
96
\end{minipage} & \begin{minipage}[b]{\linewidth}\raggedright
82.0 (0.64)
\end{minipage} & \begin{minipage}[b]{\linewidth}\raggedright
0.432 +/- 0.076
\end{minipage} & \begin{minipage}[b]{\linewidth}\raggedright
0.106
\end{minipage} & \begin{minipage}[b]{\linewidth}\raggedright
7
\end{minipage} \\
\begin{minipage}[b]{\linewidth}\raggedright
128
\end{minipage} & \begin{minipage}[b]{\linewidth}\raggedright
38
\end{minipage} & \begin{minipage}[b]{\linewidth}\raggedright
37.6 (0.29)
\end{minipage} & \begin{minipage}[b]{\linewidth}\raggedright
0.220 +/- 0.074
\end{minipage} & \begin{minipage}[b]{\linewidth}\raggedright
0.318
\end{minipage} & \begin{minipage}[b]{\linewidth}\raggedright
18
\end{minipage} \\
\begin{minipage}[b]{\linewidth}\raggedright
128
\end{minipage} & \begin{minipage}[b]{\linewidth}\raggedright
19
\end{minipage} & \begin{minipage}[b]{\linewidth}\raggedright
19.0 (0.15)
\end{minipage} & \begin{minipage}[b]{\linewidth}\raggedright
0.113 +/- 0.091
\end{minipage} & \begin{minipage}[b]{\linewidth}\raggedright
0.425
\end{minipage} & \begin{minipage}[b]{\linewidth}\raggedright
20
\end{minipage} \\
\midrule\noalign{}
\endhead
\bottomrule\noalign{}
\endlastfoot
\end{longtable}\endgroup

\emph{Table 6. Filter-count sweep, all sixteen conditions. conv.0 swept with the rest of the network at 80\%, for three first-layer widths; the 64-filter rows are those of Table 5. Connected filters (at least one surviving input weight) are five-seed means from the saved masks.}

\subsection*{5.6 What mediates the effect in the CNN: live filters}

A weight count is not what the rest of the network sees. What it sees is the set of connected channels (Section 3.7). Under magnitude pruning the number of live channels is almost exactly what uniformly random survival would give (Eq. 7) (for example 34.6 observed against 31.1 expected at 38 weights, and 10.0 against 9.5 at 10 weights), so live channels and surviving weights are nearly collinear and predict damage about equally well (Appendix S4; on the dose-sweep runs alone, a single threshold of fewer than 58 live channels separates collapsed from intact runs across both architectures with 92.7\% accuracy, as does a threshold of fewer than 96 surviving weights). Magnitude pruning cannot tell the two apart, and at the scale of a 192-weight layer it leaves survivors distributed across filters almost as random survival would (Appendix S4); whether the particular weights it keeps matter more than random ones was not tested.

Hand-built masks can. Table 7 holds the number of surviving conv.0 weights fixed at 38 or 96 and varies how they are distributed. At 96 weights, the spread mask keeps each filter\textquotesingle s largest weight (64 filters live) plus the 32 largest remaining weights, and gives loss 0.018 with no damaged class; the uniform layer-wise recipe\textquotesingle s 58 live filters give 0.055 and four; the concentrated mask keeps all three weights of the 32 filters with the largest total magnitude and gives 0.212 and twelve. At 38 weights, the concentrated mask keeps 12 whole filters plus two weights of a thirteenth (13 live) and gives 0.475 and 25 damaged classes against the uniform layer-wise recipe\textquotesingle s 0.271. Same weight count, live channels varied three-fold, damage follows the channels. One qualification belongs here: at 38 weights the fully spread mask (one weight per filter) is not better than the uniform layer-wise recipe\textquotesingle s 35 filters (0.316 against 0.271, inside the seed noise). A filter reduced to a single weight is a pointwise scalar, so at the extreme of spreading, per-filter expressivity becomes the binding constraint. Live channels mediate the collapse, subject to that floor.

\begingroup\scriptsize\setlength{\tabcolsep}{3pt}\renewcommand{\arraystretch}{1.05}\begin{longtable}[]{@{}
  >{\raggedright\arraybackslash}p{(\columnwidth - 10\tabcolsep) * \real{0.1603}}
  >{\raggedright\arraybackslash}p{(\columnwidth - 10\tabcolsep) * \real{0.2244}}
  >{\raggedright\arraybackslash}p{(\columnwidth - 10\tabcolsep) * \real{0.1389}}
  >{\raggedright\arraybackslash}p{(\columnwidth - 10\tabcolsep) * \real{0.1816}}
  >{\raggedright\arraybackslash}p{(\columnwidth - 10\tabcolsep) * \real{0.0962}}
  >{\raggedright\arraybackslash}p{(\columnwidth - 10\tabcolsep) * \real{0.1987}}@{}}
\toprule\noalign{}
\begin{minipage}[b]{\linewidth}\raggedright
\textbf{Surviving conv.0 weights}
\end{minipage} & \begin{minipage}[b]{\linewidth}\raggedright
\textbf{Mask}
\end{minipage} & \begin{minipage}[b]{\linewidth}\raggedright
\textbf{Live filters (of 64)}
\end{minipage} & \begin{minipage}[b]{\linewidth}\raggedright
\textbf{Macro-F1 (mean +/- sd)}
\end{minipage} & \begin{minipage}[b]{\linewidth}\raggedright
\textbf{Loss}
\end{minipage} & \begin{minipage}[b]{\linewidth}\raggedright
\textbf{Classes damaged}
\end{minipage} \\
\begin{minipage}[b]{\linewidth}\raggedright
38
\end{minipage} & \begin{minipage}[b]{\linewidth}\raggedright
concentrated (whole filters)
\end{minipage} & \begin{minipage}[b]{\linewidth}\raggedright
13
\end{minipage} & \begin{minipage}[b]{\linewidth}\raggedright
0.067 +/- 0.064
\end{minipage} & \begin{minipage}[b]{\linewidth}\raggedright
0.475
\end{minipage} & \begin{minipage}[b]{\linewidth}\raggedright
25
\end{minipage} \\
\begin{minipage}[b]{\linewidth}\raggedright
38
\end{minipage} & \begin{minipage}[b]{\linewidth}\raggedright
magnitude (uniform layer-wise recipe)
\end{minipage} & \begin{minipage}[b]{\linewidth}\raggedright
34.6
\end{minipage} & \begin{minipage}[b]{\linewidth}\raggedright
0.271 +/- 0.164
\end{minipage} & \begin{minipage}[b]{\linewidth}\raggedright
0.271
\end{minipage} & \begin{minipage}[b]{\linewidth}\raggedright
17
\end{minipage} \\
\begin{minipage}[b]{\linewidth}\raggedright
38
\end{minipage} & \begin{minipage}[b]{\linewidth}\raggedright
spread (one weight per filter)
\end{minipage} & \begin{minipage}[b]{\linewidth}\raggedright
38
\end{minipage} & \begin{minipage}[b]{\linewidth}\raggedright
0.225 +/- 0.142
\end{minipage} & \begin{minipage}[b]{\linewidth}\raggedright
0.316
\end{minipage} & \begin{minipage}[b]{\linewidth}\raggedright
20
\end{minipage} \\
\begin{minipage}[b]{\linewidth}\raggedright
96
\end{minipage} & \begin{minipage}[b]{\linewidth}\raggedright
concentrated
\end{minipage} & \begin{minipage}[b]{\linewidth}\raggedright
32
\end{minipage} & \begin{minipage}[b]{\linewidth}\raggedright
0.329 +/- 0.086
\end{minipage} & \begin{minipage}[b]{\linewidth}\raggedright
0.212
\end{minipage} & \begin{minipage}[b]{\linewidth}\raggedright
12
\end{minipage} \\
\begin{minipage}[b]{\linewidth}\raggedright
96
\end{minipage} & \begin{minipage}[b]{\linewidth}\raggedright
magnitude
\end{minipage} & \begin{minipage}[b]{\linewidth}\raggedright
58.4
\end{minipage} & \begin{minipage}[b]{\linewidth}\raggedright
0.487 +/- 0.063
\end{minipage} & \begin{minipage}[b]{\linewidth}\raggedright
0.055
\end{minipage} & \begin{minipage}[b]{\linewidth}\raggedright
4
\end{minipage} \\
\begin{minipage}[b]{\linewidth}\raggedright
96
\end{minipage} & \begin{minipage}[b]{\linewidth}\raggedright
spread
\end{minipage} & \begin{minipage}[b]{\linewidth}\raggedright
64
\end{minipage} & \begin{minipage}[b]{\linewidth}\raggedright
0.523 +/- 0.007
\end{minipage} & \begin{minipage}[b]{\linewidth}\raggedright
0.018
\end{minipage} & \begin{minipage}[b]{\linewidth}\raggedright
0
\end{minipage} \\
\midrule\noalign{}
\endhead
\bottomrule\noalign{}
\endlastfoot
\end{longtable}\endgroup

\emph{Table 7. Live channels at fixed weight count. Hand-built conv.0 masks with the same number of surviving weights but different numbers of live filters; conv.3 and head at 80\%.}

\includegraphics[width=0.85\linewidth]{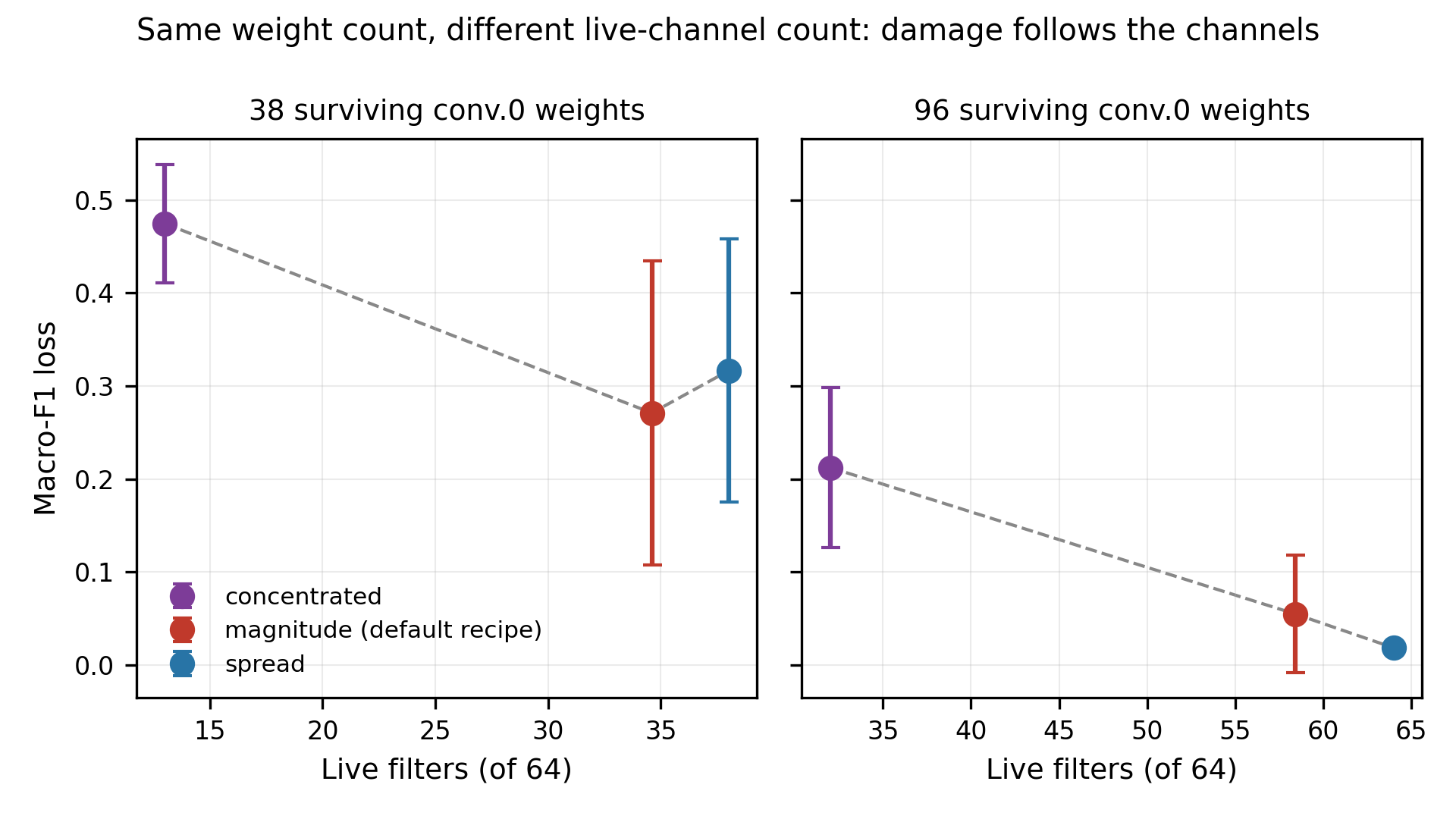}

\emph{Figure 3. Live channels at fixed weight count. Hand-built conv.0 masks with 38 (left) or 96 (right) surviving weights, distributed over few filters (concentrated), by magnitude (the uniform layer-wise recipe), or over many filters (spread).}

\subsection*{5.7 Is there one mediator? Two failed gates}

Does the same quantity mediate in the MLP, where an input weight connects one feature to one unit? Two pre-stated tests said no. A grid that held the surviving input weights fixed (96 or 192) and varied live units K and covered features F in opposite directions found that neither count governs: the loss is U-shaped in both, worst at 96 x 1 (loss 0.51) and 3 x 32 (0.54), best in the middle (0.32 to 0.35), and every block is far worse than magnitude pruning at the same count (0.06 to 0.18); what orders the blocks is the smaller of K and F (Spearman -0.79), the most independent input directions a block can pass on. A test of the effective dimensionality of the first-layer representation (Eq. 9) over all 160 input-manipulated runs found it predicts damage in the MLP (Spearman -0.91) but not in the CNN (-0.47, and +0.27 on the tap masks, where one-weight filters are shifted copies of the input). The first-layer mediator therefore stays architecture-specific: live filters in the CNN, a dimension in the MLP. The penultimate representation, included as a reference measure, correlates with loss at -0.93 in both architectures (within-family -0.94 on the tap masks, -0.72 on the grid), but nothing manipulates it directly, so it is reported as a post-hoc association. Read with Section 4.3 and the head factorial (with the input starved, an 80\%-sparse head loses 0.271 and a dense head 0.226; protected, 0.013 and 0.002), the sparse head is not the mechanism and reduced dimensionality is a correlate of the collapsed state; Section 5.9 identifies what that state is. Full tables in Appendix S4.

\subsection*{5.8 Recipe-induced and intrinsic damage}

The results above separate two components of pruning damage. The allocation-induced component is what is removed when the input layer is protected: on CICIoT2023 it is 0.271 minus 0.013, about 95\% of the total at the 80\% operating point. The residual component is what remains with the input layer intact and 80\% of everything else removed: a loss of 0.013 and two damaged classes, DoS-TCP\_Flood and DDoS-UDP\_Flood, both members of confusable flood-attack sibling pairs. That residual is the loss that remains under this protected recipe, the counterpart of the effect the disparate-impact literature describes, and which classes it hits is not random. Pairwise probes on the pruned representation show that DoS and DDoS attacks of the same protocol are barely separable (pair AUC 0.30 to 0.77) while reconnaissance and vulnerability-scan pairs remain separable (0.80 to 0.92). Starvation decides how much capacity is lost; confusability decides which classes pay for it. The two mechanisms are complementary, and protecting the input layer removes the first without touching the second. Section 6 shows that the split is dataset-dependent, and Section 9 returns to what the residual component does and does not explain.

\subsection*{5.9 The proximate failure: displaced normalisation statistics}

What is wrong with a collapsed model? Its weights are largely sound (Section 4.3) and its sparse head is not the problem (Section 5.7). The answer is in its batch-normalisation statistics. Recomputing every running mean and variance of the fine-tuned uniform models on 200 batches of training data, with no weight changed and no labels used, raises macro-F1 from 0.271 to 0.502, 85\% of the loss; the worst seed goes from 0.009 to 0.519 (Table 8). Batch statistics on shuffled test data give 0.51 in every seed. Any fresh estimate of the statistics repairs the model.

The displacement is specific and small in extent. Recalibrating only the running means recovers the entire gain; only the variances, 0.5\% of it; only the first normalisation layer, 98\%; only the second, 1\% (Appendix S7). Channels whose filter lost every input weight have exactly the right statistics. Among connected channels the median displacement (Eq. 16) is 0.001 standard deviations, the 95th percentile 0.33 and the worst 0.77. A few first-layer channels carry the collapse. Moving the statistics from saved to recalibrated values (Eq. 15) raises macro-F1 almost linearly along the path (41\% of the recovery in the first 30\%, 17\% in the last 30\%): damage is proportional to the displacement, not a threshold.

Recalibration acts where the first layer was starved. It changes the one-shot protected and global models by 0.001, improves the gradual-global model, whose schedule leaves the first layer partly starved (loss 0.029 to 0.008; benign flagged 33.8\% to 23.3\%), and recovers 83 to 92\% of the loss in every starved CNN cell (uniform 85\%, gradual per-layer 83\%, 128-filter uniform 91\%, starved input with a dense head 92\%). It recovers less where starvation also removes accuracy that statistics cannot restore: 44 to 65\% in the perceptron cells and 38\% on TON\_IoT. After recalibration the uniform CNN flags 33.3\% of benign traffic (dense 28.6\%), attributes 67.5\% of attacks exactly (dense 70.5\%) and has one damaged class instead of 17.

These interventions identify normalisation-state mismatch as a major proximal cause of the observed deployment failure. Uniform pruning leaves the first layer 38 weights and about 35 connected filters; after fine-tuning under that starvation a few of those channels carry displaced running means, and the deployed model degrades in proportion. The process linking input-layer allocation and fine-tuning to the displaced means remains incompletely characterised (Section 9.2); a plausible reason is that the momentum estimate is noisy where the rare-class signal is concentrated in few channels. Two low-overhead fixes follow: prevent the starvation by allocation, or repair the statistics with one pass of unlabelled training data. Repairing pruned networks through their normalisation state is also the mechanism behind REFLOW (Section 2.3); what is specific here is the allocation cause and the security consequence.

\begingroup\scriptsize\setlength{\tabcolsep}{3pt}\renewcommand{\arraystretch}{1.05}\begin{longtable}[]{@{}
  >{\raggedright\arraybackslash}p{(\columnwidth - 12\tabcolsep) * \real{0.2350}}
  >{\raggedright\arraybackslash}p{(\columnwidth - 12\tabcolsep) * \real{0.1175}}
  >{\raggedright\arraybackslash}p{(\columnwidth - 12\tabcolsep) * \real{0.1175}}
  >{\raggedright\arraybackslash}p{(\columnwidth - 12\tabcolsep) * \real{0.1175}}
  >{\raggedright\arraybackslash}p{(\columnwidth - 12\tabcolsep) * \real{0.1389}}
  >{\raggedright\arraybackslash}p{(\columnwidth - 12\tabcolsep) * \real{0.1389}}
  >{\raggedright\arraybackslash}p{(\columnwidth - 12\tabcolsep) * \real{0.1346}}@{}}
\toprule\noalign{}
\begin{minipage}[b]{\linewidth}\raggedright
\textbf{Cell}
\end{minipage} & \begin{minipage}[b]{\linewidth}\raggedright
\textbf{Loss before}
\end{minipage} & \begin{minipage}[b]{\linewidth}\raggedright
\textbf{Loss after}
\end{minipage} & \begin{minipage}[b]{\linewidth}\raggedright
\textbf{Recovered}
\end{minipage} & \begin{minipage}[b]{\linewidth}\raggedright
\textbf{Benign flagged before / after}
\end{minipage} & \begin{minipage}[b]{\linewidth}\raggedright
\textbf{Damaged classes before / after}
\end{minipage} & \begin{minipage}[b]{\linewidth}\raggedright
\textbf{Dense benign flagged}
\end{minipage} \\
\begin{minipage}[b]{\linewidth}\raggedright
CNN uniform 80\%
\end{minipage} & \begin{minipage}[b]{\linewidth}\raggedright
0.271
\end{minipage} & \begin{minipage}[b]{\linewidth}\raggedright
0.039
\end{minipage} & \begin{minipage}[b]{\linewidth}\raggedright
85\%
\end{minipage} & \begin{minipage}[b]{\linewidth}\raggedright
56.3\% / 33.3\%
\end{minipage} & \begin{minipage}[b]{\linewidth}\raggedright
17 / 1
\end{minipage} & \begin{minipage}[b]{\linewidth}\raggedright
28.6\%
\end{minipage} \\
\begin{minipage}[b]{\linewidth}\raggedright
CNN first layer protected
\end{minipage} & \begin{minipage}[b]{\linewidth}\raggedright
0.013
\end{minipage} & \begin{minipage}[b]{\linewidth}\raggedright
0.014
\end{minipage} & \begin{minipage}[b]{\linewidth}\raggedright
none
\end{minipage} & \begin{minipage}[b]{\linewidth}\raggedright
30.8\% / 33.0\%
\end{minipage} & \begin{minipage}[b]{\linewidth}\raggedright
2 / 0
\end{minipage} & \begin{minipage}[b]{\linewidth}\raggedright
28.6\%
\end{minipage} \\
\begin{minipage}[b]{\linewidth}\raggedright
CNN global 80\%
\end{minipage} & \begin{minipage}[b]{\linewidth}\raggedright
0.014
\end{minipage} & \begin{minipage}[b]{\linewidth}\raggedright
0.015
\end{minipage} & \begin{minipage}[b]{\linewidth}\raggedright
none
\end{minipage} & \begin{minipage}[b]{\linewidth}\raggedright
34.0\% / 31.1\%
\end{minipage} & \begin{minipage}[b]{\linewidth}\raggedright
0 / 1
\end{minipage} & \begin{minipage}[b]{\linewidth}\raggedright
28.6\%
\end{minipage} \\
\begin{minipage}[b]{\linewidth}\raggedright
CNN gradual per-layer
\end{minipage} & \begin{minipage}[b]{\linewidth}\raggedright
0.165
\end{minipage} & \begin{minipage}[b]{\linewidth}\raggedright
0.028
\end{minipage} & \begin{minipage}[b]{\linewidth}\raggedright
83\%
\end{minipage} & \begin{minipage}[b]{\linewidth}\raggedright
78.2\% / 32.1\%
\end{minipage} & \begin{minipage}[b]{\linewidth}\raggedright
8 / 2
\end{minipage} & \begin{minipage}[b]{\linewidth}\raggedright
28.6\%
\end{minipage} \\
\begin{minipage}[b]{\linewidth}\raggedright
CNN gradual global
\end{minipage} & \begin{minipage}[b]{\linewidth}\raggedright
0.029
\end{minipage} & \begin{minipage}[b]{\linewidth}\raggedright
0.008
\end{minipage} & \begin{minipage}[b]{\linewidth}\raggedright
72\%
\end{minipage} & \begin{minipage}[b]{\linewidth}\raggedright
33.8\% / 23.3\%
\end{minipage} & \begin{minipage}[b]{\linewidth}\raggedright
0 / 0
\end{minipage} & \begin{minipage}[b]{\linewidth}\raggedright
28.6\%
\end{minipage} \\
\begin{minipage}[b]{\linewidth}\raggedright
CNN starved input, dense head
\end{minipage} & \begin{minipage}[b]{\linewidth}\raggedright
0.226
\end{minipage} & \begin{minipage}[b]{\linewidth}\raggedright
0.017
\end{minipage} & \begin{minipage}[b]{\linewidth}\raggedright
92\%
\end{minipage} & \begin{minipage}[b]{\linewidth}\raggedright
72.5\% / 31.8\%
\end{minipage} & \begin{minipage}[b]{\linewidth}\raggedright
16 / 0
\end{minipage} & \begin{minipage}[b]{\linewidth}\raggedright
28.6\%
\end{minipage} \\
\begin{minipage}[b]{\linewidth}\raggedright
CNN 128 filters, uniform 80\%
\end{minipage} & \begin{minipage}[b]{\linewidth}\raggedright
0.191
\end{minipage} & \begin{minipage}[b]{\linewidth}\raggedright
0.017
\end{minipage} & \begin{minipage}[b]{\linewidth}\raggedright
91\%
\end{minipage} & \begin{minipage}[b]{\linewidth}\raggedright
79.9\% / 32.3\%
\end{minipage} & \begin{minipage}[b]{\linewidth}\raggedright
8 / 0
\end{minipage} & \begin{minipage}[b]{\linewidth}\raggedright
28.6\%
\end{minipage} \\
\begin{minipage}[b]{\linewidth}\raggedright
MLP input layer 192 weights
\end{minipage} & \begin{minipage}[b]{\linewidth}\raggedright
0.058
\end{minipage} & \begin{minipage}[b]{\linewidth}\raggedright
0.021
\end{minipage} & \begin{minipage}[b]{\linewidth}\raggedright
65\%
\end{minipage} & \begin{minipage}[b]{\linewidth}\raggedright
22.9\% / 22.2\%
\end{minipage} & \begin{minipage}[b]{\linewidth}\raggedright
4 / 0
\end{minipage} & \begin{minipage}[b]{\linewidth}\raggedright
21.1\%
\end{minipage} \\
\begin{minipage}[b]{\linewidth}\raggedright
MLP input layer 96 weights
\end{minipage} & \begin{minipage}[b]{\linewidth}\raggedright
0.178
\end{minipage} & \begin{minipage}[b]{\linewidth}\raggedright
0.079
\end{minipage} & \begin{minipage}[b]{\linewidth}\raggedright
55\%
\end{minipage} & \begin{minipage}[b]{\linewidth}\raggedright
36.1\% / 26.4\%
\end{minipage} & \begin{minipage}[b]{\linewidth}\raggedright
13 / 5
\end{minipage} & \begin{minipage}[b]{\linewidth}\raggedright
21.1\%
\end{minipage} \\
\begin{minipage}[b]{\linewidth}\raggedright
MLP input layer 38 weights
\end{minipage} & \begin{minipage}[b]{\linewidth}\raggedright
0.323
\end{minipage} & \begin{minipage}[b]{\linewidth}\raggedright
0.182
\end{minipage} & \begin{minipage}[b]{\linewidth}\raggedright
44\%
\end{minipage} & \begin{minipage}[b]{\linewidth}\raggedright
68.8\% / 41.8\%
\end{minipage} & \begin{minipage}[b]{\linewidth}\raggedright
18 / 15
\end{minipage} & \begin{minipage}[b]{\linewidth}\raggedright
21.1\%
\end{minipage} \\
\begin{minipage}[b]{\linewidth}\raggedright
TON\_IoT CNN uniform 80\%
\end{minipage} & \begin{minipage}[b]{\linewidth}\raggedright
0.256
\end{minipage} & \begin{minipage}[b]{\linewidth}\raggedright
0.160
\end{minipage} & \begin{minipage}[b]{\linewidth}\raggedright
38\%
\end{minipage} & \begin{minipage}[b]{\linewidth}\raggedright
not measured
\end{minipage} & \begin{minipage}[b]{\linewidth}\raggedright
not measured
\end{minipage} & \begin{minipage}[b]{\linewidth}\raggedright
not measured
\end{minipage} \\
\midrule\noalign{}
\endhead
\bottomrule\noalign{}
\endlastfoot
\end{longtable}\endgroup

\emph{Table 8. Normalisation recalibration across cells (five seeds each). Loss is macro-F1 loss against each architecture\textquotesingle s own dense baseline; benign flagged is the fraction of benign test records labelled as attack. Damaged classes use the criterion of Section 3.6.}

\section*{6. Generality: the same law on a second dataset}

TON\_IoT differs from CICIoT2023 in every way that could matter: 93,644 records instead of 3.7 million, 30 features, 10 classes, and a weaker split. Five CNN and five MLP baselines were trained on it, and the same two sweeps were run: conv.0 sparsity from 0\% to 95\% with the rest of the CNN at 80\%, and the MLP\textquotesingle s input layer starved to the same absolute counts of surviving weights (192, 96 and 38) with the rest at 80\%. Table 9 shows the result (curves in Appendix S5).

The pattern replicates. The CNN\textquotesingle s loss rises monotonically with conv.0 sparsity (Spearman 1.0 over dose means), and the damage at 38 surviving weights is almost the same as on CICIoT2023: 0.256 against 0.271, with 4 of 10 classes damaged. The MLP collapses at 38 surviving input weights on TON\_IoT as it did on CICIoT2023: 0.353 against 0.323, with 6 of 10 classes damaged, and its dose-response is also monotone. On both datasets and in both architectures the collapse comes at the same absolute count.

One part does not replicate, and it changes the size of the claim rather than its truth. On CICIoT2023 the MLP was intact at plain 80\% pruning (loss 0.007) and the CNN with its first layer protected lost only 0.013. On TON\_IoT the same two models lose 0.085 and 0.065. TON\_IoT\textquotesingle s models have less accuracy to spare (baseline macro-F1 about 0.46 for the CNN and 0.51 for the MLP, with seed-to-seed spreads of 0.05 to 0.08), so removing 80\% of the body is not free there. The share of damage attributable to the input layer at the collapse point is therefore about 75\% on TON\_IoT (0.256 total, of which about 0.065 is not input-related) against about 95\% on CICIoT2023. The input-layer component dominates on both datasets; how completely it dominates depends on how much slack the rest of the network has.

\begingroup\scriptsize\setlength{\tabcolsep}{3pt}\renewcommand{\arraystretch}{1.05}\begin{longtable}[]{@{}
  >{\raggedright\arraybackslash}p{(\columnwidth - 10\tabcolsep) * \real{0.1389}}
  >{\raggedright\arraybackslash}p{(\columnwidth - 10\tabcolsep) * \real{0.1603}}
  >{\raggedright\arraybackslash}p{(\columnwidth - 10\tabcolsep) * \real{0.1709}}
  >{\raggedright\arraybackslash}p{(\columnwidth - 10\tabcolsep) * \real{0.1816}}
  >{\raggedright\arraybackslash}p{(\columnwidth - 10\tabcolsep) * \real{0.0962}}
  >{\raggedright\arraybackslash}p{(\columnwidth - 10\tabcolsep) * \real{0.2521}}@{}}
\toprule\noalign{}
\begin{minipage}[b]{\linewidth}\raggedright
\textbf{Architecture}
\end{minipage} & \begin{minipage}[b]{\linewidth}\raggedright
\textbf{Input-layer sparsity}
\end{minipage} & \begin{minipage}[b]{\linewidth}\raggedright
\textbf{Surviving input weights}
\end{minipage} & \begin{minipage}[b]{\linewidth}\raggedright
\textbf{Macro-F1 (mean +/- sd)}
\end{minipage} & \begin{minipage}[b]{\linewidth}\raggedright
\textbf{Loss}
\end{minipage} & \begin{minipage}[b]{\linewidth}\raggedright
\textbf{Classes damaged (of 10)}
\end{minipage} \\
\begin{minipage}[b]{\linewidth}\raggedright
CNN1D
\end{minipage} & \begin{minipage}[b]{\linewidth}\raggedright
0\%
\end{minipage} & \begin{minipage}[b]{\linewidth}\raggedright
192
\end{minipage} & \begin{minipage}[b]{\linewidth}\raggedright
0.395 +/- 0.066
\end{minipage} & \begin{minipage}[b]{\linewidth}\raggedright
0.065
\end{minipage} & \begin{minipage}[b]{\linewidth}\raggedright
1
\end{minipage} \\
\begin{minipage}[b]{\linewidth}\raggedright
CNN1D
\end{minipage} & \begin{minipage}[b]{\linewidth}\raggedright
50\%
\end{minipage} & \begin{minipage}[b]{\linewidth}\raggedright
96
\end{minipage} & \begin{minipage}[b]{\linewidth}\raggedright
0.341 +/- 0.017
\end{minipage} & \begin{minipage}[b]{\linewidth}\raggedright
0.119
\end{minipage} & \begin{minipage}[b]{\linewidth}\raggedright
1
\end{minipage} \\
\begin{minipage}[b]{\linewidth}\raggedright
CNN1D
\end{minipage} & \begin{minipage}[b]{\linewidth}\raggedright
80\% (default)
\end{minipage} & \begin{minipage}[b]{\linewidth}\raggedright
38
\end{minipage} & \begin{minipage}[b]{\linewidth}\raggedright
0.203 +/- 0.083
\end{minipage} & \begin{minipage}[b]{\linewidth}\raggedright
0.256
\end{minipage} & \begin{minipage}[b]{\linewidth}\raggedright
4
\end{minipage} \\
\begin{minipage}[b]{\linewidth}\raggedright
CNN1D
\end{minipage} & \begin{minipage}[b]{\linewidth}\raggedright
90\%
\end{minipage} & \begin{minipage}[b]{\linewidth}\raggedright
19
\end{minipage} & \begin{minipage}[b]{\linewidth}\raggedright
0.131 +/- 0.083
\end{minipage} & \begin{minipage}[b]{\linewidth}\raggedright
0.328
\end{minipage} & \begin{minipage}[b]{\linewidth}\raggedright
5
\end{minipage} \\
\begin{minipage}[b]{\linewidth}\raggedright
CNN1D
\end{minipage} & \begin{minipage}[b]{\linewidth}\raggedright
95\%
\end{minipage} & \begin{minipage}[b]{\linewidth}\raggedright
10
\end{minipage} & \begin{minipage}[b]{\linewidth}\raggedright
0.060 +/- 0.058
\end{minipage} & \begin{minipage}[b]{\linewidth}\raggedright
0.400
\end{minipage} & \begin{minipage}[b]{\linewidth}\raggedright
5
\end{minipage} \\
\begin{minipage}[b]{\linewidth}\raggedright
MLP
\end{minipage} & \begin{minipage}[b]{\linewidth}\raggedright
80\%
\end{minipage} & \begin{minipage}[b]{\linewidth}\raggedright
1,536
\end{minipage} & \begin{minipage}[b]{\linewidth}\raggedright
0.425 +/- 0.051
\end{minipage} & \begin{minipage}[b]{\linewidth}\raggedright
0.085
\end{minipage} & \begin{minipage}[b]{\linewidth}\raggedright
2
\end{minipage} \\
\begin{minipage}[b]{\linewidth}\raggedright
MLP
\end{minipage} & \begin{minipage}[b]{\linewidth}\raggedright
97.5\%
\end{minipage} & \begin{minipage}[b]{\linewidth}\raggedright
192
\end{minipage} & \begin{minipage}[b]{\linewidth}\raggedright
0.389 +/- 0.046
\end{minipage} & \begin{minipage}[b]{\linewidth}\raggedright
0.121
\end{minipage} & \begin{minipage}[b]{\linewidth}\raggedright
3
\end{minipage} \\
\begin{minipage}[b]{\linewidth}\raggedright
MLP
\end{minipage} & \begin{minipage}[b]{\linewidth}\raggedright
98.75\%
\end{minipage} & \begin{minipage}[b]{\linewidth}\raggedright
96
\end{minipage} & \begin{minipage}[b]{\linewidth}\raggedright
0.356 +/- 0.067
\end{minipage} & \begin{minipage}[b]{\linewidth}\raggedright
0.154
\end{minipage} & \begin{minipage}[b]{\linewidth}\raggedright
3
\end{minipage} \\
\begin{minipage}[b]{\linewidth}\raggedright
MLP
\end{minipage} & \begin{minipage}[b]{\linewidth}\raggedright
99.5\%
\end{minipage} & \begin{minipage}[b]{\linewidth}\raggedright
38
\end{minipage} & \begin{minipage}[b]{\linewidth}\raggedright
0.157 +/- 0.052
\end{minipage} & \begin{minipage}[b]{\linewidth}\raggedright
0.353
\end{minipage} & \begin{minipage}[b]{\linewidth}\raggedright
6
\end{minipage} \\
\midrule\noalign{}
\endhead
\bottomrule\noalign{}
\endlastfoot
\end{longtable}\endgroup

\emph{Table 9. TON\_IoT replication. Upper rows: CNN, conv.0 swept with the rest at 80\%. Lower rows: MLP, input layer starved to matched absolute counts with the rest at 80\%. Both Spearman correlations are 1.0 over dose means. Class counts are out of 10.}

\section*{7. Security consequence}

\subsection*{7.1 Threat model}

The attacker knows one thing about the target: it is a convolutional intrusion detector pruned with the uniform layer-wise recipe, which is what the standard toolkit produces when a practitioner follows its documentation. The attacker does not know the specific trained model. From public models of the same kind, or their own training runs, they can estimate which attack types such detectors predictably fail to attribute, and they choose attack types from that set. They send ordinary traffic of those types, with no feature perturbation and no access to the model. Their goal is attribution failure, not invisibility: the attack is flagged as the wrong attack, or the wrong family, so that triage and response go to the wrong place, and it arrives into an alert stream already flooded with false positives.

\subsection*{7.2 Blind spots are small, predictable, and transferable}

We estimated the attacker\textquotesingle s blind-spot set out of sample. Over all ten ways of choosing three attacker-side seeds from the five, the blind-spot set was defined on attacker-side validation records as in Section 3.8 and evaluated on untouched test records of the two held-out deployed models the attacker never saw. The sets are small (one to five classes) and stable: MITM-ArpSpoofing appears in all ten folds and DoS-HTTP\_Flood in seven; DoS-TCP\_Flood, Mirai-greip\_flood, DDoS-SlowLoris and VulnerabilityScan recur. The random baseline draws uniformly over attack types that the dense models can detect. Two earlier versions are kept in the repository with their gates: one that used only low pruned recall selected the floored tier (classes no model detects) and measured difficulty rather than compression; one that selected on test records and evaluated on the same records gave a larger recipe-conditional ratio (2.8) than the leak-free protocol reported here, and is the version a careful reviewer should discount.

\subsection*{7.3 The advantage comes from the recipe, not from the attacker}

Table 10 gives the result. On identical blind-spot traffic, held-out uniformly pruned detectors misattribute 72.4\% of records, with 47.7\% sent to the wrong attack family; held-out detectors with the first layer protected misattribute 49.6\%, globally pruned ones 51.5\%, and dense ones 47.3\%. The blind-spot classes are hard for every model on the test partition (MITM-ArpSpoofing in particular), so the fair measure is the excess over the dense model (Eq. 12): 25 points for uniform pruning against 2 for protection and 4 for global pruning. The pre-stated ratio criterion (uniform at least twice protected) was not met (1.46) and is reported as such; the excess-over-dense reading is the one the data supports. Attack-to-benign labelling is 18 to 21\% for every recipe including dense, so it is not recipe-induced: the attacker is not gaining evasion, they are gaining misattribution. Selection adds less than the recipe does: the uniformly pruned detector already misattributes 53\% of randomly chosen detectable attack types, because 17 classes are damaged, and choosing the blind spots adds a further 19 points.

\begingroup\scriptsize\setlength{\tabcolsep}{3pt}\renewcommand{\arraystretch}{1.05}\begin{longtable}[]{@{}
  >{\raggedright\arraybackslash}p{(\columnwidth - 10\tabcolsep) * \real{0.1816}}
  >{\raggedright\arraybackslash}p{(\columnwidth - 10\tabcolsep) * \real{0.1603}}
  >{\raggedright\arraybackslash}p{(\columnwidth - 10\tabcolsep) * \real{0.1282}}
  >{\raggedright\arraybackslash}p{(\columnwidth - 10\tabcolsep) * \real{0.1603}}
  >{\raggedright\arraybackslash}p{(\columnwidth - 10\tabcolsep) * \real{0.1603}}
  >{\raggedright\arraybackslash}p{(\columnwidth - 10\tabcolsep) * \real{0.2094}}@{}}
\toprule\noalign{}
\begin{minipage}[b]{\linewidth}\raggedright
\textbf{Held-out detector}
\end{minipage} & \begin{minipage}[b]{\linewidth}\raggedright
\textbf{Blind-spot traffic: misattributed}
\end{minipage} & \begin{minipage}[b]{\linewidth}\raggedright
\textbf{Excess over dense}
\end{minipage} & \begin{minipage}[b]{\linewidth}\raggedright
\textbf{Blind-spot traffic: cross-family}
\end{minipage} & \begin{minipage}[b]{\linewidth}\raggedright
\textbf{Blind-spot traffic: labelled benign}
\end{minipage} & \begin{minipage}[b]{\linewidth}\raggedright
\textbf{Random detectable type: misattributed}
\end{minipage} \\
\begin{minipage}[b]{\linewidth}\raggedright
Uniform layer-wise 80\%
\end{minipage} & \begin{minipage}[b]{\linewidth}\raggedright
72.4\%
\end{minipage} & \begin{minipage}[b]{\linewidth}\raggedright
+25.1
\end{minipage} & \begin{minipage}[b]{\linewidth}\raggedright
47.7\%
\end{minipage} & \begin{minipage}[b]{\linewidth}\raggedright
17.8\%
\end{minipage} & \begin{minipage}[b]{\linewidth}\raggedright
53.5\%
\end{minipage} \\
\begin{minipage}[b]{\linewidth}\raggedright
First layer protected
\end{minipage} & \begin{minipage}[b]{\linewidth}\raggedright
49.6\%
\end{minipage} & \begin{minipage}[b]{\linewidth}\raggedright
+2.3
\end{minipage} & \begin{minipage}[b]{\linewidth}\raggedright
27.3\%
\end{minipage} & \begin{minipage}[b]{\linewidth}\raggedright
19.9\%
\end{minipage} & \begin{minipage}[b]{\linewidth}\raggedright
20.2\%
\end{minipage} \\
\begin{minipage}[b]{\linewidth}\raggedright
Global magnitude 80\%
\end{minipage} & \begin{minipage}[b]{\linewidth}\raggedright
51.5\%
\end{minipage} & \begin{minipage}[b]{\linewidth}\raggedright
+4.2
\end{minipage} & \begin{minipage}[b]{\linewidth}\raggedright
27.9\%
\end{minipage} & \begin{minipage}[b]{\linewidth}\raggedright
20.9\%
\end{minipage} & \begin{minipage}[b]{\linewidth}\raggedright
20.2\%
\end{minipage} \\
\begin{minipage}[b]{\linewidth}\raggedright
Dense (unpruned)
\end{minipage} & \begin{minipage}[b]{\linewidth}\raggedright
47.3\%
\end{minipage} & \begin{minipage}[b]{\linewidth}\raggedright
0
\end{minipage} & \begin{minipage}[b]{\linewidth}\raggedright
26.4\%
\end{minipage} & \begin{minipage}[b]{\linewidth}\raggedright
18.8\%
\end{minipage} & \begin{minipage}[b]{\linewidth}\raggedright
19.5\%
\end{minipage} \\
\midrule\noalign{}
\endhead
\bottomrule\noalign{}
\endlastfoot
\end{longtable}\endgroup

\emph{Table 10. Recipe-conditional evaluation on held-out detectors, blind spots selected on validation records, evaluated on test records, averaged over ten leave-two-seeds-out folds that reuse the same five models. Blind-spot traffic is the same in every row; only the deployed detector\textquotesingle s recipe changes.}

The cross-family numbers depend on the eight-family mapping, so the test-selected evaluation was repeated under three alternative mappings (supplement). Uniform pruning had the highest cross-family error under every tested mapping, while the three others were within a point of each other once DoS and DDoS were merged; the ratio of uniform to protected cross-family rates rose from 3.4 to 7.5 under that merge while the absolute gap shrank from 38 to 11 points, so most of the uniformly pruned recipe\textquotesingle s family-level errors lie between those two families.

\subsection*{7.4 The alert flood}

The second consequence needs no attacker at all. Table 11 gives the fraction of benign test records that each detector flags as an attack. The dense detector flags 28.6\%. The uniformly pruned detector flags 56.3\%, and the seed-to-seed spread of 42.7 points means that some deployments flag more than four benign flows in five while others are near the dense rate: the two fixed recipes bring the rate back to 30.8\% and 34.0\% with the spread collapsed to 6 to 8 points, and recalibrating the uniformly pruned detector\textquotesingle s statistics brings it to 33.3\% (Section 5.9). The study measures false-positive classifications, not operator behaviour; that sustained false alerts at these rates would raise triage thresholds and make misattributed attacks easier to dismiss is a hypothesis about operations, not a measured finding.

\begingroup\scriptsize\setlength{\tabcolsep}{3pt}\renewcommand{\arraystretch}{1.05}\begin{longtable}[]{@{}
  >{\raggedright\arraybackslash}p{(\columnwidth - 2\tabcolsep) * \real{0.3205}}
  >{\raggedright\arraybackslash}p{(\columnwidth - 2\tabcolsep) * \real{0.6795}}@{}}
\toprule\noalign{}
\begin{minipage}[b]{\linewidth}\raggedright
\textbf{Detector}
\end{minipage} & \begin{minipage}[b]{\linewidth}\raggedright
\textbf{Benign records flagged as attack (mean +/- sd over 5 seeds)}
\end{minipage} \\
\begin{minipage}[b]{\linewidth}\raggedright
Uniform layer-wise 80\%
\end{minipage} & \begin{minipage}[b]{\linewidth}\raggedright
56.3\% +/- 42.7\%
\end{minipage} \\
\begin{minipage}[b]{\linewidth}\raggedright
First layer protected
\end{minipage} & \begin{minipage}[b]{\linewidth}\raggedright
30.8\% +/- 6.4\%
\end{minipage} \\
\begin{minipage}[b]{\linewidth}\raggedright
Global magnitude 80\%
\end{minipage} & \begin{minipage}[b]{\linewidth}\raggedright
34.0\% +/- 8.0\%
\end{minipage} \\
\begin{minipage}[b]{\linewidth}\raggedright
Dense (unpruned)
\end{minipage} & \begin{minipage}[b]{\linewidth}\raggedright
28.6\% +/- 4.2\%
\end{minipage} \\
\midrule\noalign{}
\endhead
\bottomrule\noalign{}
\endlastfoot
\end{longtable}\endgroup

\emph{Table 11. Benign traffic flagged as attack, five seeds per recipe.}

\includegraphics[width=0.85\linewidth]{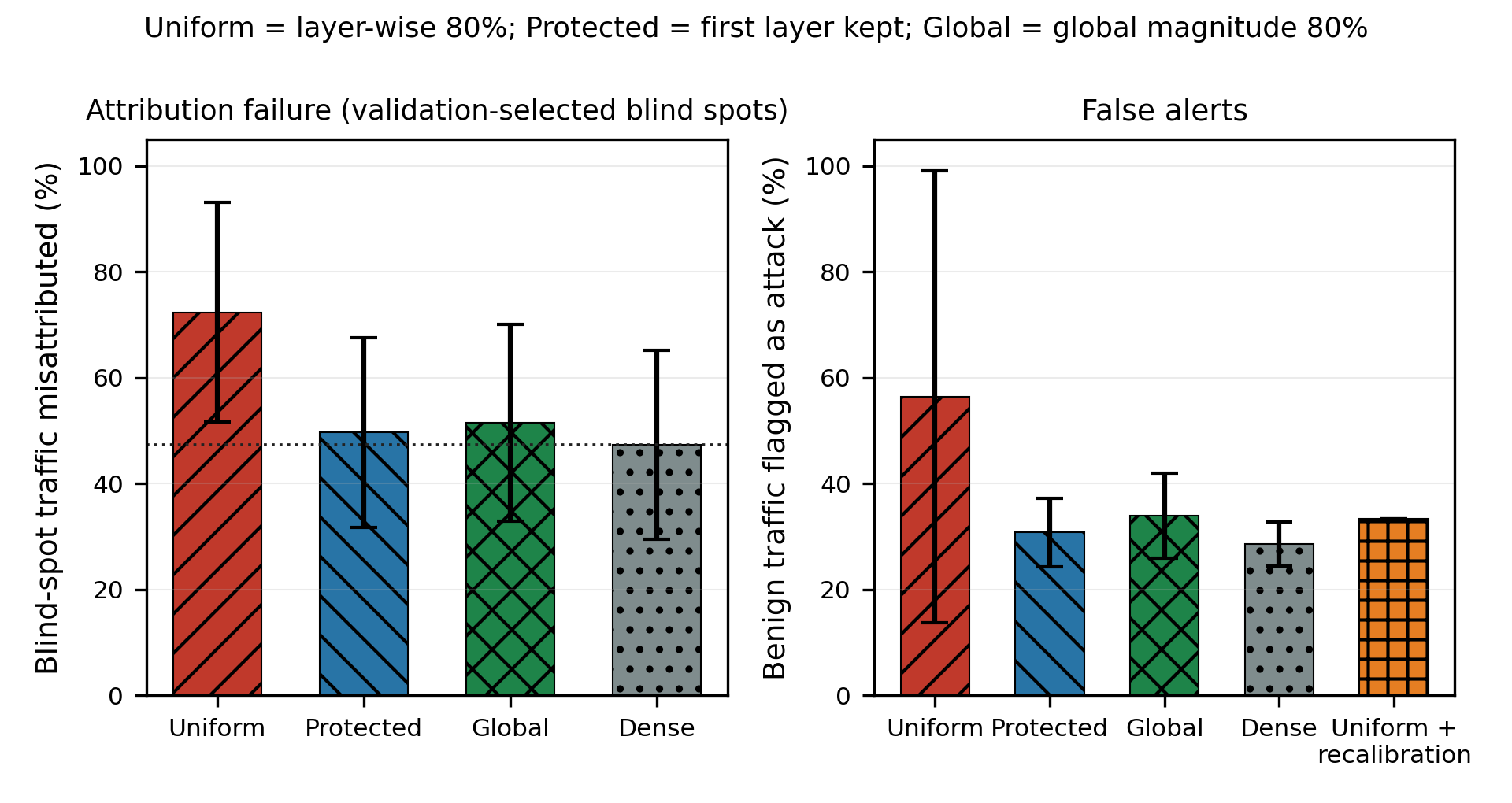}

\emph{Figure 4. Both security consequences by recipe. Left: misattribution of identical validation-selected blind-spot traffic on held-out detectors (mean and standard deviation over ten overlapping folds; dotted line is the dense rate). Right: benign traffic flagged as attack (mean and standard deviation over five seeds), with the uniformly pruned detector after recalibration.}

\subsection*{7.5 What the failure is, and is not}

It is not silent evasion; the collapsed detector misses fewer attacks than the dense one. It is not a vulnerability that needs model access or crafted inputs. It is a predictable, transferable attribution failure plus a raised false-alert rate, both created by the allocation rule rather than by pruning as such. First-layer protection removes almost all of the excess over the dense model (misattribution excess 25 points to 2; benign flagged 56.3\% to 30.8\% against 28.6\% dense), and recalibration of an already deployed model achieves nearly the same (33.3\%), at costs Section 8 measures.

\section*{8. Defence and its cost}

\subsection*{8.1 Two recipes that remove the collapse}

Section 5 established that the damage comes from starving the first layer, that two recipe changes prevent it, and that recalibration repairs it. The preventive changes are one-line edits to the standard toolkit call: the protected recipe skips conv.0 when applying the per-layer rule; the global recipe calls the toolkit\textquotesingle s global function over the same list of layers instead of its per-layer function, and in these networks the single threshold kept 71\% of the small layer without being told to (Section 5.2). The repair is one forward pass of unlabelled training data through the deployed model with its normalisation layers in training mode. None requires a new method, a new hyperparameter, or any knowledge of which classes are at risk.

\subsection*{8.2 What the fixes cost}

Table 12 reports both preventive recipes against uniform pruning and the dense model on the axes a deployment sees. Serialized size is identical in all four cases (125,683 bytes) because the dense format stores zeros. The index-value sparse payload under the convention of Section 3.9 is 49,024 bytes for uniform pruning, 49,488 for the protected recipe (+0.9\%, because its fully dense conv.0 is stored at 4 bytes per weight) and 49,032 for global; under a uniform 8-bytes-per-non-zero convention the protected figure would be 50,256 (+2.5\%). Total sparsity is 78.4\% for the uniform and global recipes and 77.8\% for the protected one. Median CPU latencies on the dense backend differ by less than the run-to-run spread at every batch size; we do not claim equivalence beyond that. Against this, macro-F1 is 0.271 under uniform pruning and 0.529 or 0.527 under the fixes, and the run-to-run standard deviation falls from 0.164 to 0.012 and 0.007. The repair (Section 5.9) applies to a model already pruned and deployed: one pass of about 800,000 unlabelled training records, no weight and no size changed, recovering 0.23 of the 0.27.

\begingroup\scriptsize\setlength{\tabcolsep}{3pt}\renewcommand{\arraystretch}{1.05}\begin{longtable}[]{@{}
  >{\raggedright\arraybackslash}p{(\columnwidth - 12\tabcolsep) * \real{0.1603}}
  >{\raggedright\arraybackslash}p{(\columnwidth - 12\tabcolsep) * \real{0.1442}}
  >{\raggedright\arraybackslash}p{(\columnwidth - 12\tabcolsep) * \real{0.0962}}
  >{\raggedright\arraybackslash}p{(\columnwidth - 12\tabcolsep) * \real{0.1175}}
  >{\raggedright\arraybackslash}p{(\columnwidth - 12\tabcolsep) * \real{0.1442}}
  >{\raggedright\arraybackslash}p{(\columnwidth - 12\tabcolsep) * \real{0.1603}}
  >{\raggedright\arraybackslash}p{(\columnwidth - 12\tabcolsep) * \real{0.1774}}@{}}
\toprule\noalign{}
\begin{minipage}[b]{\linewidth}\raggedright
\textbf{Detector}
\end{minipage} & \begin{minipage}[b]{\linewidth}\raggedright
\textbf{Macro-F1 (mean +/- sd)}
\end{minipage} & \begin{minipage}[b]{\linewidth}\raggedright
\textbf{Total sparsity}
\end{minipage} & \begin{minipage}[b]{\linewidth}\raggedright
\textbf{Serialized bytes}
\end{minipage} & \begin{minipage}[b]{\linewidth}\raggedright
\textbf{Sparse payload estimate (bytes)}
\end{minipage} & \begin{minipage}[b]{\linewidth}\raggedright
\textbf{CPU latency, batch 1 (median ms)}
\end{minipage} & \begin{minipage}[b]{\linewidth}\raggedright
\textbf{CPU latency, batch 1,024 (median ms)}
\end{minipage} \\
\begin{minipage}[b]{\linewidth}\raggedright
Dense (unpruned)
\end{minipage} & \begin{minipage}[b]{\linewidth}\raggedright
0.542 +/- 0.006
\end{minipage} & \begin{minipage}[b]{\linewidth}\raggedright
0.0\%
\end{minipage} & \begin{minipage}[b]{\linewidth}\raggedright
125,683
\end{minipage} & \begin{minipage}[b]{\linewidth}\raggedright
118,920
\end{minipage} & \begin{minipage}[b]{\linewidth}\raggedright
0.260
\end{minipage} & \begin{minipage}[b]{\linewidth}\raggedright
61.9
\end{minipage} \\
\begin{minipage}[b]{\linewidth}\raggedright
Uniform layer-wise 80\%
\end{minipage} & \begin{minipage}[b]{\linewidth}\raggedright
0.271 +/- 0.164
\end{minipage} & \begin{minipage}[b]{\linewidth}\raggedright
78.4\%
\end{minipage} & \begin{minipage}[b]{\linewidth}\raggedright
125,683
\end{minipage} & \begin{minipage}[b]{\linewidth}\raggedright
49,024
\end{minipage} & \begin{minipage}[b]{\linewidth}\raggedright
0.256
\end{minipage} & \begin{minipage}[b]{\linewidth}\raggedright
61.3
\end{minipage} \\
\begin{minipage}[b]{\linewidth}\raggedright
First layer protected
\end{minipage} & \begin{minipage}[b]{\linewidth}\raggedright
0.529 +/- 0.012
\end{minipage} & \begin{minipage}[b]{\linewidth}\raggedright
77.8\%
\end{minipage} & \begin{minipage}[b]{\linewidth}\raggedright
125,683
\end{minipage} & \begin{minipage}[b]{\linewidth}\raggedright
49,488
\end{minipage} & \begin{minipage}[b]{\linewidth}\raggedright
0.266
\end{minipage} & \begin{minipage}[b]{\linewidth}\raggedright
61.4
\end{minipage} \\
\begin{minipage}[b]{\linewidth}\raggedright
Global magnitude 80\%
\end{minipage} & \begin{minipage}[b]{\linewidth}\raggedright
0.527 +/- 0.007
\end{minipage} & \begin{minipage}[b]{\linewidth}\raggedright
78.4\%
\end{minipage} & \begin{minipage}[b]{\linewidth}\raggedright
125,683
\end{minipage} & \begin{minipage}[b]{\linewidth}\raggedright
49,032
\end{minipage} & \begin{minipage}[b]{\linewidth}\raggedright
0.252
\end{minipage} & \begin{minipage}[b]{\linewidth}\raggedright
62.3
\end{minipage} \\
\midrule\noalign{}
\endhead
\bottomrule\noalign{}
\endlastfoot
\end{longtable}\endgroup

\emph{Table 12. Deployment cost of the fixed recipes. Sizes and latency are from the same benchmark settings as the archived deployment table (single CPU thread, 25 warm-up runs, 100 timed repeats). Macro-F1 is the five-seed value from Table 4.}

\subsection*{8.3 A deployment checklist}

For a practitioner compressing a detector of this kind, the results reduce to six checks.

\begin{enumerate}
\def\labelenumi{\arabic{enumi}.}
\item
  Count the weights in the first layer. If it is a single-channel convolution, it is small; 192 in the standard CNN1D.
\item
  Do not apply a per-layer fraction to it. Either exempt it, or use a global magnitude threshold across the network.
\item
  After any pruning and fine-tuning, recompute the normalisation statistics on unlabelled training data before deployment. It is one forward pass and it repairs most of the damage if the first layer was starved.
\item
  After pruning, count first-layer filters that still have at least one input weight, as a fraction of the layer. In the conditions we tested, losses were small above about 85\% connected and large below about 65\%; uniform 80\% pruning leaves about half.
\item
  Evaluate per class, and evaluate attribution, not only detection. A binary attack-versus-benign score can rise while the detector collapses (Section 4.1).
\item
  Report the seed-to-seed spread. A large spread after pruning is itself a sign that the recipe is starving a small layer.
\end{enumerate}

\section*{9. Limitations and scope}

\subsection*{9.1 What the mechanism explains}

The mechanism explains the collapse of small tabular detectors whose input layer is tiny in absolute terms and which are pruned with a uniform layer-wise rule, one-shot or gradual (the gradual result uses a regrowth-enabled variant of the Zhu and Gupta schedule, not their masked-gradient original). It does not explain the disparate-impact results on large image networks, whose first layers have thousands of weights and are left dense by the standard implementation; those describe the residual component, which here is 0.013 with two damaged classes once the input layer is protected. We claim a decomposition, not a replacement. The allocation-induced share is about 95\% on CICIoT2023 and 75\% on TON\_IoT, whose models have less accuracy to spare; the direction and the collapse at 38 surviving weights replicate, the share does not and should not be quoted as a constant.

\subsection*{9.2 What is correlational}

The penultimate effective dimensionality that tracks damage in both architectures is a post-hoc association, and Section 5.9 makes it likely a symptom of the displaced statistics rather than a cause. Why the momentum estimate of the first-layer means ends up displaced after fine-tuning under starvation is offered as an explanation, not measured; a direct test would track per-channel batch means during fine-tuning. Recalibration repairs less in the perceptron (44 to 65\%) and on TON\_IoT (38\%) than in the CNN on CICIoT2023 (83 to 92\%), so starvation there also removes accuracy that statistics cannot restore.

\subsection*{9.3 Pre-stated criteria that were not met}

Eight headline hypothesis groups contained unmet criteria, each reported where it arose (Sections 5.5, 5.7, 5.9, 6 and 7.3); Appendix S9 reports the complete criterion-level record, 82 rows of which 38 were met, 34 were not met and 10 are reference values without a verdict, and maps each group to its constituent tests. They are of three kinds: threshold misses at the noise level (the MLP dose Spearman value, a floating-point epsilon short of 0.90; the TON\_IoT MLP at 0.085 against a bar of 0.05), genuine surprises about mechanism (the absolute-count onset; the $K \times F$ grid; first-layer rank in the CNN; the smooth rather than threshold response to displaced statistics), and criteria met in direction but not size (the threat-model ratio; the recalibration share in the perceptron and on TON\_IoT). Two earlier threat-model versions failed their gates and are retained, and one evaluation was mis-designed (batch statistics on the provenance-ordered test partition) and replaced. None changes the mechanism; all narrow a claim.

\subsection*{9.4 Data, baselines and untested settings}

The transformer is a weaker classifier than the CNN and MLP (validation macro-F1 0.524 to 0.532 against 0.561 to 0.575 and 0.567 to 0.593), so its role is limited to the capacity axis, where its weakness argues against the capacity explanation. TON\_IoT is small and its split follows file order within class, which widens its seed spreads. Both datasets are single-network benchmarks with known labelling limitations; the mechanism should transfer to any detector with a narrow input layer, the damage numbers are benchmark-specific. The width sweep covers three widths of one three-tap CNN family and leaves a gap between 65\% and 85\% connected filters, so the low-loss and high-loss regions are an empirical pattern, not a general threshold. In the threat model the ten folds reuse five models, and the validation-selected blind spots are hard for every model, which is why the excess over dense is reported. The recipe evidence in Section 2.4 rests on the toolkit\textquotesingle s documented behaviour and three papers; a systematic tally would strengthen the premise. Structured pruning and quantisation-aware training were not tested.

\section*{10. Conclusion}

Uniform layer-wise pruning of the standard IoT intrusion detector halves its macro-F1 and doubles its false-alert rate while accuracy falls by only 16 points. The cause is where the pruning lands, not how much: the recipe leaves a 192-weight first layer 38 weights, and after fine-tuning under that starvation a few first-layer normalisation channels carry wrong running means that the deployed model cannot work around. Two other architectures at equal or lower remaining weights do not collapse, and the strong increasing dose-response, the perceptron positive control and the second dataset all agree. Two low-overhead fixes follow. Protecting the first layer, or pruning globally, prevents the collapse at the same size and latency; re-estimating those statistics from unlabelled training data repairs 85\% of it in a model already deployed, without touching a weight. First-layer protection brings both selected-subset misattribution and the benign false-alert rate towards their dense-model values; recalibration recovers most of the uniformly pruned model\textquotesingle s macro-F1 loss and reduces its false-alert rate. Overall accuracy shows the loss but not its class-wise composition: per-class attribution, seed-to-seed spread, the fraction of connected first-layer filters and a recalibration pass before deployment are cheap, and any compressed detector in a security role should be reported with them.

\section*{Availability and disclosures}
\sloppy

Code, executed notebooks, configuration, result tables and pre-stated gate files underlying every reported number are in the public companion repository at \url{https://github.com/anasbiswas1/iot-trust-compression}; the version used for this paper is commit a0559c6a79d09d91c58b6bf8c95a2aa1c08f278f on the main branch (notebooks 30 to 48 produce Sections 4 to 8; notebooks 00 to 13 produce the baseline, probe and calibration results referenced in Section 4). The study uses the publicly available CICIoT2023 and TON\_IoT datasets. This work received no external funding and the author declares no competing interests. AI-based writing assistants were used to support drafting and language editing; the author reviewed and edited all content, verified all reported numbers against the released result tables, and takes full responsibility for the content.

\clearpage
\appendix
\renewcommand{\thetable}{S\arabic{table}}\setcounter{table}{0}
\renewcommand{\thefigure}{S\arabic{figure}}\setcounter{figure}{0}
\section*{Supplementary material}
Every table below is generated from a committed CSV in the companion repository (\texttt{results/tables/comnet/}), named in each caption. Values are five-seed means unless a seed column is present.
\subsection*{S1. Per-class damage by recipe (CICIoT2023, CNN)}
\begingroup\scriptsize\setlength{\tabcolsep}{2.5pt}\renewcommand{\arraystretch}{1.02}
\begin{longtable}{p{0.20\linewidth}p{0.113\linewidth} p{0.113\linewidth} p{0.113\linewidth} p{0.113\linewidth} p{0.113\linewidth} p{0.113\linewidth}}
\caption{Mean recall loss against the dense baseline and fraction of seeds in which the class was materially damaged, for the three one-shot recipes. Source: cnn\_policy\_per\_class\_summary.csv (Notebook 34).}\label{S1}\\
\toprule \textbf{Class} & \textbf{Uniform 80\%: mean recall loss} & \textbf{Uniform 80\%: affected frequency} & \textbf{Protected: mean loss} & \textbf{Protected: affected freq.} & \textbf{Global: mean loss} & \textbf{Global: affected freq.} \\ \midrule
\endfirsthead
\toprule \textbf{Class} & \textbf{Uniform 80\%: mean recall loss} & \textbf{Uniform 80\%: affected frequency} & \textbf{Protected: mean loss} & \textbf{Protected: affected freq.} & \textbf{Global: mean loss} & \textbf{Global: affected freq.} \\ \midrule
\endhead
\bottomrule\endfoot
Mirai-greip\_\allowbreak{}flood & 0.815 & 1.0 & 0.015 & 0.0 & 0.015 & 0.0 \\
DoS-HTTP\_\allowbreak{}Flood & 0.732 & 1.0 & 0.005 & 0.0 & 0.068 & 0.2 \\
DDoS-SlowLoris & 0.506 & 0.8 & 0.021 & 0.0 & 0.022 & 0.0 \\
DoS-TCP\_\allowbreak{}Flood & 0.408 & 0.8 & 0.205 & 0.6 & 0.304 & 0.4 \\
DDoS-PSHACK\_\allowbreak{}Flood & 0.405 & 0.4 & -0.0 & 0.0 & -0.0 & 0.0 \\
Recon-HostDiscovery & 0.39 & 1.0 & -0.027 & 0.0 & -0.004 & 0.0 \\
DoS-UDP\_\allowbreak{}Flood & 0.384 & 0.6 & -0.099 & 0.0 & -0.019 & 0.0 \\
DDoS-HTTP\_\allowbreak{}Flood & 0.383 & 0.6 & 0.015 & 0.0 & -0.024 & 0.0 \\
\end{longtable}\endgroup

\subsection*{S2. Per-class dose-response and the transformer arm}
\begingroup\scriptsize\setlength{\tabcolsep}{2.5pt}\renewcommand{\arraystretch}{1.02}
\begin{longtable}{p{0.20\linewidth}p{0.113\linewidth} p{0.113\linewidth} p{0.113\linewidth} p{0.113\linewidth} p{0.113\linewidth} p{0.113\linewidth}}
\caption{CNN: mean recall loss per class as conv.0 sparsity rises with the rest at 80\%. Source: input\_starvation\_per\_class\_summary.csv (Notebook 35).}\label{S2a}\\
\toprule \textbf{class} & \textbf{conv.0 0\%} & \textbf{conv.0 20\%} & \textbf{conv.0 50\%} & \textbf{conv.0 80\%} & \textbf{conv.0 90\%} & \textbf{conv.0 95\%} \\ \midrule
\endfirsthead
\toprule \textbf{class} & \textbf{conv.0 0\%} & \textbf{conv.0 20\%} & \textbf{conv.0 50\%} & \textbf{conv.0 80\%} & \textbf{conv.0 90\%} & \textbf{conv.0 95\%} \\ \midrule
\endhead
\bottomrule\endfoot
Backdoor\_\allowbreak{}Malware & 0.014 & 0.014 & 0.014 & 0.014 & 0.014 & 0.014 \\
BenignTraffic & 0.021 & 0.018 & 0.089 & 0.277 & 0.49 & 0.695 \\
BrowserHijacking & 0.008 & 0.007 & 0.001 & 0.004 & 0.011 & 0.011 \\
CommandInjection & 0.036 & 0.034 & 0.039 & 0.051 & 0.051 & 0.051 \\
DDoS-ACK\_\allowbreak{}Fragmentation & -0.036 & -0.036 & -0.022 & 0.322 & 0.555 & 0.707 \\
DDoS-HTTP\_\allowbreak{}Flood & 0.015 & -0.001 & 0.089 & 0.383 & 0.641 & 0.811 \\
DDoS-ICMP\_\allowbreak{}Flood & 0.0 & 0.0 & 0.0 & 0.19 & 0.795 & 0.998 \\
DDoS-ICMP\_\allowbreak{}Fragmentation & -0.0 & 0.001 & 0.002 & 0.371 & 0.857 & 0.97 \\
\end{longtable}\endgroup
\begingroup\scriptsize\setlength{\tabcolsep}{2.5pt}\renewcommand{\arraystretch}{1.02}
\begin{longtable}{p{0.20\linewidth}p{0.136\linewidth} p{0.136\linewidth} p{0.136\linewidth} p{0.136\linewidth} p{0.136\linewidth}}
\caption{MLP: mean recall loss per class as the input layer is starved (1,997 / 998 / 192 / 96 / 38 surviving weights) with the rest at 80\%. Source: input\_starvation\_per\_class\_summary.csv (Notebook 35).}\label{S2b}\\
\toprule \textbf{class} & \textbf{input 80.00\%} & \textbf{input 90.00\%} & \textbf{input 98.08\%} & \textbf{input 99.04\%} & \textbf{input 99.62\%} \\ \midrule
\endfirsthead
\toprule \textbf{class} & \textbf{input 80.00\%} & \textbf{input 90.00\%} & \textbf{input 98.08\%} & \textbf{input 99.04\%} & \textbf{input 99.62\%} \\ \midrule
\endhead
\bottomrule\endfoot
Backdoor\_\allowbreak{}Malware & 0.025 & 0.03 & 0.03 & 0.034 & -0.001 \\
BenignTraffic & -0.008 & -0.014 & 0.018 & 0.151 & 0.477 \\
BrowserHijacking & 0.001 & 0.005 & 0.002 & 0.019 & 0.02 \\
CommandInjection & 0.01 & 0.002 & 0.034 & 0.044 & 0.06 \\
DDoS-ACK\_\allowbreak{}Fragmentation & -0.018 & -0.046 & -0.037 & 0.321 & 0.498 \\
DDoS-HTTP\_\allowbreak{}Flood & -0.01 & -0.005 & 0.013 & 0.113 & 0.315 \\
DDoS-ICMP\_\allowbreak{}Flood & 0.0 & 0.0 & 0.002 & 0.002 & 0.007 \\
DDoS-ICMP\_\allowbreak{}Fragmentation & -0.0 & -0.0 & -0.001 & 0.078 & 0.038 \\
\end{longtable}\endgroup
\begingroup\scriptsize\setlength{\tabcolsep}{2.5pt}\renewcommand{\arraystretch}{1.02}
\begin{longtable}{p{0.20\linewidth}p{0.136\linewidth} p{0.136\linewidth} p{0.136\linewidth} p{0.136\linewidth} p{0.136\linewidth}}
\caption{TON\_IoT CNN: mean recall loss per class. Source: ton\_starvation\_per\_class\_summary.csv (Notebook 38).}\label{S2c}\\
\toprule \textbf{class} & \textbf{conv.0 0\%} & \textbf{conv.0 50\%} & \textbf{conv.0 80\%} & \textbf{conv.0 90\%} & \textbf{conv.0 95\%} \\ \midrule
\endfirsthead
\toprule \textbf{class} & \textbf{conv.0 0\%} & \textbf{conv.0 50\%} & \textbf{conv.0 80\%} & \textbf{conv.0 90\%} & \textbf{conv.0 95\%} \\ \midrule
\endhead
\bottomrule\endfoot
backdoor & 0.0 & 0.0 & 0.0 & 0.0 & 0.0 \\
ddos & 0.108 & 0.055 & 0.476 & 0.371 & 0.696 \\
dos & -0.201 & -0.147 & -0.079 & 0.105 & 0.232 \\
injection & 0.161 & 0.228 & 0.397 & 0.237 & 0.42 \\
mitm & 0.05 & 0.045 & 0.194 & 0.197 & 0.137 \\
normal & -0.002 & -0.003 & 0.189 & 0.55 & 0.761 \\
password & -0.105 & -0.138 & -0.318 & 0.23 & 0.02 \\
ransomware & 0.0 & 0.379 & 0.759 & 0.569 & 0.759 \\
scanning & 0.611 & 0.745 & 0.745 & 0.741 & 0.545 \\
xss & -0.095 & 0.016 & 0.378 & 0.244 & 0.389 \\
\end{longtable}\endgroup
\begingroup\scriptsize\setlength{\tabcolsep}{2.5pt}\renewcommand{\arraystretch}{1.02}
\begin{longtable}{p{0.20\linewidth}p{0.136\linewidth} p{0.136\linewidth} p{0.136\linewidth} p{0.136\linewidth} p{0.136\linewidth}}
\caption{FT-Transformer validation gate against the archived CNN and MLP bands. Source: transformer\_validation\_gate.csv (Notebook 32).}\label{S2d1}\\
\toprule \textbf{arch} & \textbf{seed} & \textbf{validation macro f1} & \textbf{cnn band} & \textbf{mlp band} & \textbf{overlaps archived bands} \\ \midrule
\endfirsthead
\toprule \textbf{arch} & \textbf{seed} & \textbf{validation macro f1} & \textbf{cnn band} & \textbf{mlp band} & \textbf{overlaps archived bands} \\ \midrule
\endhead
\bottomrule\endfoot
ft\_\allowbreak{}transformer & 0 & 0.5253 & (0.561, 0.575) & (0.567, 0.593) & False \\
ft\_\allowbreak{}transformer & 1 & 0.5304 & (0.561, 0.575) & (0.567, 0.593) & False \\
ft\_\allowbreak{}transformer & 2 & 0.5321 & (0.561, 0.575) & (0.567, 0.593) & False \\
ft\_\allowbreak{}transformer & 3 & 0.5294 & (0.561, 0.575) & (0.567, 0.593) & False \\
ft\_\allowbreak{}transformer & 4 & 0.5244 & (0.561, 0.575) & (0.567, 0.593) & False \\
\end{longtable}\endgroup
\begingroup\scriptsize\setlength{\tabcolsep}{2.5pt}\renewcommand{\arraystretch}{1.02}
\begin{longtable}{p{0.20\linewidth}p{0.136\linewidth} p{0.136\linewidth} p{0.136\linewidth} p{0.136\linewidth} p{0.136\linewidth}}
\caption{FT-Transformer capacity sweep macro-F1 by cell. Source: transformer\_capacity\_sweep\_macro\_f1\_summary.csv (Notebook 32).}\label{S2d2}\\
\toprule \textbf{cell} & \textbf{n} & \textbf{mean} & \textbf{sd} & \textbf{min} & \textbf{max} \\ \midrule
\endfirsthead
\toprule \textbf{cell} & \textbf{n} & \textbf{mean} & \textbf{sd} & \textbf{min} & \textbf{max} \\ \midrule
\endhead
\bottomrule\endfoot
M0 & 5 & 0.4927 & 0.0091 & 0.4793 & 0.501 \\
prune90 & 5 & 0.4584 & 0.0052 & 0.4496 & 0.4629 \\
prune95 & 5 & 0.4371 & 0.0073 & 0.4255 & 0.4432 \\
prune97 & 5 & 0.4226 & 0.017 & 0.3994 & 0.4398 \\
prune98p5 & 5 & 0.3966 & 0.0177 & 0.3704 & 0.4151 \\
\end{longtable}\endgroup
\begingroup\scriptsize\setlength{\tabcolsep}{2.5pt}\renewcommand{\arraystretch}{1.02}
\begin{longtable}{p{0.28\linewidth}p{0.207\linewidth} p{0.207\linewidth} p{0.207\linewidth}}
\caption{FT-Transformer leakage-safe probes at the matched-capacity cell on the CNN's collapsed classes. Source: transformer\_prune97\_leakage\_safe\_probe.csv.}\label{S2d3}\\
\toprule \textbf{label} & \textbf{auc M0} & \textbf{auc compressed} & \textbf{auc drop} \\ \midrule
\endfirsthead
\toprule \textbf{label} & \textbf{auc M0} & \textbf{auc compressed} & \textbf{auc drop} \\ \midrule
\endhead
\bottomrule\endfoot
BenignTraffic & 0.9439 & 0.9286 & 0.0153 \\
DDoS-SlowLoris & 0.9932 & 0.9853 & 0.0078 \\
DNS\_\allowbreak{}Spoofing & 0.9529 & 0.9332 & 0.0197 \\
DoS-TCP\_\allowbreak{}Flood & 0.9847 & 0.9828 & 0.0019 \\
DoS-UDP\_\allowbreak{}Flood & 0.9915 & 0.9896 & 0.0019 \\
Mirai-greip\_\allowbreak{}flood & 0.9981 & 0.9923 & 0.0058 \\
VulnerabilityScan & 0.9464 & 0.936 & 0.0104 \\
\end{longtable}\endgroup
\begingroup\scriptsize\setlength{\tabcolsep}{2.5pt}\renewcommand{\arraystretch}{1.02}
\begin{longtable}{p{0.20\linewidth}p{0.085\linewidth} p{0.085\linewidth} p{0.085\linewidth} p{0.085\linewidth} p{0.085\linewidth} p{0.085\linewidth} p{0.085\linewidth} p{0.085\linewidth}}
\caption{FT-Transformer head refit at prune97. Source: transformer\_prune97\_head\_refit\_calibration.csv.}\label{S2d4}\\
\toprule \textbf{cell} & \textbf{n} & \textbf{accuracy} & \textbf{macro f1} & \textbf{nll} & \textbf{multiclass brier} & \textbf{ece 10} & \textbf{mean confidence} & \textbf{conf. wrong >0.5} \\ \midrule
\endfirsthead
\toprule \textbf{cell} & \textbf{n} & \textbf{accuracy} & \textbf{macro f1} & \textbf{nll} & \textbf{multiclass brier} & \textbf{ece 10} & \textbf{mean confidence} & \textbf{conf. wrong >0.5} \\ \midrule
\endhead
\bottomrule\endfoot
transformer\_\allowbreak{}prune97\_\allowbreak{}head\_\allowbreak{}refit & 549271 & 0.6255 & 0.4529 & 0.9965 & 0.4572 & 0.0371 & 0.6412 & 0.1436 \\
\end{longtable}\endgroup

\subsection*{S3. Alert-family mapping}
\begingroup\scriptsize\setlength{\tabcolsep}{2.5pt}\renewcommand{\arraystretch}{1.02}
\begin{longtable}{ll}
\caption{Fine-grained class to alert family. Source: ciciot2023\_alert\_family\_mapping.csv (Notebook 09).}\label{S3}\\
\toprule \textbf{fine label} & \textbf{alert family} \\ \midrule
\endfirsthead
\toprule \textbf{fine label} & \textbf{alert family} \\ \midrule
\endhead
\bottomrule\endfoot
Backdoor\_\allowbreak{}Malware & malware\_\allowbreak{}botnet \\
BenignTraffic & benign \\
BrowserHijacking & web\_\allowbreak{}application \\
CommandInjection & web\_\allowbreak{}application \\
DDoS-ACK\_\allowbreak{}Fragmentation & ddos \\
DDoS-HTTP\_\allowbreak{}Flood & ddos \\
DDoS-ICMP\_\allowbreak{}Flood & ddos \\
DDoS-ICMP\_\allowbreak{}Fragmentation & ddos \\
DDoS-PSHACK\_\allowbreak{}Flood & ddos \\
DDoS-RSTFINFlood & ddos \\
DDoS-SYN\_\allowbreak{}Flood & ddos \\
DDoS-SlowLoris & ddos \\
DDoS-SynonymousIP\_\allowbreak{}Flood & ddos \\
DDoS-TCP\_\allowbreak{}Flood & ddos \\
DDoS-UDP\_\allowbreak{}Flood & ddos \\
DDoS-UDP\_\allowbreak{}Fragmentation & ddos \\
DNS\_\allowbreak{}Spoofing & spoofing\_\allowbreak{}mitm \\
DictionaryBruteForce & credential\_\allowbreak{}attack \\
DoS-HTTP\_\allowbreak{}Flood & dos \\
DoS-SYN\_\allowbreak{}Flood & dos \\
DoS-TCP\_\allowbreak{}Flood & dos \\
DoS-UDP\_\allowbreak{}Flood & dos \\
MITM-ArpSpoofing & spoofing\_\allowbreak{}mitm \\
Mirai-greeth\_\allowbreak{}flood & malware\_\allowbreak{}botnet \\
Mirai-greip\_\allowbreak{}flood & malware\_\allowbreak{}botnet \\
Mirai-udpplain & malware\_\allowbreak{}botnet \\
Recon-HostDiscovery & reconnaissance \\
Recon-OSScan & reconnaissance \\
Recon-PingSweep & reconnaissance \\
Recon-PortScan & reconnaissance \\
SqlInjection & web\_\allowbreak{}application \\
Uploading\_\allowbreak{}Attack & malware\_\allowbreak{}botnet \\
VulnerabilityScan & reconnaissance \\
XSS & web\_\allowbreak{}application \\
\end{longtable}\endgroup

\subsection*{S4. Live channels, the K x F grid and effective-rank correlations}
\begingroup\scriptsize\setlength{\tabcolsep}{2.5pt}\renewcommand{\arraystretch}{1.02}
\begin{longtable}{p{0.20\linewidth}p{0.113\linewidth} p{0.113\linewidth} p{0.113\linewidth} p{0.113\linewidth} p{0.113\linewidth} p{0.113\linewidth}}
\caption{Observed live channels under magnitude pruning against the random-survival coverage model (Eq. 7). Source: live\_channel\_coverage\_model.csv (Notebook 36).}\label{S4a}\\
\toprule \textbf{arch} & \textbf{dose} & \textbf{surviving} & \textbf{observed live} & \textbf{expected live random} & \textbf{expected covered features random} & \textbf{observed minus expected} \\ \midrule
\endfirsthead
\toprule \textbf{arch} & \textbf{dose} & \textbf{surviving} & \textbf{observed live} & \textbf{expected live random} & \textbf{expected covered features random} & \textbf{observed minus expected} \\ \midrule
\endhead
\bottomrule\endfoot
cnn1d & 0.0 & 192 & 64.0 & 64.0 &  & 0.0 \\
cnn1d & 0.2 & 154 & 63.6 & 63.54 &  & 0.06 \\
cnn1d & 0.5 & 96 & 58.4 & 56.13 &  & 2.27 \\
cnn1d & 0.8 & 38 & 34.6 & 31.1 &  & 3.5 \\
cnn1d & 0.9 & 19 & 18.6 & 17.26 &  & 1.34 \\
cnn1d & 0.95 & 10 & 10.0 & 9.54 &  & 0.46 \\
mlp & 0.8 & 1997 & 256.0 & 255.96 & 39.0 & 0.04 \\
mlp & 0.9 & 998 & 254.8 & 251.82 & 39.0 & 2.98 \\
mlp & 0.9808 & 192 & 140.4 & 136.13 & 38.75 & 4.27 \\
mlp & 0.9904 & 96 & 80.8 & 80.5 & 35.82 & 0.3 \\
mlp & 0.9962 & 38 & 35.0 & 35.44 & 24.49 & -0.44 \\
\end{longtable}\endgroup
\begingroup\scriptsize\setlength{\tabcolsep}{2.5pt}\renewcommand{\arraystretch}{1.02}
\begin{longtable}{p{0.20\linewidth}p{0.113\linewidth} p{0.113\linewidth} p{0.113\linewidth} p{0.113\linewidth} p{0.113\linewidth} p{0.113\linewidth}}
\caption{MLP K x F grid at fixed surviving input weights. Source: mlp\_grid\_comparison.csv (Notebook 42).}\label{S4b}\\
\toprule \textbf{T} & \textbf{live units K} & \textbf{covered features F} & \textbf{mean macro f1} & \textbf{sd} & \textbf{mean macro f1 loss} & \textbf{classes affected ge3of5} \\ \midrule
\endfirsthead
\toprule \textbf{T} & \textbf{live units K} & \textbf{covered features F} & \textbf{mean macro f1} & \textbf{sd} & \textbf{mean macro f1 loss} & \textbf{classes affected ge3of5} \\ \midrule
\endhead
\bottomrule\endfoot
96 & 96 & 1 & 0.0449 & 0.0305 & 0.5081 & 26 \\
96 & 32 & 3 & 0.1144 & 0.0471 & 0.4386 & 21 \\
96 & 12 & 8 & 0.2 & 0.0736 & 0.3529 & 19 \\
96 & 8 & 12 & 0.1185 & 0.1173 & 0.4344 & 24 \\
96 & 3 & 32 & 0.0174 & 0.0212 & 0.5355 & 26 \\
192 & 192 & 1 & 0.0691 & 0.0454 & 0.4839 & 25 \\
192 & 64 & 3 & 0.14 & 0.054 & 0.413 & 22 \\
192 & 24 & 8 & 0.2334 & 0.0977 & 0.3196 & 19 \\
192 & 16 & 12 & 0.2209 & 0.1411 & 0.3321 & 19 \\
192 & 6 & 32 & 0.0729 & 0.0983 & 0.4801 & 23 \\
\end{longtable}\endgroup
\begingroup\scriptsize\setlength{\tabcolsep}{2.5pt}\renewcommand{\arraystretch}{1.02}
\begin{longtable}{p{0.20\linewidth}p{0.097\linewidth} p{0.097\linewidth} p{0.097\linewidth} p{0.097\linewidth} p{0.097\linewidth} p{0.097\linewidth} p{0.097\linewidth}}
\caption{Spearman correlations between candidate mediators and macro-F1 loss across 160 runs, pooled and by family. Source: effective\_rank\_correlations.csv (Notebook 43).}\label{S4c}\\
\toprule \textbf{runs} & \textbf{surviving input weights} & \textbf{live channels} & \textbf{rep1 participation ratio} & \textbf{rep1 rank99} & \textbf{rep1 numerical rank} & \textbf{penult participation ratio} & \textbf{penult rank99} \\ \midrule
\endfirsthead
\toprule \textbf{runs} & \textbf{surviving input weights} & \textbf{live channels} & \textbf{rep1 participation ratio} & \textbf{rep1 rank99} & \textbf{rep1 numerical rank} & \textbf{penult participation ratio} & \textbf{penult rank99} \\ \midrule
\endhead
\bottomrule\endfoot
pooled & -0.549 & -0.682 & -0.781 & -0.794 & -0.696 & -0.868 & -0.895 \\
cnn1d & -0.871 & -0.893 & -0.468 & -0.525 & -0.771 & -0.868 & -0.926 \\
mlp & -0.705 & -0.758 & -0.911 & -0.931 & -0.86 & -0.9 & -0.929 \\
mlp\_\allowbreak{}grid\_\allowbreak{}only & -0.234 & -0.128 & -0.605 & -0.753 & -0.48 & -0.614 & -0.718 \\
cnn\_\allowbreak{}tap\_\allowbreak{}masks\_\allowbreak{}only & -0.763 & -0.776 & 0.271 & 0.186 & -0.456 & -0.904 & -0.944 \\
\end{longtable}\endgroup

\subsection*{S5. TON\_IoT dose-response}
\begin{figure}[h]\centering\includegraphics[width=0.8\linewidth]{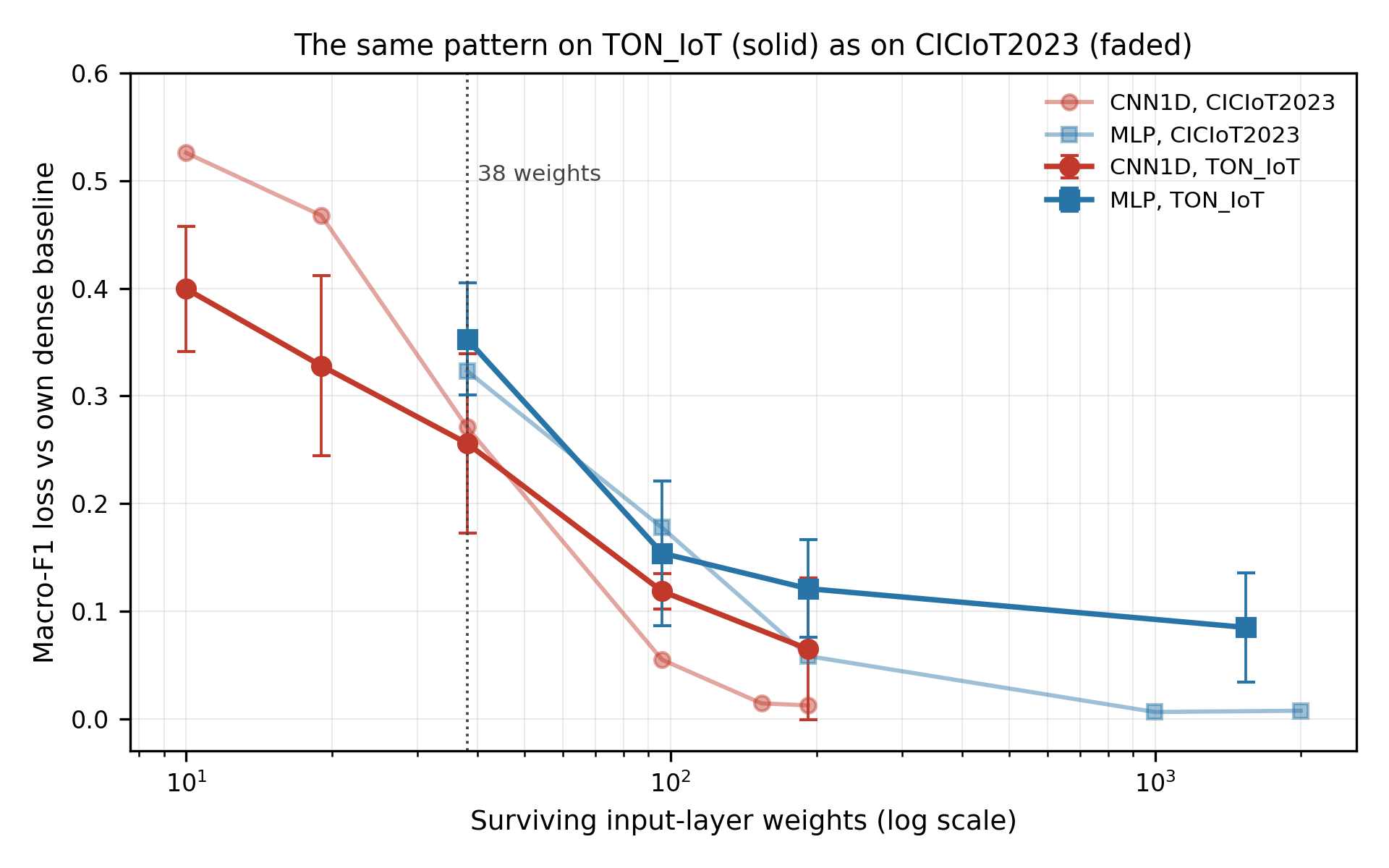}\caption{TON\_IoT dose-response (solid) over the CICIoT2023 curves (faded). Source: ton\_starvation\_dose\_response.csv (Notebook 38).}\end{figure}
\begingroup\scriptsize\setlength{\tabcolsep}{2.5pt}\renewcommand{\arraystretch}{1.02}
\begin{longtable}{p{0.20\linewidth}p{0.113\linewidth} p{0.113\linewidth} p{0.113\linewidth} p{0.113\linewidth} p{0.113\linewidth} p{0.113\linewidth}}
\caption{TON\_IoT dose-response by cell.}\label{S5}\\
\toprule \textbf{arch} & \textbf{input layer sparsity} & \textbf{surviving input weights} & \textbf{mean macro f1} & \textbf{sd macro f1} & \textbf{mean macro f1 loss} & \textbf{classes affected ge3of5} \\ \midrule
\endfirsthead
\toprule \textbf{arch} & \textbf{input layer sparsity} & \textbf{surviving input weights} & \textbf{mean macro f1} & \textbf{sd macro f1} & \textbf{mean macro f1 loss} & \textbf{classes affected ge3of5} \\ \midrule
\endhead
\bottomrule\endfoot
cnn1d & 0.0 & 192 & 0.3945 & 0.0659 & 0.0648 & 1 \\
cnn1d & 0.5 & 96 & 0.3408 & 0.0167 & 0.1185 & 1 \\
cnn1d & 0.8 & 38 & 0.2034 & 0.0832 & 0.256 & 4 \\
cnn1d & 0.9 & 19 & 0.1313 & 0.0834 & 0.3281 & 5 \\
cnn1d & 0.95 & 10 & 0.0598 & 0.0583 & 0.3995 & 5 \\
mlp & 0.8 & 1536 & 0.4248 & 0.051 & 0.0849 & 2 \\
mlp & 0.975 & 192 & 0.3887 & 0.0455 & 0.121 & 3 \\
mlp & 0.9875 & 96 & 0.3558 & 0.0672 & 0.1539 & 3 \\
mlp & 0.9951 & 38 & 0.1569 & 0.0521 & 0.3528 & 6 \\
\end{longtable}\endgroup

\subsection*{S6. Threat model}
\begingroup\scriptsize\setlength{\tabcolsep}{2.5pt}\renewcommand{\arraystretch}{1.02}
\begin{longtable}{p{0.26\linewidth}p{0.46\linewidth}p{0.20\linewidth}}
\caption{Blind-spot sets per attacker-side fold, selected on validation records (reported in the main text). Source: controls\_D\_threat\_val\_selection\_blind\_spots.csv (Notebook 46).}\label{S6a}\\
\toprule \textbf{attacker seeds} & \textbf{n blind} & \textbf{blind spots} \\ \midrule
\endfirsthead
\toprule \textbf{attacker seeds} & \textbf{n blind} & \textbf{blind spots} \\ \midrule
\endhead
\bottomrule\endfoot
(0, 1, 2) & 3 & DoS-HTTP\_\allowbreak{}Flood;\allowbreak{}DoS-TCP\_\allowbreak{}Flood;\allowbreak{}MITM-ArpSpoofing \\
(0, 1, 3) & 2 & DoS-TCP\_\allowbreak{}Flood;\allowbreak{}MITM-ArpSpoofing \\
(0, 1, 4) & 2 & DoS-HTTP\_\allowbreak{}Flood;\allowbreak{}MITM-ArpSpoofing \\
(0, 2, 3) & 2 & DoS-TCP\_\allowbreak{}Flood;\allowbreak{}MITM-ArpSpoofing \\
(0, 2, 4) & 5 & DDoS-SlowLoris;\allowbreak{}DDoS-TCP\_\allowbreak{}Flood;\allowbreak{}DoS-HTTP\_\allowbreak{}Flood;\allowbreak{}DoS-UDP\_\allowbreak{}Flood;\allowbreak{}MITM-ArpSpoofing \\
(0, 3, 4) & 1 & MITM-ArpSpoofing \\
(1, 2, 3) & 5 & DNS\_\allowbreak{}Spoofing;\allowbreak{}DoS-HTTP\_\allowbreak{}Flood;\allowbreak{}DoS-TCP\_\allowbreak{}Flood;\allowbreak{}MITM-ArpSpoofing;\allowbreak{}Mirai-greip\_\allowbreak{}flood \\
(1, 2, 4) & 5 & DDoS-UDP\_\allowbreak{}Flood;\allowbreak{}DoS-HTTP\_\allowbreak{}Flood;\allowbreak{}MITM-ArpSpoofing;\allowbreak{}Mirai-greip\_\allowbreak{}flood;\allowbreak{}VulnerabilityScan \\
(1, 3, 4) & 3 & DoS-HTTP\_\allowbreak{}Flood;\allowbreak{}MITM-ArpSpoofing;\allowbreak{}Mirai-greip\_\allowbreak{}flood \\
(2, 3, 4) & 5 & DDoS-HTTP\_\allowbreak{}Flood;\allowbreak{}DoS-HTTP\_\allowbreak{}Flood;\allowbreak{}MITM-ArpSpoofing;\allowbreak{}Mirai-greip\_\allowbreak{}flood;\allowbreak{}VulnerabilityScan \\
\end{longtable}\endgroup
\begingroup\scriptsize\setlength{\tabcolsep}{2.5pt}\renewcommand{\arraystretch}{1.02}
\begin{longtable}{p{0.28\linewidth}p{0.155\linewidth} p{0.155\linewidth} p{0.155\linewidth} p{0.155\linewidth}}
\caption{Recipe-conditional evaluation on test records, validation-selected blind spots; standard deviations over the ten overlapping folds are in the source file. Source: controls\_D\_threat\_val\_selection\_summary.csv (Notebook 46).}\label{S6b}\\
\toprule \textbf{recipe} & \textbf{strategy} & \textbf{misattribution rate} & \textbf{attack to benign rate} & \textbf{cross family rate} \\ \midrule
\endfirsthead
\toprule \textbf{recipe} & \textbf{strategy} & \textbf{misattribution rate} & \textbf{attack to benign rate} & \textbf{cross family rate} \\ \midrule
\endhead
\bottomrule\endfoot
default\_\allowbreak{}layerwise80 & blind\_\allowbreak{}spot\_\allowbreak{}set & 0.7235 & 0.1776 & 0.4766 \\
default\_\allowbreak{}layerwise80 & random\_\allowbreak{}over\_\allowbreak{}detectable & 0.5345 & 0.0496 & 0.3007 \\
dense & blind\_\allowbreak{}spot\_\allowbreak{}set & 0.473 & 0.1881 & 0.2641 \\
dense & random\_\allowbreak{}over\_\allowbreak{}detectable & 0.1945 & 0.0371 & 0.1252 \\
global80 & blind\_\allowbreak{}spot\_\allowbreak{}set & 0.5148 & 0.2091 & 0.2794 \\
global80 & random\_\allowbreak{}over\_\allowbreak{}detectable & 0.2023 & 0.0395 & 0.1274 \\
protect\_\allowbreak{}conv0 & blind\_\allowbreak{}spot\_\allowbreak{}set & 0.4962 & 0.1985 & 0.2729 \\
protect\_\allowbreak{}conv0 & random\_\allowbreak{}over\_\allowbreak{}detectable & 0.2019 & 0.038 & 0.1278 \\
\end{longtable}\endgroup
\begingroup\scriptsize\setlength{\tabcolsep}{2.5pt}\renewcommand{\arraystretch}{1.02}
\begin{longtable}{p{0.28\linewidth}p{0.207\linewidth} p{0.207\linewidth} p{0.207\linewidth}}
\caption{Test-selected blind-spot sets (leaked version, retained for the record). Source: threat\_model\_v2\_blind\_spot\_sets.csv (Notebook 39b).}\label{S6c}\\
\toprule \textbf{attacker seeds} & \textbf{n detectable} & \textbf{n blind} & \textbf{blind spots} \\ \midrule
\endfirsthead
\toprule \textbf{attacker seeds} & \textbf{n detectable} & \textbf{n blind} & \textbf{blind spots} \\ \midrule
\endhead
\bottomrule\endfoot
(0, 1, 2) & 20 & 3 & DoS-HTTP\_\allowbreak{}Flood;\allowbreak{}DoS-TCP\_\allowbreak{}Flood;\allowbreak{}Recon-HostDiscovery \\
(0, 1, 3) & 19 & 2 & DoS-HTTP\_\allowbreak{}Flood;\allowbreak{}DoS-TCP\_\allowbreak{}Flood \\
(0, 1, 4) & 20 & 1 & DoS-HTTP\_\allowbreak{}Flood \\
(0, 2, 3) & 18 & 4 & DNS\_\allowbreak{}Spoofing;\allowbreak{}DoS-HTTP\_\allowbreak{}Flood;\allowbreak{}DoS-TCP\_\allowbreak{}Flood;\allowbreak{}Recon-HostDiscovery \\
(0, 2, 4) & 19 & 4 & DDoS-SlowLoris;\allowbreak{}DDoS-SynonymousIP\_\allowbreak{}Flood;\allowbreak{}DoS-HTTP\_\allowbreak{}Flood;\allowbreak{}DoS-UDP\_\allowbreak{}Flood \\
(0, 3, 4) & 18 & 1 & DoS-HTTP\_\allowbreak{}Flood \\
(1, 2, 3) & 19 & 4 & DNS\_\allowbreak{}Spoofing;\allowbreak{}DoS-HTTP\_\allowbreak{}Flood;\allowbreak{}DoS-TCP\_\allowbreak{}Flood;\allowbreak{}Mirai-greip\_\allowbreak{}flood \\
(1, 2, 4) & 19 & 4 & DDoS-UDP\_\allowbreak{}Flood;\allowbreak{}DoS-HTTP\_\allowbreak{}Flood;\allowbreak{}Mirai-greip\_\allowbreak{}flood;\allowbreak{}VulnerabilityScan \\
(1, 3, 4) & 19 & 2 & DoS-HTTP\_\allowbreak{}Flood;\allowbreak{}Mirai-greip\_\allowbreak{}flood \\
(2, 3, 4) & 19 & 5 & DDoS-HTTP\_\allowbreak{}Flood;\allowbreak{}DNS\_\allowbreak{}Spoofing;\allowbreak{}DoS-HTTP\_\allowbreak{}Flood;\allowbreak{}Mirai-greip\_\allowbreak{}flood;\allowbreak{}VulnerabilityScan \\
\end{longtable}\endgroup
\begingroup\scriptsize\setlength{\tabcolsep}{2.5pt}\renewcommand{\arraystretch}{1.02}
\begin{longtable}{p{0.20\linewidth}p{0.113\linewidth} p{0.113\linewidth} p{0.113\linewidth} p{0.113\linewidth} p{0.113\linewidth} p{0.113\linewidth}}
\caption{Recipe-conditional evaluation with test-selected blind spots, the ratio the main text discounts. Source: threat\_model\_v2\_summary.csv (Notebook 39b).}\label{S6d}\\
\toprule \textbf{recipe} & \textbf{strategy} & \textbf{misattribution} & \textbf{attack to benign} & \textbf{cross family} & \textbf{wrong family or benign} & \textbf{n types} \\ \midrule
\endfirsthead
\toprule \textbf{recipe} & \textbf{strategy} & \textbf{misattribution} & \textbf{attack to benign} & \textbf{cross family} & \textbf{wrong family or benign} & \textbf{n types} \\ \midrule
\endhead
\bottomrule\endfoot
default\_\allowbreak{}layerwise80 & blind\_\allowbreak{}spot\_\allowbreak{}set & 0.6456 & 0.0342 & 0.5375 & 0.5718 & 3.0 \\
default\_\allowbreak{}layerwise80 & random\_\allowbreak{}over\_\allowbreak{}detectable & 0.5134 & 0.029 & 0.2833 & 0.3123 & 19.0 \\
dense & blind\_\allowbreak{}spot\_\allowbreak{}set & 0.1931 & 0.0161 & 0.1307 & 0.1468 & 3.0 \\
dense & random\_\allowbreak{}over\_\allowbreak{}detectable & 0.1622 & 0.0155 & 0.098 & 0.1135 & 19.0 \\
global80 & blind\_\allowbreak{}spot\_\allowbreak{}set & 0.2599 & 0.0156 & 0.196 & 0.2117 & 3.0 \\
global80 & random\_\allowbreak{}over\_\allowbreak{}detectable & 0.1647 & 0.0153 & 0.1079 & 0.1233 & 19.0 \\
protect\_\allowbreak{}conv0 & blind\_\allowbreak{}spot\_\allowbreak{}set & 0.227 & 0.0146 & 0.1588 & 0.1734 & 3.0 \\
protect\_\allowbreak{}conv0 & random\_\allowbreak{}over\_\allowbreak{}detectable & 0.1634 & 0.0151 & 0.104 & 0.1191 & 19.0 \\
\end{longtable}\endgroup
\begingroup\scriptsize\setlength{\tabcolsep}{2.5pt}\renewcommand{\arraystretch}{1.02}
\begin{longtable}{p{0.20\linewidth}p{0.097\linewidth} p{0.097\linewidth} p{0.097\linewidth} p{0.097\linewidth} p{0.097\linewidth} p{0.097\linewidth} p{0.097\linewidth}}
\caption{First attempt with the low-pruned-recall definition (failed gate). Source: threat\_model\_summary.csv (Notebook 39).}\label{S6e}\\
\toprule \textbf{recipe} & \textbf{strategy} & \textbf{misattribution} & \textbf{attack to benign} & \textbf{cross family} & \textbf{wrong family or benign} & \textbf{n types} & \textbf{exposed lift vs random} \\ \midrule
\endfirsthead
\toprule \textbf{recipe} & \textbf{strategy} & \textbf{misattribution} & \textbf{attack to benign} & \textbf{cross family} & \textbf{wrong family or benign} & \textbf{n types} & \textbf{exposed lift vs random} \\ \midrule
\endhead
\bottomrule\endfoot
default\_\allowbreak{}layerwise80 & exposed\_\allowbreak{}set & 0.7693 & 0.1685 & 0.4592 & 0.6277 & 14.4 & 1.1198 \\
default\_\allowbreak{}layerwise80 & oracle\_\allowbreak{}per\_\allowbreak{}model & 0.9744 & 0.1412 & 0.4658 & 0.607 & 19.8 & 1.1198 \\
default\_\allowbreak{}layerwise80 & random\_\allowbreak{}attack\_\allowbreak{}type & 0.687 & 0.0999 & 0.4052 & 0.505 & 33.0 & 1.1198 \\
dense & exposed\_\allowbreak{}set & 0.6091 & 0.161 & 0.2864 & 0.4474 & 14.4 & 1.343 \\
dense & oracle\_\allowbreak{}per\_\allowbreak{}model & 0.8712 & 0.2238 & 0.4028 & 0.6265 & 10.6 & 1.343 \\
dense & random\_\allowbreak{}attack\_\allowbreak{}type & 0.4535 & 0.0788 & 0.2735 & 0.3523 & 33.0 & 1.343 \\
global80 & exposed\_\allowbreak{}set & 0.6484 & 0.1698 & 0.2974 & 0.4672 & 14.4 & 1.4093 \\
global80 & oracle\_\allowbreak{}per\_\allowbreak{}model & 0.8733 & 0.2107 & 0.4377 & 0.6484 & 11.4 & 1.4093 \\
\end{longtable}\endgroup
\begingroup\scriptsize\setlength{\tabcolsep}{2.5pt}\renewcommand{\arraystretch}{1.02}
\begin{longtable}{p{0.20\linewidth}p{0.136\linewidth} p{0.136\linewidth} p{0.136\linewidth} p{0.136\linewidth} p{0.136\linewidth}}
\caption{Family-mapping sensitivity (test-selected evaluation). Source: mapping\_sensitivity\_summary.csv (Notebook 44).}\label{S6f}\\
\toprule \textbf{mapping} & \textbf{recipe} & \textbf{misattribution rate} & \textbf{attack to benign rate} & \textbf{cross family rate} & \textbf{wrong family or benign rate} \\ \midrule
\endfirsthead
\toprule \textbf{mapping} & \textbf{recipe} & \textbf{misattribution rate} & \textbf{attack to benign rate} & \textbf{cross family rate} & \textbf{wrong family or benign rate} \\ \midrule
\endhead
\bottomrule\endfoot
M0\_\allowbreak{}original & default\_\allowbreak{}layerwise80 & 0.6456 & 0.0342 & 0.5375 & 0.5718 \\
M0\_\allowbreak{}original & dense & 0.1931 & 0.0161 & 0.1307 & 0.1468 \\
M0\_\allowbreak{}original & global80 & 0.2599 & 0.0156 & 0.196 & 0.2117 \\
M0\_\allowbreak{}original & protect\_\allowbreak{}conv0 & 0.227 & 0.0146 & 0.1588 & 0.1734 \\
M1\_\allowbreak{}dos\_\allowbreak{}ddos\_\allowbreak{}merged & default\_\allowbreak{}layerwise80 & 0.6456 & 0.0342 & 0.1234 & 0.1577 \\
M1\_\allowbreak{}dos\_\allowbreak{}ddos\_\allowbreak{}merged & dense & 0.1931 & 0.0161 & 0.0172 & 0.0332 \\
M1\_\allowbreak{}dos\_\allowbreak{}ddos\_\allowbreak{}merged & global80 & 0.2599 & 0.0156 & 0.0209 & 0.0365 \\
M1\_\allowbreak{}dos\_\allowbreak{}ddos\_\allowbreak{}merged & protect\_\allowbreak{}conv0 & 0.227 & 0.0146 & 0.0166 & 0.0311 \\
\end{longtable}\endgroup

\subsection*{S7. Normalisation recalibration}
\begingroup\scriptsize\setlength{\tabcolsep}{2.5pt}\renewcommand{\arraystretch}{1.02}
\begin{longtable}{p{0.14\linewidth}p{0.078\linewidth} p{0.078\linewidth} p{0.078\linewidth} p{0.078\linewidth} p{0.078\linewidth} p{0.078\linewidth} p{0.078\linewidth} p{0.078\linewidth} p{0.078\linewidth}}
\caption{Per-seed macro-F1 and benign-flagged rate before and after recalibration, all cells. Source: recal\_scope\_wide.csv (Notebook 47).}\label{S7a}\\
\toprule \textbf{arch} & \textbf{cell} & \textbf{seed} & \textbf{M0} & \textbf{before} & \textbf{after} & \textbf{loss before} & \textbf{loss after} & \textbf{benign to attack before} & \textbf{benign to attack after} \\ \midrule
\endfirsthead
\toprule \textbf{arch} & \textbf{cell} & \textbf{seed} & \textbf{M0} & \textbf{before} & \textbf{after} & \textbf{loss before} & \textbf{loss after} & \textbf{benign to attack before} & \textbf{benign to attack after} \\ \midrule
\endhead
\bottomrule\endfoot
cnn1d & uniform80 & 0 & 0.5451 & 0.3377 & 0.5047 & 0.2075 & 0.0404 & 0.8215 & 0.3746 \\
cnn1d & uniform80 & 1 & 0.5494 & 0.4052 & 0.475 & 0.1442 & 0.0744 & 0.0532 & 0.3518 \\
cnn1d & uniform80 & 2 & 0.5361 & 0.009 & 0.519 & 0.527 & 0.0171 & 0.9852 & 0.2842 \\
cnn1d & uniform80 & 3 & 0.5406 & 0.3849 & 0.5004 & 0.1558 & 0.0402 & 0.1548 & 0.2919 \\
cnn1d & uniform80 & 4 & 0.5369 & 0.2163 & 0.5123 & 0.3206 & 0.0246 & 0.8021 & 0.3615 \\
cnn1d & protected & 0 & 0.5451 & 0.508 & 0.5148 & 0.0372 & 0.0303 & 0.3947 & 0.3437 \\
cnn1d & protected & 1 & 0.5494 & 0.5286 & 0.5257 & 0.0208 & 0.0237 & 0.3459 & 0.4464 \\
cnn1d & protected & 2 & 0.5361 & 0.5387 & 0.5407 & -0.0027 & -0.0046 & 0.2279 & 0.2409 \\
\end{longtable}\endgroup
\begingroup\scriptsize\setlength{\tabcolsep}{2.5pt}\renewcommand{\arraystretch}{1.02}
\begin{longtable}{p{0.28\linewidth}p{0.155\linewidth} p{0.155\linewidth} p{0.155\linewidth} p{0.155\linewidth}}
\caption{Diagnosis on the uniform CNN: saved statistics, batch statistics on the ordered test partition (an artefact; see Table~\ref{S7f}), cumulative and EMA recalibration. Source: recal\_diagnosis\_uniform.csv (Notebook 47).}\label{S7b}\\
\toprule \textbf{seed} & \textbf{saved running stats} & \textbf{batch stats on test} & \textbf{recal cumulative} & \textbf{recal ema 0.1} \\ \midrule
\endfirsthead
\toprule \textbf{seed} & \textbf{saved running stats} & \textbf{batch stats on test} & \textbf{recal cumulative} & \textbf{recal ema 0.1} \\ \midrule
\endhead
\bottomrule\endfoot
0 & 0.3377 & 0.0502 & 0.5047 & 0.512 \\
1 & 0.4052 & 0.0389 & 0.475 & 0.4891 \\
2 & 0.009 & 0.0454 & 0.519 & 0.5193 \\
3 & 0.3849 & 0.0501 & 0.5004 & 0.5077 \\
4 & 0.2163 & 0.0408 & 0.5123 & 0.5098 \\
\end{longtable}\endgroup
\begingroup\scriptsize\setlength{\tabcolsep}{2.5pt}\renewcommand{\arraystretch}{1.02}
\begin{longtable}{p{0.26\linewidth}p{0.46\linewidth}p{0.20\linewidth}}
\caption{Batch statistics on shuffled test batches. Source: sensitivity\_shuffled\_batchstats.csv (Notebook 48).}\label{S7f}\\
\toprule \textbf{seed} & \textbf{saved running stats} & \textbf{batch stats shuffled test} \\ \midrule
\endfirsthead
\toprule \textbf{seed} & \textbf{saved running stats} & \textbf{batch stats shuffled test} \\ \midrule
\endhead
\bottomrule\endfoot
0 & 0.3377 & 0.5126 \\
1 & 0.4052 & 0.5114 \\
2 & 0.009 & 0.5155 \\
3 & 0.3849 & 0.5006 \\
4 & 0.2163 & 0.5122 \\
\end{longtable}\endgroup
\begingroup\scriptsize\setlength{\tabcolsep}{2.5pt}\renewcommand{\arraystretch}{1.02}
\begin{longtable}{p{0.26\linewidth}p{0.46\linewidth}p{0.20\linewidth}}
\caption{Gain from recalibrating one layer or one moment at a time. Source: sensitivity\_partial\_recal.csv (Notebook 48).}\label{S7c}\\
\toprule \textbf{variant} & \textbf{mean} & \textbf{std} \\ \midrule
\endfirsthead
\toprule \textbf{variant} & \textbf{mean} & \textbf{std} \\ \midrule
\endhead
\bottomrule\endfoot
all & 0.2317 & 0.1771 \\
bn0\_\allowbreak{}only & 0.2281 & 0.1778 \\
bn1\_\allowbreak{}only & 0.0033 & 0.0089 \\
means\_\allowbreak{}only & 0.2342 & 0.1728 \\
vars\_\allowbreak{}only & 0.0011 & 0.0054 \\
\end{longtable}\endgroup
\begingroup\scriptsize\setlength{\tabcolsep}{2.5pt}\renewcommand{\arraystretch}{1.02}
\begin{longtable}{p{0.26\linewidth}p{0.46\linewidth}p{0.20\linewidth}}
\caption{Macro-F1 along the interpolation path from saved (t = 0) to recalibrated (t = 1) statistics (Eq. 15). Source: sensitivity\_interpolation.csv (Notebook 48).}\label{S7d}\\
\toprule \textbf{t} & \textbf{mean} & \textbf{std} \\ \midrule
\endfirsthead
\toprule \textbf{t} & \textbf{mean} & \textbf{std} \\ \midrule
\endhead
\bottomrule\endfoot
0.0 & 0.2706 & 0.1636 \\
0.1 & 0.3032 & 0.1531 \\
0.2 & 0.333 & 0.1352 \\
0.3 & 0.3654 & 0.1169 \\
0.4 & 0.3955 & 0.0981 \\
0.5 & 0.4288 & 0.0583 \\
0.6 & 0.4494 & 0.0401 \\
0.7 & 0.463 & 0.0226 \\
0.8 & 0.4778 & 0.0153 \\
0.9 & 0.4962 & 0.015 \\
1.0 & 0.5023 & 0.0168 \\
\end{longtable}\endgroup
\begingroup\scriptsize\setlength{\tabcolsep}{2.5pt}\renewcommand{\arraystretch}{1.02}
\begin{longtable}{p{0.20\linewidth}p{0.136\linewidth} p{0.136\linewidth} p{0.136\linewidth} p{0.136\linewidth} p{0.136\linewidth}}
\caption{Per-channel displacement summary (Eq. 16) by normalisation layer. Source: sensitivity\_per\_channel\_stats.csv (Notebook 48).}\label{S7e}\\
\toprule \textbf{bn layer} & \textbf{channels} & \textbf{abs mean shift median} & \textbf{abs mean shift p95} & \textbf{abs mean shift max} & \textbf{var ratio median} \\ \midrule
\endfirsthead
\toprule \textbf{bn layer} & \textbf{channels} & \textbf{abs mean shift median} & \textbf{abs mean shift p95} & \textbf{abs mean shift max} & \textbf{var ratio median} \\ \midrule
\endhead
\bottomrule\endfoot
0 & 320 & 0.0008 & 0.3264 & 0.7675 & 1.0 \\
1 & 640 & 0.0058 & 0.0295 & 0.0812 & 0.9933 \\
\end{longtable}\endgroup

\subsection*{S8. Deployment benchmark}
\begingroup\scriptsize\setlength{\tabcolsep}{2.5pt}\renewcommand{\arraystretch}{1.02}
\begin{longtable}{p{0.20\linewidth}p{0.113\linewidth} p{0.113\linewidth} p{0.113\linewidth} p{0.113\linewidth} p{0.113\linewidth} p{0.113\linewidth}}
\caption{Parameters, non-zero parameters, serialized bytes and sparse payload per recipe and seed. Source: defence\_cost\_sizes.csv (Notebook 40).}\label{S8b}\\
\toprule \textbf{recipe} & \textbf{seed} & \textbf{parameters} & \textbf{nonzero parameters} & \textbf{whole model sparsity} & \textbf{torch state dict bytes} & \textbf{index value sparse payload estimate bytes} \\ \midrule
\endfirsthead
\toprule \textbf{recipe} & \textbf{seed} & \textbf{parameters} & \textbf{nonzero parameters} & \textbf{whole model sparsity} & \textbf{torch state dict bytes} & \textbf{index value sparse payload estimate bytes} \\ \midrule
\endhead
\bottomrule\endfoot
dense & 0 & 29730 & 29730 & 0.0 & 125683 & 118920 \\
dense & 1 & 29730 & 29730 & 0.0 & 125683 & 118920 \\
dense & 2 & 29730 & 29730 & 0.0 & 125683 & 118920 \\
dense & 3 & 29730 & 29730 & 0.0 & 125683 & 118920 \\
dense & 4 & 29730 & 29730 & 0.0 & 125683 & 118920 \\
default\_\allowbreak{}layerwise80 & 0 & 29730 & 6433 & 0.7836 & 125683 & 49024 \\
default\_\allowbreak{}layerwise80 & 1 & 29730 & 6433 & 0.7836 & 125683 & 49024 \\
default\_\allowbreak{}layerwise80 & 2 & 29730 & 6433 & 0.7836 & 125683 & 49024 \\
\end{longtable}\endgroup
\begingroup\scriptsize\setlength{\tabcolsep}{2.5pt}\renewcommand{\arraystretch}{1.02}
\begin{longtable}{p{0.20\linewidth}p{0.136\linewidth} p{0.136\linewidth} p{0.136\linewidth} p{0.136\linewidth} p{0.136\linewidth}}
\caption{CPU latency and throughput per recipe, seed and batch size (single thread, 25 warm-ups, 100 repeats). Source: defence\_cost\_cpu\_benchmark.csv (Notebook 40).}\label{S8a}\\
\toprule \textbf{recipe} & \textbf{seed} & \textbf{batch size} & \textbf{median latency ms} & \textbf{p95 latency ms} & \textbf{throughput flows per s} \\ \midrule
\endfirsthead
\toprule \textbf{recipe} & \textbf{seed} & \textbf{batch size} & \textbf{median latency ms} & \textbf{p95 latency ms} & \textbf{throughput flows per s} \\ \midrule
\endhead
\bottomrule\endfoot
dense & 0 & 1 & 0.248 & 0.284 & 4037.949 \\
dense & 0 & 32 & 2.176 & 2.428 & 14705.399 \\
dense & 0 & 256 & 16.226 & 17.745 & 15777.345 \\
dense & 0 & 1024 & 61.48 & 65.926 & 16655.904 \\
dense & 1 & 1 & 0.258 & 0.422 & 3879.322 \\
dense & 1 & 32 & 2.351 & 2.662 & 13611.238 \\
dense & 1 & 256 & 15.681 & 17.026 & 16325.464 \\
dense & 1 & 1024 & 61.76 & 67.882 & 16580.275 \\
\end{longtable}\endgroup

\subsection*{S9. Pre-stated gates and notebook map}
\begingroup\scriptsize\setlength{\tabcolsep}{2.5pt}\renewcommand{\arraystretch}{1.02}
\begin{longtable}{p{0.28\linewidth}p{0.207\linewidth} p{0.207\linewidth} p{0.207\linewidth}}
\caption{Every pre-stated criterion in notebooks 30 to 48 with its value and whether it was met, as written by each notebook's verdict file.}\label{S9}\\
\toprule \textbf{Notebook} & \textbf{Criterion} & \textbf{Value} & \textbf{Met} \\ \midrule
\endfirsthead
\toprule \textbf{Notebook} & \textbf{Criterion} & \textbf{Value} & \textbf{Met} \\ \midrule
\endhead
\bottomrule\endfoot
30 & T1\_\allowbreak{}compensation\_\allowbreak{}pairs & 2.0 of 4 & False \\
30 & T2\_\allowbreak{}exact\_\allowbreak{}p & 0.5333 of 105 & False \\
30 & T3\_\allowbreak{}within\_\allowbreak{}pair\_\allowbreak{}share & 0.5145 of 1 & False \\
31 & i\_\allowbreak{}mean\_\allowbreak{}prune80\_\allowbreak{}macro\_\allowbreak{}f1\_\allowbreak{}loss\_\allowbreak{}gt\_\allowbreak{}0.15 & 0.0187 & False \\
31 & ii\_\allowbreak{}classes\_\allowbreak{}affected\_\allowbreak{}in\_\allowbreak{}ge3of5\_\allowbreak{}seeds\_\allowbreak{}ge\_\allowbreak{}4 & 1.0 & False \\
31 & iii\_\allowbreak{}min\_\allowbreak{}probe\_\allowbreak{}auc\_\allowbreak{}collapsed\_\allowbreak{}ge\_\allowbreak{}0.85 & 0.9902 & True \\
31 & iv\_\allowbreak{}refit\_\allowbreak{}recovers\_\allowbreak{}ge\_\allowbreak{}half\_\allowbreak{}of\_\allowbreak{}loss & 0.4927 & False \\
32 & matched\_\allowbreak{}cell & prune97 &  \\
\end{longtable}\endgroup
\begingroup\scriptsize\setlength{\tabcolsep}{2.5pt}\renewcommand{\arraystretch}{1.02}
\begin{longtable}{p{0.26\linewidth}p{0.46\linewidth}p{0.20\linewidth}}
\caption{Where each manuscript table, figure and section draws its numbers (figure sources are listed here rather than in the captions).}\label{MAP}\\
\toprule \textbf{Manuscript item} & \textbf{Committed table(s)} & \textbf{Notebook(s)} \\ \midrule
\endfirsthead
\toprule \textbf{Manuscript item} & \textbf{Committed table(s)} & \textbf{Notebook(s)} \\ \midrule
\endhead
\bottomrule\endfoot
Table 1, 2 & Methods & Notebook 34 structural check;\allowbreak{} src/\allowbreak{}models.py \\
Table 3, Fig. 1 & capacity\_\allowbreak{}axis\_\allowbreak{}combined.csv & 31, 32, 33 \\
Table 4 & cnn\_\allowbreak{}policy\_\allowbreak{}comparison.csv, gradual\_\allowbreak{}comparison.csv & 34, 41 \\
Table 5, Fig. 2 & input\_\allowbreak{}starvation\_\allowbreak{}dose\_\allowbreak{}response.csv & 35 \\
Table 6 & filter\_\allowbreak{}sweep\_\allowbreak{}dose\_\allowbreak{}response.csv, controls\_\allowbreak{}A\_\allowbreak{}f128\_\allowbreak{}80\_\allowbreak{}wide.csv & 45, 46 \\
Table 7, Fig. 3 & tap\_\allowbreak{}distribution\_\allowbreak{}comparison.csv & 37 \\
Table 8 & recal\_\allowbreak{}scope\_\allowbreak{}summary.csv, recal\_\allowbreak{}ton\_\allowbreak{}uniform.csv & 47 \\
Table 9 & ton\_\allowbreak{}starvation\_\allowbreak{}dose\_\allowbreak{}response.csv & 38 \\
\end{longtable}\endgroup

\end{document}